\documentclass{aa}  
\usepackage{amsmath}
\usepackage[normalem]{ulem}
\usepackage{graphicx}
\usepackage{txfonts}
\usepackage[usenames, dvipsnames]{xcolor}
\usepackage[colorlinks=true,allcolors=blue]{hyperref}
\usepackage[thinc]{esdiff}

\usepackage[varvw]{newtxmath}

\titlerunning{Multimessenger view of PTA black holes using Horizon-AGN}
\authorrunning{Quelquejay Leclere et al.}

\def\kms{\rm\, km\,s^{-1}}
\def\hagn{\mbox{{\sc \small Horizon-AGN}}}
\def\msun{\,{\rm M_\odot}}
\def\kms{\,{\rm km\,s^{-1}}}
\def\Egw{E_{\mathrm{GW}}}
\def\forb{f_{\mathrm{orb}}}
\def\fgw{f_{\mathrm{GW}}}
\def\Lgwcirc{L_{\mathrm{GW}}^{\mathrm{(circ)}}}
\def\Ayr{A_{\mathrm{yr}}}
\def\resPSD{S_{\rm r}^{(\mathrm{GWB})}}
\def\fyr{f_{\mathrm{yr}}}
\def\GWRMS{\rho_{\rm r}^{(\mathrm{GWB})}}
\newcommand{\Nensemb}[2]{\mathcal{N}_{#1}\left(#2\right)}
\def\Vsim{V_{\mathrm{sim}}}
\def\Dlnf{\Delta \ln f}
\def\Tobs{T_{\rm obs}}

\begin{document}

   \title{
   % The multi-messenger view of a pulsar timing array: Black holes with the Horizon-AGN simulation
   The multi-messenger view of pulsar timing array black holes with the Horizon-AGN simulation
   }

   \author{Hippolyte Quelquejay Leclere\inst{1}
          \and
          Kunyang Li\inst{2}
          \and
          Marta Volonteri\inst{2}
          \and Stanislav Babak\inst{1}
          \and Ricarda S. Beckmann \inst{4}
          \and Yohan Dubois\inst{2}
          \and Clotilde Laigle\inst{2}
          \and Natalie A. Webb\inst{3}
          }

   \institute{Universit\'e Paris Cit\'e, CNRS, Astroparticule et Cosmologie, F-75013 Paris, France\\
              \email{quelquejay@apc.in2p3.fr}
         \and
             Institut d’Astrophysique de Paris, UMR 7095, CNRS and Sorbonne Universit\'e, 98 bis boulevard Arago, 75014 Paris, France
        \and
            IRAP, Universit\'e de Toulouse, CNRS, UPS, CNES, 9 Avenue du Colonel Roche, BP 44346, 31028 Toulouse Cedex 4, France
        % \and
        %     LERMA, Sorbonne Universit\'e, Observatoire de Paris, PSL research university, CNRS, 75014 Paris, France
        \and
            Institute for Astronomy, University of Edinburgh, Royal Observatory, Edinburgh EH9 3HJ, UK
            }
 
  \abstract
   {
   We used the $\hagn$ cosmological simulation to study the properties of supermassive black hole binaries (MBHBs) with the largest contribution to the gravitational wave background (GWB) signal expected for the pulsar timing array (PTA) band. 
   We developed a pipeline to generate realistic populations of MBHBs, which enabled us to estimate both the characteristic strain and GWB time series observable by PTA experiments.
   We identified potential continuous wave (CW) candidates standing above the background noise, using toy PTA sensitivities representing the current EPTA and future SKA. We estimated the probability of detecting at least one CW with a signal-to-noise ratio of S/N $>3$ to be $4\%$ ($20\%$) for EPTA (SKA)-like sensitivities, assuming a ten-year baseline. 
   We found that the GWB is dominated by hundreds to thousands of binaries at redshifts in the range $0.05-1$, with chirp masses of $10^{8.5} - 10^{9.5}\,\msun$, primarily hosted in quiescent massive galaxies residing in halos of mass $\sim 10^{13}\,\msun$. 
   CW candidates have larger masses and lower redshifts, and they tend to be found in even more massive halos, typical of galaxy groups and clusters. 
   The majority of these systems would appear as active galactic nuclei (AGNs), rather than quasars, because of their low Eddington ratios.
   Nevertheless, CW candidates with $f_{\rm Edd} > 10^{-3}$ can still outshine their hosts, particularly in radio and X-ray bands, suggesting that they  could serve as the most promising route for identification. 
   Our findings imply that optical and near-infrared (NIR) searches based on light curve variability are challenging and biased toward more luminous systems. 
   Finally, we highlight important caveats in the common method used to compare PTA observations with theoretical models. 
   We find that GWB spectral inferences used by PTAs could be biased toward shallower slopes and higher amplitudes at $f=1/\rm yr$, thereby reducing the apparent tension between astrophysical expectations and PTA observations.
    }


   \maketitle

\section{Introduction}

The origin and evolution of supermassive black holes (MBHs, 10$^6$ -- 10$^{10}$ $\msun$) remain unclear. A number of MBHs have been detected very early in the evolution of the Universe, up to $z\sim 10-11$ \citep{2024Natur.627...59M,2024NatAs...8..126B}, while the peak of MBH activity occurs at $z\sim 2-3$ \citep{2004MNRAS.354L..37M} and many local galaxies host quiescent MBHs \citep{1998Natur.395A..14R}. Their evolution must have started at early cosmic times, plausibly from  intermediate mass black holes (10$^2$ -- 10$^{5}$ M$_\odot$), growing through long or repeated periods of sub- (and perhaps even super-) Eddington accretion and/or via multiple mergers, such as \cite{2021NatRP...3..732V}. Whilst mHz gravitational wave (GW) detections with {\it LISA} \citep{2017arXiv170200786A} will probe the existence and mergers of massive black holes out to z$\sim$20, the importance of MBH mergers in the Local Universe is currently being probed by the pulsar timing array (PTA) experiments \citep{1990ApJ...361..300F}, which are able to detect GW signals in the 1--100 nHz frequency range.

Recently, various PTA experiments have presented the first observational evidence for a GW signal (NANOGrav, \citealt{2023ApJ...951L...8A}; EPTA \& InPTA, \citealt{2023A&A...678A..50E}; PPTA, \citealt{2023ApJ...951L...6R}; CPTA, \citealt{2023RAA....23g5024X}; and MeerKAT, \citealt{10.1093/mnras/stae2571}). 
This signal shows both spatial and temporal correlations among pulsars, consistent with those expected from a stochastic GW background
 \citep{1983ApJ...265L..39H}. 
 The low signal-to-noise ratio (S/N) of the detection makes the interpretation difficult, but one possibility is the incoherent superposition of GWs from a population of massive black hole binaries (MBHBs) with separations on the milliparsec scale. 
 Indeed, inference of the GW background spectrum using PTA data confirms that the observed signal is broadly consistent with expectations from the MBHB population \citep{2023ApJ...952L..37A, 2024A&A...685A..94E}, although the data tend to favor a larger amplitude than what has been theoretically predicted. 
 One possibility is that the signal emanates from individually resolvable MBHBs that are particularly close and heavy, therefore standing above the background signal \citep[e.g.,][]{2017NatAs...1..886M, 2023ApJ...951L..50A}.
 However, so far, this scenario is not supported by current PTA datasets \citep{2024A&A...690A.118E,2023ApJ...951L..50A}.
 Meanwhile, unmodeled noise or overly agnostic priors for the noise parameters could also explain part of the tension \citep{2022MNRAS.516..410Z,Goncharov:2024htb,2025MNRAS.537L...1V}.
 In this work, we show that an overly direct comparison between the theoretical power spectrum predicted by cosmological simulations \citep{2017MNRAS.464.3131K,2022MNRAS.511.5241S,2025ApJ...991L..19C} or semi-analytical models \citep{2003ApJ...590..691W, 2014ApJ...789..156M, 2018MNRAS.477.2599B} and the data-inferred power spectra could lead to misleading conclusions about potential tensions between the two. 
 In particular, the impact of the finite duration of the datasets on spectral inference should be considered more carefully\footnote{{After the submission of this paper, \citet{2025arXiv250613866C} also investigated the impact of improperly accounting for data windowing in PTA spectral inference.}}.

 In the absence of a stronger GW detection, complementary data are necessary to understand the origin of the signal. 
 Cosmological simulations can be used to understand the prevalence and nature of MBHBs in the Universe. These would also help  in identifying which of these would be detectable with the PTAs and contribute to the GW background and/or be detectable as individual objects. 
 Electromagnetic observations of these objects would help confirm the nature of the GW sources, but simulations can help us better understand where to search and provide the best strategy for the wavelength ranges to target \citep[see e.g.][]{2025CQGra..42b5021C}. With such prior knowledge, the overall challenging multi-messenger detection of MBHBs using PTAs \citep{2022MNRAS.510.5929C, 2024ApJ...976..129P} could be facilitated \citep{2021ApJ...921..178L, 2025arXiv250401074T}.

In this paper, we use the Horizon-AGN cosmological simulation to identify and study the MBHBs that contribute most to the GW signal in the PTA band, either by contributing to the background signal or as resolvable continuous wave sources.
We study the properties of these MBHBs along with those of their host galaxies.
We determine the expected electromagnetic emission from radio to X-ray wavelengths, which we validate by comparing with the properties of MBHB candidates. We also identify the types of searches that would have the best chance of detecting the electromagnetic counterparts of the PTA sources.
Finally, we  investigate the commonly used methodology for comparing PTA observations with the predicted characteristic strain signal, highlighting its potential limitations when used to interpret the origin of the observed signal.

The paper is organized as follows.
Section~\ref{HAGN} presents the \hagn~ simulation used in our study.
In Sect.~\ref{catalogs}, we describe how we built our catalog of merging MBHBs from the simulation and modeled their sub-parsec dynamics in post-processing.
In Sect.~\ref{GWB}, we present how we generated realizations of the inspiralling MBHB population and inferred the associated characteristic strain signal.
We also present the properties of the most contributing binaries and estimate their number in the PTA band.
In Sect.~\ref{PTA-implications}, we compare the standard methodology used to convert the GW strain signal into timing residuals measured by PTAs, with a more accurate method, and highlight the potential limitations of the former.
Using two toy PTA sensitivities, we also identified potentially resolvable individual binaries, for which we analyzed their properties and those of their host galaxies in Sect.~\ref{MBH_gal_props}, along with their electromagnetic properties, and evaluate their potential detectability in Sect.~\ref{EM}.
We summarize our findings in Sect.~\ref{Ccl}.

\section{The Horizon-AGN simulation}\label{HAGN}

\hagn~\citep{Dubois14} is a hydrodynamical cosmological simulation with a large volume (142 comoving Mpc)$^3$, assuming a standard $\Lambda$ cold dark matter ($\Lambda$CDM) cosmology  with total matter density of $\Omega_{\rm m}=0.272$, dark energy density of $\Omega_{\Lambda}=0.728$, baryon density of $ \Omega_{\rm b}=0.045$, Hubble constant of $H_{0}=70.4 \rm \ km\,s^{-1}\, Mpc^{-1}$, and an amplitude of the matter power spectrum and power-law index of the primordial power spectrum of $\sigma_8=0.81$ and $n_{\rm s}=0.967,$ respectively. In this simulation, refinement is permitted down to $\Delta x=1\,\rm kpc$, the stellar particle mass is $2\times 10^6 \msun$, the dark matter particle mass is $8\times 10^7 \msun$, and the MBH seed mass is $10^5 \msun$. 

The gas in \hagn~has an equation of state for an ideal monoatomic gas with an adiabatic index of $\gamma_{\rm ad}=5/3$. 
Gas cooling can be modeled down to a floor temperature of $10^4\, \rm K$ using the curves from \cite{sutherland&dopita93}. Heating of the gas from a uniform UV background starts after redshift $z_{\rm reion} = 10,$ following the prescription in \cite{haardt&madau96}. In regions that exceed a gas hydrogen number density threshold of $n_0=0.1\, \rm H\, cm^{-3}$,  star formation is triggered in a Poisson random process \citep{rasera&teyssier06, dubois&teyssier08winds}, following the Schmidt relation with a constant star formation efficiency of $\varepsilon_*=0.02$ \citep{kennicutt98, krumholz&tan07}. A mechanical energy injection from Type Ia supernovae (Type Ia SNe), Type II SNe, and stellar winds is included assuming a \citet{1955ApJ...121..161S} initial mass function (IMF) with cutoffs at $0.1 \msun$  and $100 \msun$. 

Massive black holes and their feedback follow the numerical implementation of~\cite{2012MNRAS.420.2662D}. They are seeded in cells where the gas density is larger than $n_0$ and where the gas velocity dispersion is larger than $100\, \rm km\,s^{-1}$. The MBH seeding is stopped at $z=1.5$. To avoid formation of multiple MBHs in the same galaxy, an exclusion radius of 50 comoving kpc is imposed in the seeding process. To compensate for the inability to capture the multiphase nature of the interstellar gas, the MBH accretion rate is set to be the Bondi-Hoyle-Littleton rate multiplied by a factor $\alpha_{\rm b}=(n/n_0)^2$ when $n>n_0$ and $\alpha_{\rm b}=1$ otherwise \citep{Booth2009}. The effective accretion rate onto MBHs is capped at the Eddington luminosity with a radiative efficiency of 0.1. 15\% of the MBH emitted energy is isotropically coupled to the gas within $4\Delta x$ as thermal energy for luminosities above 1\% of the Eddington luminosity. If the 1\% of the Eddington luminosity threshold is not reached, then the feedback takes a mechanical form instead, with 100\% of the power injected into a bipolar cylindrical jet with a radius of $\Delta x$ and a height of $2 \Delta x$, at a velocity of $10^4\,\rm km\, s^{-1}$.

Instead of constantly repositioning MBHs at the minimum of the local potential, which causes unnatural dynamics \citep{2015MNRAS.451.1868T},  gas dynamical friction is exerted on the MBH to avoid spurious motions due to finite force resolution effects (see \citealp{Dubois2013} for additional details). The boost factor used here is the same as the boost factor $\alpha_{\rm b}$ for accretion. The equation of gas dynamical friction is $F_{\rm DF}= f_{\rm gas} 4 \pi \alpha_{\rm b} \rho_{\rm gas} (G M_{\rm BH}/\bar c_s)^2$, where $G$ is the gravitational constant, $\rho_{\rm gas}$ is the mass-weighted mean gas density within a sphere with a radius of $4 \, \Delta x$, and $f_{\rm gas}$ is a factor function of the Mach number, ${\mathcal M}=\bar u/\bar c_s$, which accounts for the extension and shape of the wake \citep{Ostriker1999}. Finally, $f_{\rm gas}$ is in a range between 0 and 2 for an assumed Coulomb logarithm of 3 \citep{chaponetal13,lescaudronetal23}. 

In \hagn~the dark matter halos and galaxies are identified with the AdaptaHOP halo finder \citep{aubertetal04}. The density field used in AdaptaHOP is smoothed over 20 particles. The identification density threshold is set to be 178 times the average total matter density. On top of the density criteria, either 50 dark matter particles or 50 star particles are required for a dark matter halo or galaxy identification. The centers of halos and galaxies are estimated using the shrinking sphere approach proposed by \cite{poweretal03}.

\section{Selection and modeling of MBH binaries in Horizon-AGN}
\label{catalogs}

We generated a catalog of MBH mergers using the post-processed simulation data. To find merging MBHs, we searched through the data of all MBHs at each coarse time step ($\sim 0.6-0.7$ Myr) of the \hagn~simulation. We note that this time step is much shorter than the  time step over which a full output is saved, which is $\sim 150$ Myr. Each MBH was assigned  a unique ID number when seeded. This ID is carried by the MBH through the simulation until it numerically merges with another MBH at a separation of $4 \Delta x$. The merged MBH inherits the ID of the primary black hole in the merger, while the secondary MBH ID is erased after the merger. By searching for disappearing MBH IDs, we can find the corresponding MBH mergers. Possible spurious numerical mergers are filtered out by selecting only MBH pairs within $\max(2R_{\rm eff},4\Delta x)$ from the galaxy centers, where $R_{\rm eff}$ is the effective radius of the galaxy measured as its projected two-dimensional (2D) radius containing half of the galaxy stellar mass. 
This check was performed for the primary MBH at the outputs before the merger and for the merged MBH at the output after the merger. The reason is that the MBHs are likely to move relative to the galaxy center in the $150$ Myr between outputs.

Although numerical mergers occur when the separation of two MBHs is about $4\,{\rm kpc}$ in $\hagn$, physical mergers actually take place later on when the separation is of the order of the black hole's gravitational radius. 
We define `delayed mergers' as the outcome of adding delays on top of the numerical mergers calculated in post-processing. We followed the methodology described in \cite{2020MNRAS.498.2219V}, which we briefly summarize here. 

First, we added a dynamical friction phase from the position of the  MBHs when they are numerically merged down to the point when they become gravitationally bound. The dynamical friction timescale is estimated assuming the MBH is in an isothermal sphere, considering only the stellar component of the galaxy and including a factor of $0.3$ to account for typical orbits being noncircular,
\begin{equation}
t_{\rm df}=0.67 \left(\frac{d}{4\, {\rm kpc}}\right)^2 \left(\frac{\sigma_{\star}}{100 \, \kms} \right) \left(\frac{M_{\rm BH}}{10^8 \, \msun}\right)^{-1}\Lambda^{-1}\, {\rm Gyr},
\label{eq:tdf}
\end{equation}
where $M_{\rm BH}$ is the black hole mass, $\sigma_{\star}$ is the central stellar velocity dispersion approximated as $(0.25GM_{\star}/R_{\rm eff})^{1/2}$, $\Lambda=\ln(1+M_{\star}/M_{\rm BH})$, with $M_{\star}$ being the total stellar mass of the galaxy hosting the MBH at the output before the numerical merger and $d$ its distance from the galactic center. As confirmed by \cite{2020ApJ...896..113L}, \cite{2020ApJ...905..123L}, and \cite{2022MNRAS.510..531C}, the stellar dynamical friction dominates over the dynamical friction from gas or that from dark matter. We calculated the dynamical friction delay time of both MBHs in a pair and used the longer, which is normally associated with $M_2$.

We also accounted for the shrinking of the binary orbit until coalescence via stellar hardening, viscous torques in a circumbinary disc, and emission of gravitational waves, following \cite{2015MNRAS.454L..66S} and \cite{2015MNRAS.448.3603D}. The binary evolution timescale to coalescence was taken to be the minimum between the two following equations:
\begin{align}
&t_{\rm bin,h}=1.5 \times 10^{-3} \left(\frac{\sigma_{\rm inf}}{\rm 100 \,km\,s^{-1}}\right)\nonumber\\
&\times \left(\frac{\rho_{\rm inf}}{10^6 \msun \,{\rm pc}^{-3}}\right)^{-1}\left(\frac{a_{\rm gw}}{10^{-3}\,{\rm pc}}\right)^{-1} \, {\rm Gyr}, 
\label{eq:tbinh}
\end{align}
and
\begin{equation}
t_{\rm bin,d}=1.5\times10^{-2} \,\varepsilon_{0.1} \,f_{\rm Edd}^{-1} \frac{q}{(1+q)^2}\ln\left(\frac{a_{\rm i}}{a_{\rm c}}\right) \, {\rm Gyr}. 
\label{eq:tbind}
\end{equation}
Here, $q=M_{2}/M_{1}$ is the mass ratio of the MBHB. 
In Eq. \ref{eq:tbinh}, $\sigma_{\rm inf}$ and $\rho_{\rm inf}$ are the velocity dispersion and stellar density at the sphere of influence, defined as the sphere containing twice the binary mass in stars: 
\begin{equation}
    r_{\rm inf}=R_{\rm eff}\left(\frac{4M_{\rm 12}}{M_{\rm gal}}\right),
    \label{eq:rinf}
\end{equation}
where $M_{12}$ is the total mass of the binary and 
\begin{equation}
\rho_{\rm inf}=\frac{M_{\rm gal}r_{\rm inf}^{-2}}{8\pi R_{\rm eff}},
\label{eq:rhoinf}
\end{equation}
where we continue to assume a singular isothermal sphere power-law density profile with slope $-2$, and $a_{\rm gw}$ is the separation at which the binary spends most of the time \citep[see][]{2015MNRAS.454L..66S}:
\begin{multline}
a_{\rm gw}=3.87\times 10^{-3}  \, \times \\ 
\left[ \frac{q}{(1+q)^2} \left(\frac{M_{\rm 12}}{10^8 \msun}\right)^3  \left(\frac{\sigma_{\rm inf}}{\rm 100 \,km\,s^{-1}}\right)
\left(\frac{\rho_{\rm inf}}{10^6 \msun \,{\rm pc}^{-3}}\right)^{-1} \right]^{1/5}  \, {\rm pc}. 
\label{eq:agw}
\end{multline}
{The spatial resolution of the Horizon-AGN simulation does not provide reliable information on the density profile at such small orbital separations. However, it was demonstrated in \citep{2020MNRAS.498.2219V} that the choice of a single isothermal sphere is in very good agreement with the densities measured in observations of local galaxies. }
In Eq. \ref{eq:tbind}, $\varepsilon_{0.1}$ is the radiative efficiency normalized to 0.1. We followed \cite{2015MNRAS.448.3603D} in selecting $a_{\rm i}=G M_{12}/2\sigma_{\star}^2$ and $a_{\rm c}=1.9\times10^{-3} (M_{12}/10^8 \msun)^{3/4}\, {\rm pc}$. 

To model the eccentricity evolution in post-processing, we first assigned an eccentricity to each binary when reaching the influence radius, $r_{\rm inf}$,
according to the eccentricity distribution shown in the right panel of Fig. 6 in \cite{2020ApJ...896..113L}.  The eccentricity is then evolved under loss-cone scattering, viscous drag from the circumbinary disk, and GW emission, from the influence radius to the final coalescence. The orbit hardening and eccentricity evolution due to the loss-cone scattering is described by
\begin{equation}
\label{eq:LC1}
\left(\frac{{\rm d}f_{\rm orb}}{{\rm d}t}\right )_{\mathrm{LC}}=\frac{3G^{4/3}}{2(2\pi)^{2/3}}\frac{H \rho_{\rm inf}}{\sigma_{\rm inf}}M_{\rm 12}^{1/3}f_{\rm orb}^{1/3}
\end{equation}
and
\begin{equation}
\label{eq:LC2}
\left(\frac{{\rm d}e}{{\rm d}t}\right )_{\mathrm{LC}}=\frac{G^{4/3}}{(2\pi)^{2/3}}\frac{HK \rho_{\rm inf}}{\sigma_{\rm inf}}M_{\rm 12}^{1/3}f_{\rm orb}^{-2/3},
\end{equation}
where $f_{\rm orb}$ is the orbital frequency, and $H$ and $K$ are numerical factors resulting from the three-body scattering experiments \citep{Q1996, Sesana2006}.

The viscous drag due to the circumbinary disk is included when the separation between the binary is below $1\,\rm pc$. \cite{Haiman2009} described how the binary orbit embedded in a
circumbinary \citep{SS1973} $\alpha$-disk evolves due to viscous drag and how this evolution depends on the different physical conditions within the disk.

According to \cite{Haiman2009}, there are different regimes for an MBHB in a gap-opened, $\alpha$-disk depending on: (1) whether the radiation pressure or gas pressure balance the vertical gravity ($r^{\rm gas/rad}$); (2) whether the opacity is dominated by electron scattering or free-free absorption ($r^{\rm es/ff}$); (3) whether the binary is massive enough compared to the local disk mass ($M_{\rm 2}$-dominated or disk-dominated). The characteristic radii are defined as
\begin{equation}
\label{eq:r_gas_rad}
r^{\rm gas/rad}=5.15\times  10^{\rm 2}\,M_7^{\rm 2/21} R_{\rm sch}
\end{equation}
and 
\begin{equation}
\label{eq:r_es_ff}
r^{\rm es/ff}=4.10\times 10^{\rm 3} R_{\rm sch}\, ,
\end{equation}
where $M_{\rm 7}$ is the binary mass in units of $10^{\rm 7} \, {\msun}$ and $R_{\rm sch} = 2G M_{\rm 12}/c^{\rm 2}$ is the Schwarzschild radius corresponding to the binary mass (we denote the speed of light as $c$ in the following).

According to \cite{Shapiro1983}, a disk can be divided into three regions: (i) inner region ($r<r^{\rm gas/rad}$) with radiation-dominated pressure and electron scattering-dominated opacity; (ii) middle region ($r^{\rm es/ff}>r>r^{\rm gas/rad}$) with gas-dominated pressure and electron scattering-dominated opacity; and (iii) outer region ($r>r^{\rm es/ff}$) with gas-dominated pressure and free-free scattering-dominated opacity. Within each of those three region, there are two possibilities: 1) an $M_{\rm 2}$-dominated region ($r<r^{\rm \nu /s}$) and 2) a disk-dominated region ($r>r^{\rm \nu /s}$). The $r^{\rm \nu /s}$ in three regions of an $\alpha$-disk are defined in \cite{Haiman2009} as
\begin{eqnarray}
\label{eq:r_v_s1}
r_{\rm in}^{\rm \nu/s}=3.61\times 10^{\rm 3} \,M_{\rm 7}^{\rm -2/7}\, q_{\rm s}^{\rm 2/7}\, R_{\rm sch}\; &{\rm if}& r\lesssim r^{\rm gas/rad}\, , \\
\label{eq:r_v_s2}r_{\rm mid}^{\rm \nu/s}=1.21\times 10^{\rm 5}\,M_{\rm 7}^{\rm -6/7}\,q_{\rm s}^{\rm 5/7}\,  R_{\rm sch}\; &{\rm if}& r^{\rm gas/rad}\lesssim r\lesssim r^{\rm es/ff}\, , \\
\label{eq:r_v_s3}r_{\rm out}^{\rm \nu/s}=1.82\times 10^{\rm 5}\,M_{\rm 7}^{\rm -24/25}\,q_{\rm s}^{\rm 4/5}\,  R_{\rm sch}\; &{\rm if}& r\gtrsim r^{\rm es/ff}\, .
\end{eqnarray}
Thus, there are six regimes and their orbital frequency evolution rates are listed below.

\noindent (1) Disk-dominated, inner region:
%%%%%%%%%%%%%%%%%%%%%% inner region %%%%%%%%%%%%%%
\begin{equation}
\label{eq:gas_drag1} \left ( \frac{{\rm d} f_{\rm orb}}{{\rm d} t} \right)_{\mathrm{VD}}=6.0 \times 10^{-8} M_{\rm 7}^{-2} r_{\rm 3}^{-5} \,\rm yr^{-2}
\end{equation}
if $r_{\rm in}^{\rm \nu/s}<r<r^{\rm gas/rad}$;

\noindent(2) $M_{\rm 2}$-dominated, inner region:
\begin{equation}
\label{eq:gas_drag2}
\left ( \frac{{\rm d} f_{\rm orb}}{{\rm d} t} \right)_{\mathrm{VD}}=2.8 \times 10^{-7} M_{\rm 7}^{-13/8} r_{\rm 3}^{-59/16} q_{\rm s}^{-3/8} \,\rm yr^{-2}
\end{equation}
if $r<r^{\rm gas/rad}$ and $r<r_{\rm in}^{\rm \nu/s}$;

\noindent(3) Disk-dominated, middle region:
%%%%%%%%%%%%%%%%%%%%%% mid region %%%%%%%%%%%%%%
\begin{equation}
\label{eq:gas_drag}
\left ( \frac{{\rm d} f_{\rm orb}}{{\rm d} t} \right)_{\mathrm{VD}}=2.9 \times 10^{-5} M_{\rm 7}^{-11/5} r_{\rm 3}^{-29/10} \,\rm yr^{-2}\, ,
\end{equation}
if $r^{\rm gas/rad}<r<r^{\rm es/ff}$ and $r>r_{\rm mid}^{\rm \nu/s}$;
where $r_{\rm 3}$ is the orbital semi-major axis in units of $10^{\rm 3}R_{\rm sch}$, $q_{\rm s}=4q/(1+q)^{\rm 2}$ is the symmetric mass ratio. We note that this prescription implies that the MBHB orbit always shrinks under the influence of viscous drag, especially in the presence of stellar hardening suggested by some most recent simulations \citep{Cuadra2009, Roedig2012, Bortolas2021, Franchini2021,Amaro2023}

\noindent(4) $M_{\rm 2}$-dominated, middle region:
\begin{equation}
\label{eq:gas_drag4}
\left ( \frac{{\rm d} f_{\rm orb}}{{\rm d} t} \right)_{\mathrm{VD}}=2.3 \times 10^{-6} M_{\rm 7}^{-7/4} r_{\rm 3}^{-19/8} q_{\rm s}^{-3/8} \,\rm yr^{-2}\, ,
\end{equation}
if $r^{\rm gas/rad}<r<r^{\rm es/ff}$ and $r\leq r_{\rm mid}^{\rm \nu/s}$;

%%%%%%%%%%%%%%%%%%%%%% outer region %%%%%%%%%%%%%%
\noindent(5) Disk-dominated, outer region: 
\begin{equation}
\label{eq:gas_drag5}
\left ( \frac{{\rm d} f_{\rm orb}}{{\rm d} t} \right)_{\mathrm{VD}}=2.3 \times 10^{-5} M_{\rm 7}^{-11/5} r_{\rm 3}^{-11/4} \,\rm yr^{-2}\, ,
\end{equation}
if $r>r^{\rm es/ff}$ and $r>r_{\rm out}^{\rm \nu/s}$;

\noindent(6) $M_{\rm 2}$-dominated, outer region:
\begin{equation}
\label{eq:gas_drag6}
\left ( \frac{{\rm d} f_{\rm orb}}{{\rm d} t} \right)_{\mathrm{VD}}=1.6 \times 10^{-6} M_{\rm 7}^{-29/17} r_{\rm 3}^{-76/34} q_{\rm s}^{-3/8} \,\rm yr^{-2}\, ,
\end{equation}
if $r>r^{\rm es/ff}$ and $r\leq r_{\rm out}^{\rm \nu/s}$.

The eccentricity evolution due to viscous drag can be complex and cannot be trivially reduced to a prescription for a single dominant regime. The simulation results in \cite{Roedig2011} show that if the incoming eccentricity of the MBHB on a prograde orbit is $>0.04$ then there is a saturation
eccentricity in the range $(0.6,0.8)$. Following this result, we randomly assign an eccentricity between $0.6$ and $0.8$ after one viscous timescale (measured at the separation where viscous drag begins to dominate the evolution). If the eccentricity of the orbit is less than $0.04$ when viscous drag takes over the orbital decay, the eccentricity remains fixed until GW emission
takes over the orbital evolution.  

By numerically solving these equations, we can determine the dynamics (given an initial eccentricity at $r_{\rm inf}$) of each delayed merger in $\hagn$, tracking both the orbital frequency and eccentricity of the binaries. In the following section, we discuss how to use this information to compute a realistic GW background signal from the population of MBHBs.

\section{Estimating the gravitational wave background with Horizon-AGN}
\label{GWB}

In this section, we present the methodology for estimating the GW strain spectrum from the population of MBHBs extracted from the simulation $\hagn$. 
We computed it by assuming both circular and eccentric ensembles of MBHBs and investigated the properties of the resulting backgrounds.

\subsection{Methodology}

\subsubsection{The analytic average}

The gravitational wave background (GWB) can be described at each observed frequency, $f$, by the characteristic strain, $h_{\rm c}(f)$, which can be evaluated by integrating the comoving density of coalescing binaries  \citep{2001astro.ph..8028P}, expressed as
\begin{equation}
    \label{eq: characteristic strain Phinney}
    h_{\rm c}^2(f) = \frac{4 G}{\pi c^2} \frac{1}{f^2} \int {\rm d}z\,{\rm d}\vec\xi\, \frac{{\rm d}^2 n}{{\rm d}z{\rm d}\vec\xi} \frac{1}{1+z}  \frac{{\rm d}\Egw}{{\rm d}\ln f_{\rm s}}(f_{\rm s}),
\end{equation}
where $z$ is the merger redshift, $n$ is the number of MBHB mergers per comoving volume and where we grouped the binary parameters ($M_1$, $q$, $\sigma_{\mathrm{star}}$, ...) in the $\vec\xi$ vector. 
The characteristic strain is related to the energy released in GWs over the entire binary history (${\rm d}\Egw$) per logarithmic frequency interval in the source rest frame, where $f_{\rm s} = (1+z)f$.

An MBHB on an (almost) circular orbit emits GWs at twice its orbital frequency, as higher order modes can be safely ignored since they are suppressed by a factor of $v_{\rm{orb}}/c \ll 1$, where $v_{\rm{orb}}$ is the orbital velocity.
In the single dominant mode case, we have
a one-to-one relationship between the GW frequency in the source frame $f_{\rm s}$ and the binary orbital frequency $\forb$: ${\rm d}\Egw/{\rm d}\ln f_{\rm s} = {\rm d}\Egw/{\rm d}\ln \forb$.
However, this is no longer true for binaries in eccentric orbits. In this case, the emitted GW power is distributed across a set of orbital frequency harmonics, $m$, and the distribution depends on the orbital eccentricity of the binary. 
Loss of orbital momentum through gravitational radiation leads to orbital circularization. 
As a result,  the total GW energy $\text{d}\Egw$ emitted within an observer frequency band centered on $f_{\rm s}$ will include contributions from different harmonics $m$ reflecting different stages of the binary evolution, where $\forb = f_{\rm s} / m$. In most general (eccentric orbits) case the energy release can be written as \citep{2007PThPh.117..241E}, expressed as
\begin{equation}
    \label{eq:general-dEgwdlnf}
    \frac{{\rm d}\Egw}{{\rm d} \ln f_{\rm s}}(f) = \sum_{m=1}^{+\infty} \left\{ \frac{{\rm d}\tau_{\rm c}}{{\rm d}\ln\forb} \Lgwcirc(\forb) g\left[m,e(\forb)\right] \right\}_{\forb = \frac{f_{\rm s}}{m}}.
\end{equation}
Here, we introduce the time to coalescence, $\tau_{\rm c}$, measured in the binary rest frame, and the GW luminosity\footnote{The influence of MBH spins on the GW luminosity can be safely neglected for such broad orbits and over the observational period of several decades
\citep{2012PhRvL.109h1104M}.} of a binary in a circular orbit corresponding to the mean eccentric motion:
\begin{equation}
    \label{eq: GW luminosity circ}
    \Lgwcirc(\vec{\xi}, \forb) = \frac{32}{5} \frac{G^{7/3}}{c^5} \mathcal{M}_{\rm c}^{10/3} \left(2\pi\forb\right)^{10/3},
\end{equation}
where $\mathcal{M}_{\rm c} = (M_1M_2)^{3/5} / (M_{12})^{1/5}$ is the binary chirp mass. 
The GW spectrum is then obtained by including the $g(m, e)$ function, which gives the relative average (over one orbit) GW power emitted at a given harmonic $m$ by a binary with orbital eccentricity $e$ (see Eq. A1 in \citealp{1963PhRv..131..435P} for an explicit expression of $g(m,e)$ in terms of Bessel functions). 

The energy emitted per logarithmic frequency band is given as a product of GW luminosity and the time the binary spent in that band, expressed as $\frac{{\rm d}\tau_{\rm c}}{{\rm d}\ln\forb}$.
As mentioned, we use the energy release averaged over the orbit; however, for eccentric motion, it varies significantly over the orbit, where most of the GW power is emitted during the periapse passage. 
The uneven energy release over time in highly eccentric binaries with long orbital periods could induce nonstationarity in the GWB signal \citep{PhysRevD.111.023047}. In this work we neglect this effect and consider the averaged (over period) GWB spectrum.

For circular orbits, as described in Sect.~\ref{sec: circular case} below, we use the $e \to 0 $ limit of Eq. \ref{eq:general-dEgwdlnf} directly. For eccentric binaries, we approximate this expression by computing the MBHB dynamics over a finite-width orbital log-frequency grid using the method presented in Sect.~\ref{sec: eccen case}, namely, 
\begin{equation}
    \label{eq: discretization dEgw on forb grid}
    \frac{\Delta \Egw}{\Dlnf_{\rm s}}(f) \simeq \sum_k \left[\frac{\Delta \tau_{\rm c}}{\Delta \ln\forb} \Lgwcirc\right]_{\forb^{(k)}} \times \sum_{m \in \Nensemb{\forb^{(k)}}{f}} g\left[m,e(\forb^{(k)})\right].
\end{equation}

We  introduce
\begin{equation}
    \label{eq: set of harmonics}
    \Nensemb{\forb^{(k)}}{f} = \left\{m \in \mathbb{N}^{*} \left\lvert\, \frac{m \forb^{(k)}}{1+z} \in \left[fe^{-\Dlnf/2}, f e^{\Dlnf/2}\right]\right.\right\},
\end{equation}
which is the set of orbital frequency harmonics, $m$, that would lead to an observed GW frequency within the log-frequency band $[\ln f - \Dlnf /2, \ln f + \Dlnf /2]$ for a binary at a redshift, $z$.

Next, we use the estimation for the merger comoving density given by the $\hagn$ merger catalog,
\begin{equation}
    \label{eq: HAGN merger density}
    \frac{{\rm d}^2n}{{\rm d}z{\rm d}\vec\xi} \simeq \frac{1}{\Vsim} \sum_j \delta(z-z_j)\delta(\vec\xi - \vec\xi_j),
\end{equation}
where $\Vsim$ is the comoving volume of the simulation, to obtain the analytic expression for the average GWB spectrum:
\begin{equation}
    \label{eq: Phinney with catalog}
    h_{\rm c}^2(f) = \frac{4 G}{\pi c^2} \frac{1}{f^2} \frac{1}{\Vsim} \sum_j \frac{1}{1+z_j}  \frac{{\rm d}\Egw^{(j)}}{{\rm d}\ln f_{\rm s}}(f_{\rm s}).
\end{equation}

The sum is over the total energy emitted in GWs by the simulation mergers labeled by $j$, $\Egw^{(j)}$, which can be computed either directly using Eq. \ref{eq:general-dEgwdlnf}, if an analytic expression for ${\rm d}\tau_{\rm c}/{\rm d}\ln\forb$ is available, or approximately by Eq. \ref{eq: discretization dEgw on forb grid}.
Let us emphasize that this expression is fully deterministic, since we compute only one dynamical evolution for each delayed merger in $\hagn$.

\subsubsection{Building realizations of the Universe}

The analytic computation {of the characteristic strain spectrum} presented in the previous section is, by definition, deterministic and does not introduce any cosmic variance {inherent to the discreteness of the binary population} (see \citealp{2008MNRAS.390..192S}).  
To estimate the variation in the strain signal due to different realizations\footnote{For a given Universe, this variance also arises depending on the observer's location.} of the Universe, one must consider the population of inspiralling MBHBs instead of mergers, which are the actual GW sources in the PTA band. 
In this case, the GWB characteristic strain can be computed as the integral over the redshift of the distribution of inspiralling MBHBs,
\begin{equation}
    \label{eq: GWB as a sum over sources}
    h_{\rm c}^2(f) = \int {\rm d}z {\rm d}\vec\xi {\rm d} \ln\forb \frac{{\rm d}^3 N_{\rm i}}{{\rm d}z {\rm d}\vec\xi {\rm d} \ln\forb} h_{\rm c,1}^2(f; z, \vec\xi, \forb),
\end{equation}
where $N_{\rm i}(z, \vec\xi, \forb)$ is the number of inspiralling sources and $h_{\rm c,1}^2(f; z, \vec\xi, \forb)$ is the characteristic strain of one source with given properties. 
The distribution of inspiralling MBHBs can be derived from the comoving density of mergers using
\begin{equation}
    \label{eq: relation Ni to n}
    \frac{{\rm d}^3 N_{\rm i}}{{\rm d}z {\rm d}\vec\xi {\rm d} \ln\forb} = \frac{{\rm d}^2 n}{{\rm d}\vec\xi {\rm d}z} \frac{{\rm d}z}{{\rm d}\tau_{\rm c}}\frac{{\rm d}\tau_{\rm c}}{{\rm d}\ln\forb} \frac{{\rm d}V_{\rm c}}{{\rm d}z},
\end{equation}
where $V_{\rm c}$ is the comoving volume (see Appendix \ref{sec: appendix link Ni n} and also \citealp{2008MNRAS.390..192S} and \citealp{2023JCAP...08..034B} for details).
As a result, for each merger $j$ of the simulation, a PTA should observe, on average, $\langle N_{\rm i}^{(j)} \rangle (\forb^{(k)})$ inspiralling binaries with an orbital frequency in $\left[\forb^{(k)} e^{ - \Delta \ln\forb /2}, \forb^{(k)} e^{ \Delta \ln\forb /2}\right]$,
\begin{equation}
    \label{eq: mean occupation number}
    \langle N_{\rm i}^{(j)} \rangle (\forb^{(k)}) = \frac{1}{\Vsim} \left[\frac{{\rm d}z}{{\rm d}\tau_{\rm c}} \frac{{\rm d}V_{\rm c}}{{\rm d}z}\right]_{z_j} \Delta \tau^{(j)}_{\rm c}(\forb^{(k)}).
\end{equation}

For the $\Lambda$CDM cosmology used in the $\hagn$ simulation, the expression in the square brackets equals $4\pi d_{\rm M}^2 c (1+z_j)$, where $d_{\rm M}$ is the comoving distance to the source at redshift $z_j$ \citep{1999astro.ph..5116H} and $\Delta \tau_{\rm c}^{(j)}(\forb^{(k)})$ is the residence time of the binary $j$ in the $k$-th log-orbital frequency bin.
To construct a Universe realization, we follow the approach of \cite{2017MNRAS.471.4508K} and draw a number of inspiralling binaries from a Poisson distribution with mean $\langle N_{\rm i}^{(j)} \rangle (\forb^{(k)})$, for each $\hagn$ merger $j$ and each orbital frequency bin $k$. 
To obtain the total induced characteristic strain for a finite-width observer's log-frequency band centered on $f$, we sum the contributions from each inspiralling MBHB using Eq. \ref{eq: GWB as a sum over sources}, Eq. \ref{eq: HAGN merger density} and
\begin{align}
    \label{eq: single binary hc2}
    h_{c,1}^2\left(f;z_j, \vec\xi_j, \forb^{(k)}\right) = \frac{4G}{\pi c^2} \frac{1}{f^2}& \frac{1}{\Dlnf} 
 \frac{\Lgwcirc\left(\forb^{(k)}\right)}{c4\pi d_{\rm L}^2(z_j)} \\
 &\times \sum_{m \in \Nensemb{\forb^{(k)}}{f}} g\left[m, e\left(\forb^{(k)}\right)\right], \nonumber
\end{align}
where $d_{\rm L}(z) = (1+z)d_{\rm M}$ is the luminosity distance to the source. The dependence on $\vec\xi_j$ on the right-hand side is hidden in the GW luminosity function of the binary and its orbital eccentricity evolution function. 
This expression is only valid if the orbital frequency evolution of the binary is slow enough to consider that all of the GW emission over the observation period is within the considered observer frequency band centered on $f$. 
This condition is met for the current duration of the PTA observations.
Indeed, for the $\hagn$ MBHBs, we checked that
the frequency evolution is negligible
for a PTA with 15 years of observing duration and for an observer frequency below $\fyr \equiv 1 / \text{year} \simeq 32$ nHz.

\subsubsection{The circular ensemble modeling} \label{sec: circular case}

In this section, we describe our derivation of the GWB spectrum for the ensemble of circular MBHBs with the orbital dynamics driven by GW emission and loss-cone scattering. 
With this simplification, we can carry out all the derivations analytically and use them as a reference for comparison with a more realistic model that allows for eccentric MBHBs.

For circular orbits ($e=0$), the GW power is emitted at the harmonic $m=2$, for which $g(2,0) = 1$ and $f_{\rm s} = 2 \forb = (1 + z) f$. This significantly simplifies Eq. \ref{eq:general-dEgwdlnf}, as the sum over $m$ is now reduced to a single term ($m=2$). The remaining term is inherently dependent on the binary orbital dynamics, which can be characterized by the evolution of the semi-major axis $a$:
\begin{equation}
    \label{eq: time spent function of freq}
    \frac{{\rm d}\tau_{\rm c}}{{\rm d}\ln \forb} = \frac{{\rm d}\tau_{\rm c}}{{\rm d}a}\frac{{\rm d}a}{{\rm d}\ln \forb}.
\end{equation}

In addition to the GW emission, we incorporate the orbit hardening due to loss-cone scattering (see Eq. \ref{eq:LC1}). In the binary rest frame, the dynamics can be modeled as
\begin{equation}
    \label{eq: binary dynamics}
    \frac{{\rm d}a}{{\rm d}\tau_{\rm c}} = \left. \frac{{\rm d}a}{{\rm d}\tau_{\rm c}}\right|_{\mathrm{LC}} + \left. \frac{{\rm d}a}{{\rm d}\tau_{\rm c}}\right|_{\mathrm{GW}} = -\frac{G H \rho_{\rm inf}}{\sigma_{\rm inf}} a^2 - \frac{64 G^3 M_1 M_2 M_{12}}{5c^5a^3}.
\end{equation} 

Here, we neglected the impact of loss-cone scattering on the binary eccentricity evolution, as we forced the binaries to remain circular\footnote{As shown in figure 1 of \cite{2017MNRAS.471.4508K}, loss-cone scattering does not raise eccentricity above 0.2 for most MBHBs in the PTA band, if the binaries were initially circular.}. 
We can compute the second term of Eq. \ref{eq: time spent function of freq} by assuming that the binaries follow Keplerian orbits.

\begin{figure}
   \centering
   \includegraphics[width=0.49\textwidth]{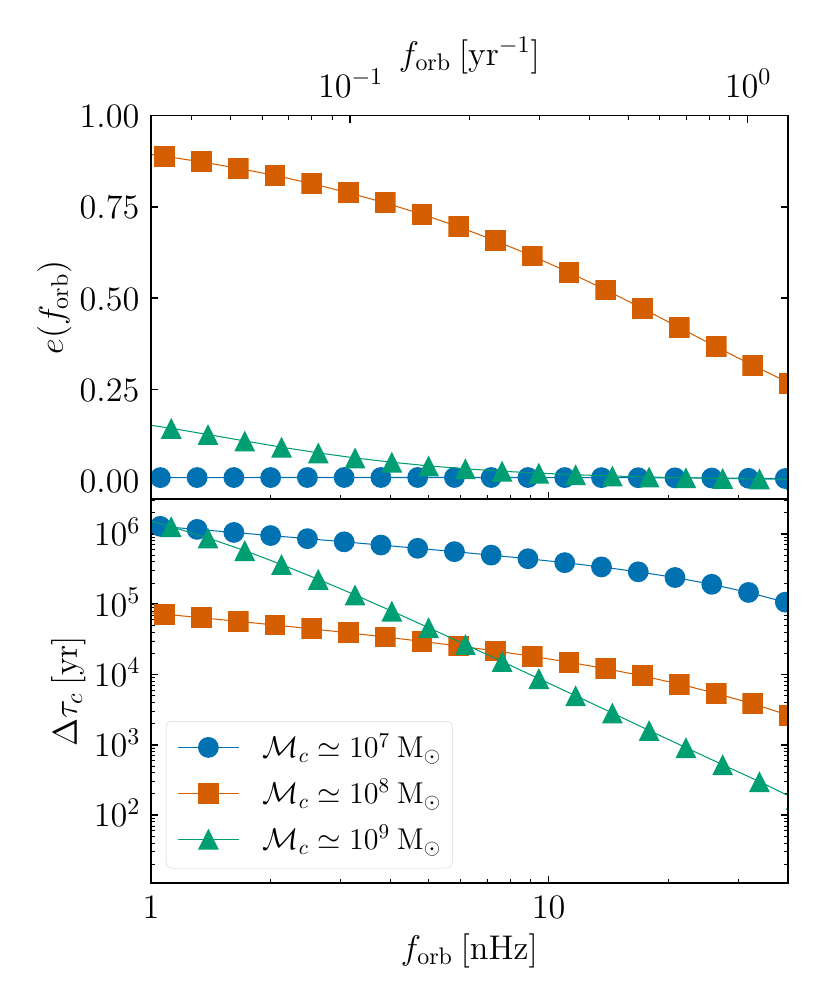}
   \caption{Examples of the residence time (bottom panel), $\Delta \tau_{\rm c}$, per $\log_{10}$ orbital frequency bin are shown for three binaries from the $\hagn$ catalog, with chirp masses of $10^7$, $10^8$, and $10^9 \msun$.
   We also show their respective eccentricity evolution in the top panel. 
   Eccentric systems are more efficient in dissipating the energy through GWs and this can be seen in the residence time (low panel) as compared to (nearly) circular binaries.
   The evolution of binaries at high frequencies (starting from a few nHz) is determined by the gravitational radiation, which is more efficient in energy dissipation for heavy and eccentric binaries. 
   We also observe a fast circularization of eccentric systems in the top panel.}
  \label{Fig:Dt_evol}
\end{figure}

As a result, Eq. \ref{eq: time spent function of freq} can be integrated on a given finite-width orbital frequency grid and plugged in Eq. \ref{eq: mean occupation number} to build Universe realizations. 
In the following, we used log-orbital frequency grids corresponding, for each merger, to 125 log-observer frequency bins equally spaced between $f_{\rm min} = 0.8$ nHz and $f_{\rm max} = 60$ nHz. 
The upper bound is constrained by neglecting the frequency evolution (over PTA observing duration) of binaries in our study.

\subsubsection{The eccentric ensemble modeling}
\label{sec: eccen case}

As mentioned in Sect. \ref{catalogs}, a more detailed description of the MBHB dynamics requires the inclusion of additional physics (viscous drag, eccentricity evolution, etc.) which makes it technically  unfeasible to obtain a derivation of the analytical expressions for the orbital parameters ($\forb, e, ...$) as a function of time. 
To overcome this limitation (as detailed in Sect. \ref{catalogs}), we numerically integrated the dynamics of each merger, $j$, and stored their orbital parameters on a logarithmic (observer) orbital frequency grid $\log_{10}\forb^{(k)} = \log_{10} (0.01 \, \mathrm{nHz}) + k\log_{10} (50 \, \mathrm{nHz}/0.01 \, \mathrm{nHz})$, for $k\in[0, 40]$. 
One must go to such a low orbital frequency to have a precise estimate of the GW signal in the PTA band because the GWs emitted by eccentric binaries could have a significant power 
at frequencies as high as $100\;\forb$ for $e \gtrsim 0.9$ \citep{1963PhRv..131..435P}. 
We can extract the residence time in each orbital frequency bin $k$ from the numerical integration and evaluate Eq. \ref{eq: mean occupation number} for each merger, $j$. 
An example for three $\hagn$ binaries is shown in Fig. \ref{Fig:Dt_evol}. 
The eccentricity evolution function $e(\forb^{(k)})$ for each merger can then be used to compute both Eq. \ref{eq: discretization dEgw on forb grid} and Eq. \ref{eq: single binary hc2} to obtain the analytic average GW spectrum using Eq. \ref{eq: Phinney with catalog} and the Universe realization spectrum using Eq. \ref{eq: GWB as a sum over sources}.

\subsection{Characterizing GW signal from a population of MBHBs}

\subsubsection{Properties of the GW background}

\begin{figure}
   \centering
   \includegraphics[width=0.49\textwidth]{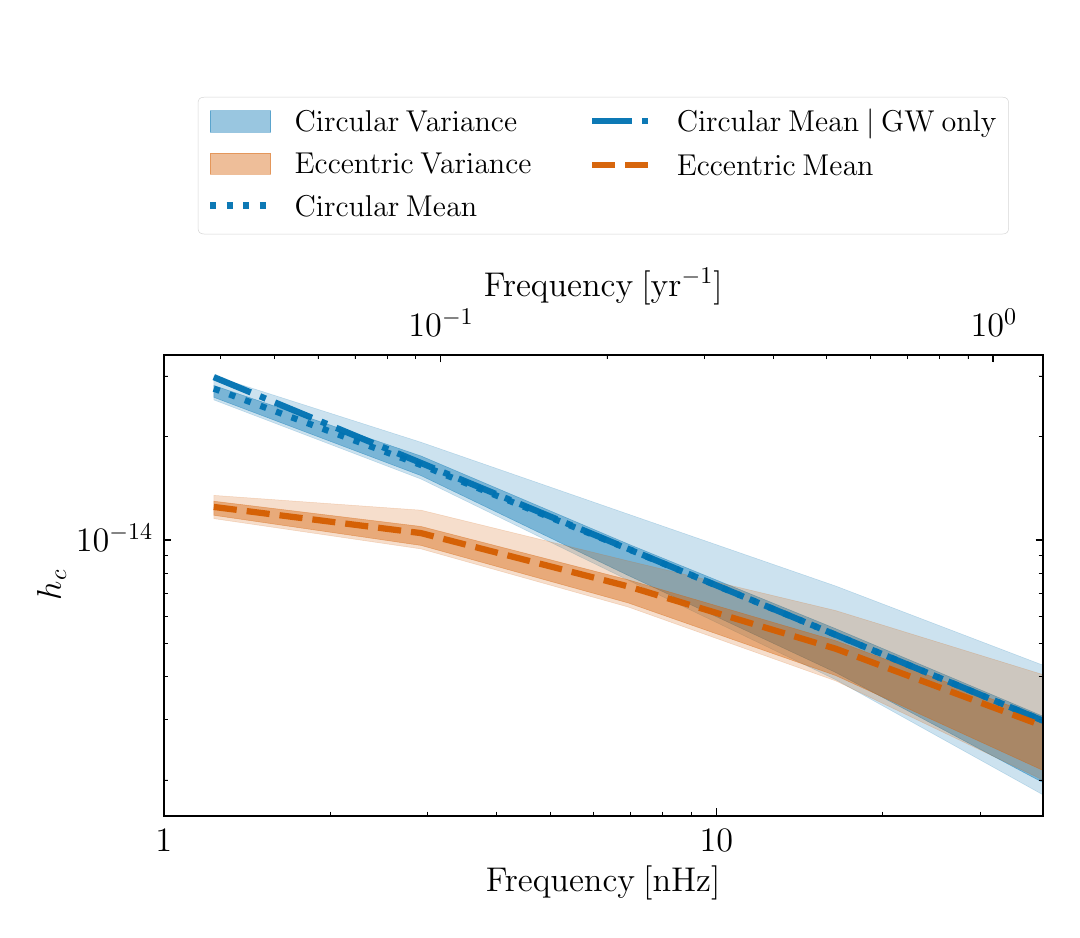}
   \caption{68\% (dark) and 90\% (light) confidence regions for the characteristic strain $h_{\rm c}$ are shown for 2\,000 Universe realizations drawn from the $\hagn$ MBHB population, assuming either circular orbits (blue) or noncircular orbits (orange). We also plot their respective deterministic mean expectation values in dotted and dashed lines. For the circular population, we overlay as a dashed-dotted line the average spectrum, assuming the binary dynamics are driven by GW emission only. The inclusion of eccentricity introduces a significant signal reduction at frequencies lower than $10\,\rm nHz$, due to the reduction of the MBHBs residence time at low orbital frequency.}
  \label{Fig:hc_circ_ecc}
\end{figure}

The characteristic strain spectra for the circular and eccentric ensembles are shown in Fig. \ref{Fig:hc_circ_ecc}.
As expected, assuming circular binaries driven by GW emission, we find a mean characteristic strain spectrum that follows the power-law behavior of Eq. \ref{eq:power-law-strain}, with an amplitude at $\fyr$ of $\Ayr = 3.4 \times 10^{-15}$ and a spectral index of $\alpha=-2/3$ as derived in \cite{2001astro.ph..8028P}.
The background amplitude for $\hagn$ is higher than for many other theoretical models. This can be seen in Appendix A of \citet{2023ApJ...952L..37A} for pre-2023 models, as well as in new models developed by the EPTA collaboration to explain the detected amplitude \citep{2024A&A...685A..94E}. 
In fact, \citet{2022MNRAS.509.3488I} already noted that models that reproduce the background amplitude exceed electromagnetic constraints, such as the bright end of the low-redshift AGN luminosity function. 
In fact, $\hagn$ overpredicts the AGN luminosity function \citep{2016MNRAS.460.2979V}, explaining why the background amplitude of the simulation is closer to the value measured by PTAs compared to other models. 
We stress that in this paper we are not trying to ``fit'' the amplitude of the background, but we are trying to understand what factors influence the background, and explore the properties of the sources within a unified framework.

Since we included interactions with unbound stars when modeling the binary dynamics for the circular population, we did observe a slight bending of the GW spectrum at low frequencies due to the reduction of the residence time at those frequencies \citep{2014MNRAS.442...56R}.
Overall, the bend in the $\hagn$ population starts to significantly affect the spectrum at observer frequencies lower than 2 nHz, which corresponds to an observation duration of around $15.8$ years, in good agreement with previous studies \citep{2013CQGra..30v4014S, 2017MNRAS.471.4508K}. 
The current PTA experiments might already be sensitive to this effect \citep{2024arXiv241105906C}.

As already demonstrated in \cite{2008MNRAS.390..192S} and recently investigated on the basis of PTA data in \cite{2025ApJ...978...31A}, the realistic Universe realization spectra from the circular population could significantly deviate from the power-law
behavior with the mean spectral index corresponding to  $\alpha=-2/3$. 
This is related to the stochasticity of the GW signal from the MBHB population: it is stochastic in the limit of a large number of sources emitting in the same frequency bin during the observation period. 
The number of binaries per frequency bin decreases with an increase in the frequency due to the reduction in their residence time (see Fig. \ref{Fig:Dt_evol} and Eq. \ref{eq: mean occupation number}), leading to a complete loss of stochasticity above 100-$200 \,\rm nHz$. 
{In the intermediate frequency range of 10-$50\,\rm nHz$, the decrease in the number of contributing sources increases both the variance and skewness\footnote{{The skewness (third moment) of the strain distribution at a given observer frequency, across Universe realizations, quantifies the asymmetry of the distribution about its mean.}} of the total characteristic strain distribution across Universe realizations.}
This is a well-known effect, {already discussed in \cite{2003ApJ...583..616J,2008MNRAS.390..192S}}, and later elaborated in \cite{2024ApJ...971L..10L}. 
{In Fig. \ref{Fig:hc_circ_ecc}, the skewness of the strain distribution above $10\,\rm nHz$ is evident for the circular population, where approximately 84\% of the GWB spectra have amplitudes below the square root of the average squared strain value. This renders the strain distribution at these frequencies strongly non-Gaussian.}
As shown in the following, this ultimately results in GWB power spectra with steeper power-law slopes compared to the average spectrum. 

When eccentricity is included, we observe two main effects on the GWB spectra. 
First, we observe the expected bending of the GW spectrum for observer frequencies $\lesssim 10\,\rm nHz$, significantly decreasing the background amplitude at a few nanohertz (nHz). This occurs because eccentricity spreads the GW power across multiple orbital frequency harmonics, combined with the fact that the MBHBs gradually circularize due to the emission of GW \citep{2007PThPh.117..241E}.
The second effect of eccentricity is a slight reduction in the variance and skewness of the GW strain amplitude around $\fyr$ compared to the circular population. 
This is caused by the input of highly eccentric binaries with a low orbital frequency, which effectively increases the number of contributions and maintains the stochasticity at higher frequencies.

\begin{figure*}
   \centering
   \includegraphics[width=0.99\textwidth]{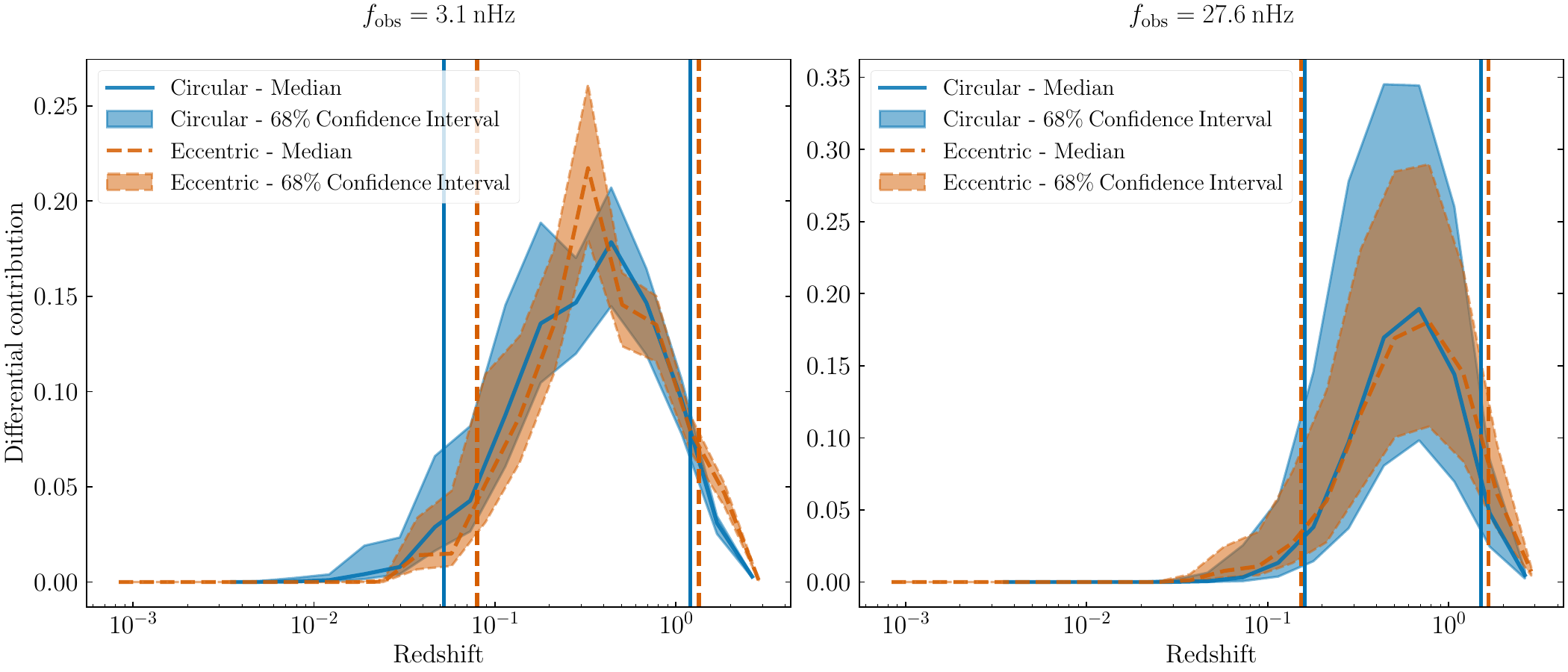}
   \caption{Median and 68\% confidence interval of the differential contribution to the total characteristic strain signal, based on 2\,000 Universe realizations, shown across MBHB redshift bins for both circular and eccentric populations. 
   We compare the results at two observer frequencies: 3.1 nHz (left panel) and 27.6 nHz (right panel). 
   The vertical lines represent the median redshift values at which the cumulative contribution (integrating from low and high redshifts, respectively) reaches 5\% of the total strain signal for the circular (solid line) and eccentric (dashed line) populations.
   }
  \label{Fig:z_10yr_comp}
\end{figure*}

\begin{figure*}
   \centering
   \includegraphics[width=0.99\textwidth]{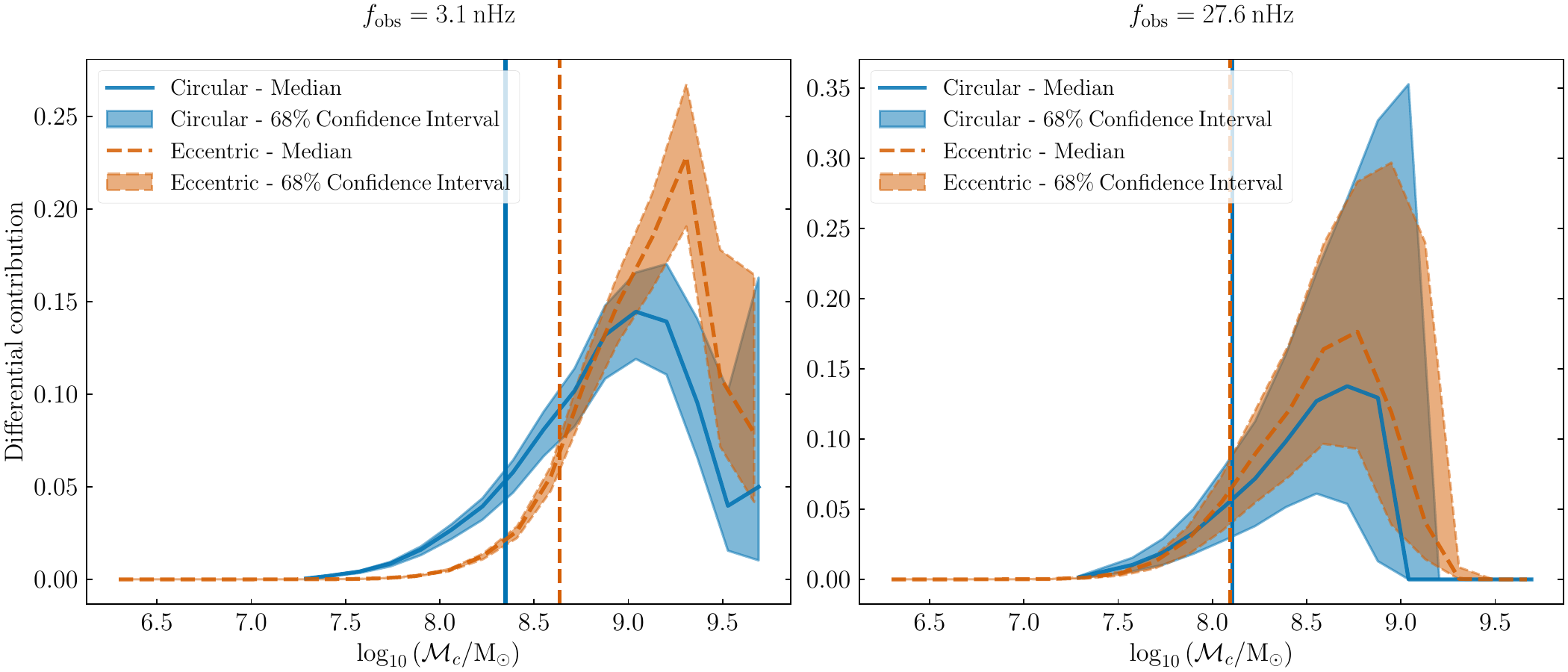}
   \caption{Median and 68\% confidence interval of the differential contribution to the total characteristic strain signal, based on 2\,000 Universe realizations, shown across MBHB $\log_{10}$-chirp mass bins for both circular and eccentric populations. 
   We compare the results at two observer frequencies: 3.1 nHz (left panel) and 27.6 nHz (right panel). 
   Vertical lines represent the median chirp mass value at which the cumulative contribution (integrating from lower masses) reaches 10\% of the total strain signal for the circular (solid line) and eccentric (dashed line) populations.}
  \label{Fig:Mc_contrib_comp}%
\end{figure*}

\subsubsection{Properties of the contributing MBHBs}
\label{Props_contr_MBHBs}

We now investigate the properties of the MBHBs that contribute the most to the GWB across the PTA frequency band. 
In particular, we compute the contribution of binaries to the total characteristic strain as a function of their redshift and chirp mass. 
We consider a PTA experiment with observation span $\Tobs = T_{\rm EPTA} \simeq 10.33$ years.  
As mentioned above, the properties of the total characteristic GW strain vary significantly between the low and high frequency end of the PTA sensitivity band, so we take two representative frequencies: $1/T_{\rm EPTA} \simeq 3.1\, \rm nHz$ and $9/T_{\rm EPTA} \simeq 27.6\,\rm nHz$. 
In Fig. \ref{Fig:z_10yr_comp}, we show the differential contribution of the MBHB population divided into redshift slices.
The solid line represents the median values across Universe realizations for each redshift slice, while the shaded region indicates the 68\% confidence interval, defined here by the 16$^{\rm th}$ and 84$^{\rm th}$ percentiles of the distributions. 
At low frequencies, for circular and eccentric populations, the peak of the distribution is found around $z \simeq 0.3-0.4$. 
At lower redshifts, we observe a larger variance in the contribution compared to higher redshifts. 
Since the variance (width of the confidence interval) is primarily determined by the average number of sources contributing to the GW strain in each bin, this reflects the smaller number of MBHBs expected at low redshift.
At high frequency, the peak of distribution for both populations shifts toward a higher redshift, around $z\simeq 0.6-0.7$.  
The heavy nearby systems, which dominate the GW signal at low frequencies, spend very little time at high frequencies; therefore, the signal accumulates over a larger volume. 
This also explains the very large variance compared to the low-frequency distribution: we have a lower number of binaries that contribute at high frequencies.
Again, at high frequencies, the eccentric population exhibits smaller variance due to the increased number of contributing binaries.
The typical redshift ranges of the MBHBs contributing to 90\% of the total characteristic strain signal are given by the vertical lines in Fig. \ref{Fig:z_10yr_comp}.
These correspond to the median redshift values (across Universe realizations) at which the cumulative contribution — integrated from low and high redshifts, respectively — reaches 5\% of the total strain signal. 
For the low frequency of 3.1 nHz considered here, where the GW signal is expected to be seen as a stochastic background, we find that most of the contribution comes from the range $0.05 < z < 1.2$ for the circular population and $0.08 < z < 1.3$ for the eccentric population.

Similar distributions as a function of MBHB chirp mass are shown in Fig. \ref{Fig:Mc_contrib_comp}. 
These distributions confirm the findings noted above. 
The GW signal at low frequencies is mainly formed by heavy systems with chirp mass in the range $10^{8.5} - 10^{9.5} \msun$, while most of the contribution at high frequencies comes from binaries in the range $10^{8} - 10^{9} \msun$.
At low frequency, we find that the contribution to the total strain signal extends to higher masses for the eccentric population relative to the circular population. 
The main reason is the contribution of the massive and \emph{eccentric} binaries with orbital periods around and even below the $0.5/\Tobs$ frequency, emitting at higher orbital harmonics.
The typical lower bound on the chirp mass of MBHBs contributing 90\% of the total characteristic strain signal is indicated by the vertical lines in Fig. \ref{Fig:Mc_contrib_comp}. 
These represent the median chirp mass values (across Universe realizations) at which the cumulative contribution (integrated from the lowest masses) reaches 10\% of the total strain signal, for each population.
We find $\log_{10} (\mathcal{M}_{\rm c} / \rm M_\odot) \gtrsim 8.35$ for the circular population and $\log_{10} (\mathcal{M}_{\rm c} / {\rm M_\odot}) \gtrsim 8.63$ for the eccentric population.

\subsubsection{Number of contributing MBHBs}

In this section, we aim to quantify the number of MBHBs that effectively contribute to the strain power spectrum of the GWB. 
This number is particularly important as it quantifies the validity of the stochasticity and Gaussianity assumption for the GWB signal in the PTA band.
To estimate this number, we use two complementary metrics.
The first, suggested in \cite{2022ApJ...941..119B}, computes how many binary signals must be summed, starting from the brightest, to reach a fraction $\alpha_{\rm tot}$ of the total GWB power in a given observer frequency bin centered at $f$. This can be expressed as
\begin{equation}
    \label{eq: N_thresh definition}
    N_{\alpha_{\rm tot}}(f) = \min\left\{N \left| \frac{\sum_{l=1}^Nh_{{\rm c},l}^2(f)}{\sum_{l} h_{{\rm c},l}^2(f)} > \alpha_{\rm tot}\right.\right\},
\end{equation}
where the summation in the numerator is performed over the merger population in decreasing order of $h_{\rm c}^2$. We use Eq. \ref{eq: single binary hc2} to compute each binary's contribution. In this analysis, we adopt $\alpha_{\rm tot} = 0.75$, which corresponds to a S/N of 3 for the brightest sources relative to the rest of the MBHB population, defined as $\sum_{l=1}^{N} h_{{\rm c},l}^2(f) / \sum_{l \ge N+1}h_{{\rm c},l}^2(f) = 3$. 

However, this metric does not quantify the relative contribution of the "brightest sources". For example, assume that 74\% of the GWB power is produced by a few binaries and the remaining 1\% needed to reach 75\% comes from 999 MBHBs, in this case, we would obtain $N_{75}=1000$ even though the GWs from the population are dominated by one system. 
To capture this potential problem, we suggest another measure, $N_{\rm eff}$, inspired by the effective sample size of a Markov Chain, derived in \cite{ESS}, which can be written as
\begin{equation}
    \label{eq: Neff definiton}
    N_{\rm eff}(f) = N_{\rm b} \frac{ \left[ \frac1{N_{\rm b}}\sum_l h_{{\rm c},l}^2(f) \right]^2} {\frac1{N_{\rm b}} \sum_{l} \left[h_{{\rm c},l}^2(f) \right]^2},
\end{equation}
where $N_{\rm b}$ is the total number of inspiralling binaries contributing to the strain signal at frequency $f$. One can verify that $N_{\rm eff}$ equals $N_{\rm b}$ if the GW signal originates from a population of equally contributing binaries.
However, if 99\% of the GW signal from a population is produced by a single binary, $N_{\rm eff}$ will be much closer to 1, as desired.

\begin{figure}
    \centering
    \includegraphics[width=0.95\linewidth]{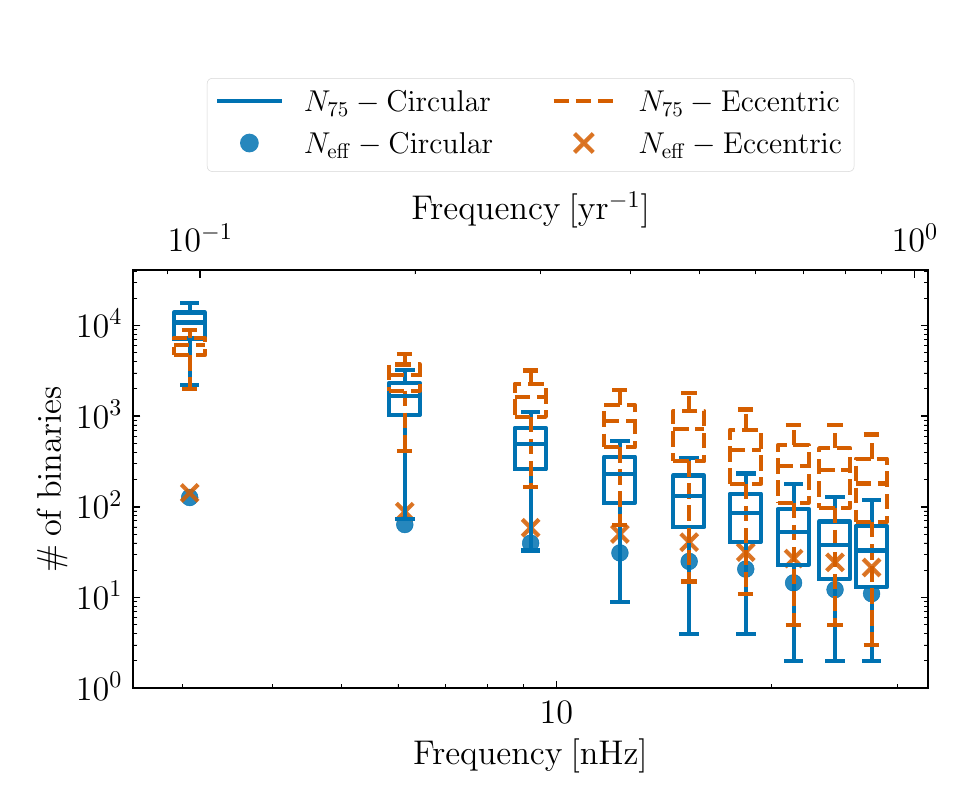}
    \caption{Comparison of the effective number of binaries in terms of $N_{75}$ and $N_{\rm eff}$ contributing to the GWB at each observed frequency bin of a PTA with an observing duration of 10.3 years as for \cite{2023A&A...678A..50E}. At each observer frequency, the two extreme horizontal lines represent the 5th and 95th quartiles of the $N_{75}$ distribution, derived from 2\,000 Universe realizations. The edges and the central line of the box represent the 25th, 50th, and 75th quartiles, respectively. For $N_{\rm eff}$, we only show the median of the distribution at each observer frequency with blue dots for the circular population and orange crosses for the eccentric population.}
    \label{Fig:Neff_comp}
\end{figure}

The (effective) number of binaries that contribute the most to the total GW strain across the PTA frequency band is presented in Fig. \ref{Fig:Neff_comp}. 
The results for $N_{75}$ are given as error boxes and median $N_{\rm eff}$ values are presented by circles for the circular population and crosses for the eccentric population, for our $2\;000$ Universe realizations.
Below 10 nHz, for both populations, the typical number of binary signals that must be added to reach 75\% of the background signal is greater than 300, indicating that the stochastic approximation seems justifiable at those frequencies.  
However, around and above $\fyr$, for 25\% of the Universe realizations, $N_{75}$ is lower than 10, which is not sufficiently high to apply the central limit theorem. 
The eccentric background has a factor $\approx 10$ greater number of contributors ($N_{75}$) around $\fyr$ due to the distributed spectrum of GW emission and therefore maintains stochasticity to higher frequencies.

As the effective number of binaries, $N_{\rm eff}$, does not have a direct physical interpretation, we only compare the median values of $N_{75}$ and $N_{\rm eff}$ at each frequency to quantify the variation in the relative contributions of the binaries counted in $N_{75}$.
We find that at low frequencies, $N_{75}$ is about two orders of magnitude larger than $N_{\rm eff}$ for both populations. 
This suggests that the contribution to the total GW strain is highly uneven among binaries.
In the lowest frequency bin, although around 10\,000 binaries produce 75\% of the background signal, only hundreds of them are significantly contributing to the signal.
At frequencies below $10\,\rm nHz$, even the $N_{\rm eff}$ metric indicates the presence of at least tens of sources per bin, suggesting that it could be well described as a stochastic GWB.

At higher frequencies, the contributions are more balanced and point to a number of contributing binaries from a few up  to several dozen binaries. This implies that at higher frequency, the background is less likely to be well approximated by a Gaussian noise and therefore is less similar to the conventional definition of a background. 
Taking the frequency range as a whole, one might conclude that detecting individual sources is unlikely due to the large number of contributing binaries, especially at low frequencies. 
However, even if the fractional contribution of an individual source is small, the cumulative effect of observing multiple pulsars within a given PTA can enhance its detectability. 
Furthermore, the detectability of an individual binary by a PTA must consider both pulsar noise and the detector response to GWs.
This is examined in the next section, where we investigate how to pass from the properties of the strain signal to the PTA observables.

\section{\label{PTA-implications} Implications for pulsar timing array observations}

To accurately compare the theoretical signal computed from the $\hagn$ population with the one currently measured by PTA experiments, we must study the properties of the GWB-induced timing residuals.
Timing residuals are the time difference between the measured times of arrival (ToAs) of radio pulses from Galactic millisecond pulsars and the ToAs predicted by a theoretical model, referred to as the timing model \citep{2006MNRAS.372.1549E}.  
The residuals also carry an imprint from GWs; in particular, a GWB would appear as a common correlated signal with a spatial correlation pattern first derived in \cite{1983ApJ...265L..39H}.

\subsection{Methodology}

In this section we present two methods for estimating the spectral properties of the timing residuals induced by the MBHB population.
We emphasize the importance of considering what affects the spectral inference of the detector (here the PTA), to avoid potential biases when interpreting its results.
We also outline our methodology for identifying potentially resolvable individual binaries within the population.

\subsubsection{The Gaussian ensemble approach}

This approach considers the GWB signal generated by an MBHBs population as a Gaussian ensemble (see e.g., \citealp{2015MNRAS.451.2417R,2017LRR....20....2R,2019PhRvD.100j4028H, 2023ApJ...952L..37A, 2024A&A...685A..94E}). 
In this case, the GW strain is considered as a random variable, fully characterized by its one-sided power spectral density (PSD), $S_{\rm h}(f) = h_{\rm c}^2(f) / f$. 
{If one assumes that the GWB signal is stationary, isotropic and un-polarized, the average timing residuals response is given, in terms of PSD, by \citep{1983MNRAS.203..945B,2006ApJ...653.1571J}}
\begin{equation}
    \label{eq: Gaussian strain to residual PSD}
    \resPSD(f) = \frac{S_{\rm h}(f)}{12 \pi^2 f^2}.
\end{equation}

The timing residual PSDs from the different Universe realizations obtained using this method have two limitations. 
First, they do not account for the fact that the signal is produced by a discrete number of sources, each with its own response to the PTA. 
Second, the derivation of $h_{\rm c}^2$ that leads to the timing residuals PSD $\resPSD$, assumes that each MBHB contributes a Dirac delta function at its orbital frequency harmonics as in Eq. \ref{eq: single binary hc2}. 
We recall below that this is only the case in the limit of an infinitely long dataset\footnote{It also assumes that the GW sources exhibit no frequency evolution.}.
For realistic PTA data, the effect of the finite observation duration on spectral estimation needs to be considered a priori.
We use a second method for evaluating
$\resPSD$ that accounts for both of these issues in the following subsection.

\subsubsection{\label{sec: population approach}The population approach}

The principle of this approach is to compute the GWB-induced timing residuals by summing the contributions of each individual MBHB  \citep[see e.g.][]{2022ApJ...941..119B}. 
This requires knowledge of the waveform for any set of orbital parameters.
For this approach, we consider only circular binaries.

The exact residuals induced for a given pulsar by an inspiralling MBHB depend on their relative position, the binary inclination angle $\iota$ to the line of sight and the polarization angle $\psi$. 
The time series of timing residuals is composed of two terms, corresponding to the impact of the GW strain at the time of emission of the radio pulses (referred to as the Pulsar term) and at the time of their reception (referred to as the Earth term; \citealt{1979ApJ...234.1100D}). 
The correlation between GWB timing residuals among pulsars is entirely due to the Earth term. 
Thus, in the following, we only consider the GW strain signal of the Earth term when computing the GWB timing residuals power spectrum. 
We assume GWB to be isotropic, reflecting the distribution of MBHBs across the sky. 
As a result, we expect the GWB to imprint, on average, the same signal in every pulsar \citep{1983ApJ...265L..39H}; therefore, we compute the GWB spectrum for only one pulsar and assume that it is the same for the other pulsars in the array.

As shown in \cite{2019PhRvD.100j4028H}, for a finite observing duration $\Tobs$, the Fourier transform of the GW residuals induced by a circular MBHB located at a given sky position $\hat k$ at redshift $z$ and emitting GW at $\fgw = 2 \forb / (1+z)$ in the observer frame, takes the form
\begin{equation}
    \label{eq: r(f) ET finite duration}
    \tilde{r}_{\rm GW}(f) = R_0 \left[ 
        G_{\hat k} e^{i\phi_0}\delta_{\Tobs}(f - \fgw) + G_{\hat k}^*e^{-i\phi_0} \delta_{\Tobs}(f + \fgw)
    \right],
\end{equation}
where $\phi_0$ is an (arbitrary) initial GW phase and the amplitude $R_0$ depends on the intrinsic properties of the binary:
\begin{equation}
    \label{eq: R0 expression}
    R_0 = \frac{2 \left(G\mathcal{M}_{{\rm c},z}\right)^{5/3} \left(\pi\fgw\right)^{-1/3}}{c^4 d_{\rm L}},
\end{equation}
where we introduced the redshifted chirp mass $\mathcal{M}_{{\rm c},z} = (1+z) \mathcal{M}_{\rm c}$.
The residuals amplitude is modulated by the complex geometric factor $G_{\hat k}$ given as
\begin{equation}
    \label{eq: Gk factor}
    G_{\hat k} = -\frac{1}{2}\left[\cos\iota F^\times(\hat k, \psi)+i\frac{1 + \cos^2\iota}{2}F^+(\hat k, \psi) \right],
\end{equation}
where we introduced the antenna pattern response function for both $+$ and $\times$ polarization modes that depends on the sky position of the pulsar and binary as well as on the polarization angle $\psi$. Their exact expressions can be found in \cite{2012ApJ...756..175E}.
Finally, we must correctly take into account the finite duration of the PTA observations ($\Tobs$) by introducing the edge effect into the Dirac delta function:
\begin{equation}
    \label{eq: delta T def}
    \delta_{\Tobs}(f) = \Tobs \text{sinc}(\pi f\Tobs) . 
\end{equation}

Contrary to the previous case where we considered Dirac delta functions as a contribution of each MBHB, the finite width of the sinc function  leads to spectral leakage in the non-white PSD.  As a result, every MBHB can potentially contribute to several frequencies, which makes frequency bins correlated. 
It is common practice in signal processing to taper the time series to minimize the effect of spectral leakage \citep{1993sapa.book.....P}. 
However, in current PTA data analysis, no such pre-processing is applied to the timing residuals time series, so it might, a priori, affects the inference of the PSD.
This suggests that even MBHBs emitting GWs at frequencies much lower than the PTA frequency ($1/\Tobs$) could contribute to the GWB power inferred in the PTA band due to spectral leakage.

The spectral leakage could (at least partially) be mitigated by the timing model. 
A part of the timing model fits and removes linear and quadratic trends in the PTA residuals that could be caused by the spin-down in the millisecond pulsars. 
Obviously, it also removes some very low-frequency GWs covariant with such a trend \citep{2010arXiv1010.3785C, 2013MNRAS.428.1147V}. 
As a result, it can significantly reduce the dynamical range in the expected PSD and decrease the effect of spectral leakage.
{An effect already discussed in \citet{2013PhRvD..87d4035T,2013PhRvD..87j4021L}.}
However, the timing model cannot be a replacement for a well-designed taper that controls the leakage. 
Nonetheless, the timing model acts as a high-pass filter that removes \emph{very} low-frequency GWB
($\lesssim 1/\Tobs$) from the pulsar timing residuals \citep{2019PhRvD.100j4028H}. 
As a result, we consider only MBHBs that emit GWs with $\fgw > 1/(2\Tobs)$ in order to reduce the numerical cost of the GWB computation without affecting its inferred properties in the PTA frequency band.

We produce realistic GWB timing residuals by summing the contributions from all MBHBs of a given Universe realization, uniformly drawing the sky location, polarization angle, and initial phase for each binary. 
However, since the number of MBHBs contributing to the PTA band typically exceeds millions, we employed a numerical simplification, which allowed us to carry out this procedure more efficiently; the procedure is described in Appendix \ref{sec: appendix fasten rf sum}.
Once the GWB timing residuals waveform $\tilde{r}_{\rm GWB}$ is obtained, we can return to the time domain and infer the GWB PSD using the data analysis procedure commonly employed in the PTA community. 
In the following, we briefly discuss the PSD inference procedure used for PTA data. 

\subsubsection{The GWB spectral inference in PTA}

The GWB analysis performed by PTA collaborations is based on the modeling of the timing residuals data (intrinsic red noise, dispersion measure variations, GWB, etc.) using Gaussian processes. 
Each noise component $X$ is modeled using a truncated Fourier basis where the $2 N_{\rm f}^{(X)}$ coefficients $(a_k^{(X)})_{1\leq k \leq 2N_{\rm f}^{(X)} }$, are distributed according to a multivariate Gaussian with a covariance matrix reflecting the underlying process, characterized by its PSD $S_{\rm r}^{(X)}$:
\begin{equation}
    \label{eq:GP-prior}
    \langle a_k^{(X)} a_l^{(X)}\rangle = S_{\rm r}^{(X)}(f_k;\theta_X) \times \frac{1}{\Tobs} \delta_{kl}.
\end{equation}
This defines a prior for the noise Fourier coefficients, that depends on some parameters $\theta_X$, which characterize the PSD of the process.
It is then usual procedure to marginalize over these coefficients so that the PTA likelihood depends only on the PSD parameters $\theta_X$ \citep{2013PhRvD..87j4021L, 2014PhRvD..90j4012V}.
The observation time $\Tobs$, in Eq. \ref{eq:GP-prior}, is defined depending on whether the noise component, $X$, is intrinsic to a specific pulsar or not. 
If it is intrinsic, $\Tobs$ is the observation duration of that pulsar. 
If $X$ is a common process, such as a GWB, $\Tobs$ is the total observation duration of the PTA.

There are two main methods for inferring the timing residuals PSD, $S_{\rm r}(f)$. 
The first introduces the parametrized description resulting from a particular physical model. 
For example, the population of inspiralling binaries should produce a characteristic strain described by a power law \citep{2001astro.ph..8028P}:
\begin{equation}
    \label{eq:power-law-strain}
    h_{\rm c}(f) = \Ayr \left(\frac{f}{\fyr}\right)^\alpha,
\end{equation}
which implies a power-law form $\resPSD \propto f^{-\gamma}$ for the timing residuals PSD with $\gamma = 3 - 2\alpha$, using Eq. \ref{eq: Gaussian strain to residual PSD}. The spectral index $\gamma$ and an amplitude at fiducial frequency $\fyr = 1/\rm{yr}$ are inferred from the data using a Bayesian framework.
This requires a choice of prior for the GWB parameters $\theta_{\rm GWB} = \left(A_{\rm yr}, \gamma\right)$.
We use a log-uniform prior for the GWB amplitude, $\log_{10} \Ayr \in [-18, -10]$ and a uniform prior for its spectral index, $\gamma \in [0,7]$.

In the second approach, often referred to as "free spectrum", we infer directly the values $S_{\rm r}(f_k)$ at a set of harmonics $f_k=k/\Tobs$ assuming that they are independent {\citep{2013PhRvD..87j4021L,2013PhRvD..87d4035T}}. 
More specifically, the inferred quantities are the residuals root-mean-square (RMS), $\GWRMS(f_k) = \sqrt{\resPSD(f_k) / \Tobs}$.
Here, we only sample the RMS coefficients up to $\fyr$, choosing a log-uniform prior for each $\log_{10} \left(\GWRMS(f_k) / \text{s}\right)\in [-10, -4]$.
This method is model-agnostic, but constrained by the assumption of independence of estimate at each frequency bin. 
Given the current poor sensitivity of PTAs, both approaches give a consistent estimate of the signal PSD (see \citealp{2024ApJ...966..105A}).

In the following, we apply this PTA Bayesian spectral estimation method to the GWB-induced timing residuals of a single pulsar, obtained using the population approach described above. 
Since we are only interested in the accuracy of the spectral inference and its potential limitations (e.g., regarding spectral leakage), we simplify simulated PTA data and include/consider only white noise residuals in addition to the GWB timing residuals associated with each Universe realization\footnote{To obtain the same S/N for the GWB despite using only one pulsar, we scale its white noise residuals RMS as $\sigma_{\rm WN} = \sigma_{\rm PTA}/\sqrt{N_{\rm psr}}$, where $\sigma_{\rm PTA}$ denotes the characteristic ToA measurement error of the considered PTA made of $N_{\rm psr}$ pulsars.}. 

\subsubsection{Searching for bright individual binaries}\label{sec: CW methodo}

In the previous sections, we refer to the GW signal from the MBHB population as a GWB. 
However, the population modeling approach emphasizes that the signal is, in reality, a sum of individual sources. 
As a result, it could be that the GW signal consists of a few resolvable bright binaries, referred to as continuous wave (CW) in the following, on top of the stochastic GWB \citep{2008MNRAS.390..192S}. 
To quantify the potential detectability of individual sources, we evaluate the S/N of the brightest binaries for our 2\,000 Universe realizations, using two toy PTA datasets. 
We then consider a CW with an S/N greater than 3 as a potentially detectable individual binary.
This calculation applies only to the population modeling, and thus, for this study, to the circular ensemble.

We identify the CW candidates as follows. First, for each Universe realization, we obtain a list of potential CW candidates by identifying the individual binaries with the highest residuals amplitude $R_0$ across a frequency grid covering the PTA sensitivity band. 
For each of these brightest candidates, we then compute the associated GWB `noise' excluding the CW candidate waveform, $\tilde{r}_{\rm GWBn}(f_k) = \tilde{r}_{\rm GWB}(f_k) - \tilde{r}_{\rm CW}(f_k)$. 
We then apply a Hanning window function $w_{\rm Hanning}$ to limit spectral leakage \citep{1978IEEEP..66...51H} and then compute its periodogram to estimate its PSD. 
This effectively gives
\begin{equation}
    \label{eq: GWB PSD estimation}
    S_{\rm r}^{\rm (GWBn)}(f) =\frac{2 |(\tilde{w}_{\rm Hanning} * \tilde{r}_{\rm GWBn})(f)|^2}{\Tobs}.
\end{equation}

\begin{figure*}
   \centering
   \includegraphics[width=0.49\textwidth]{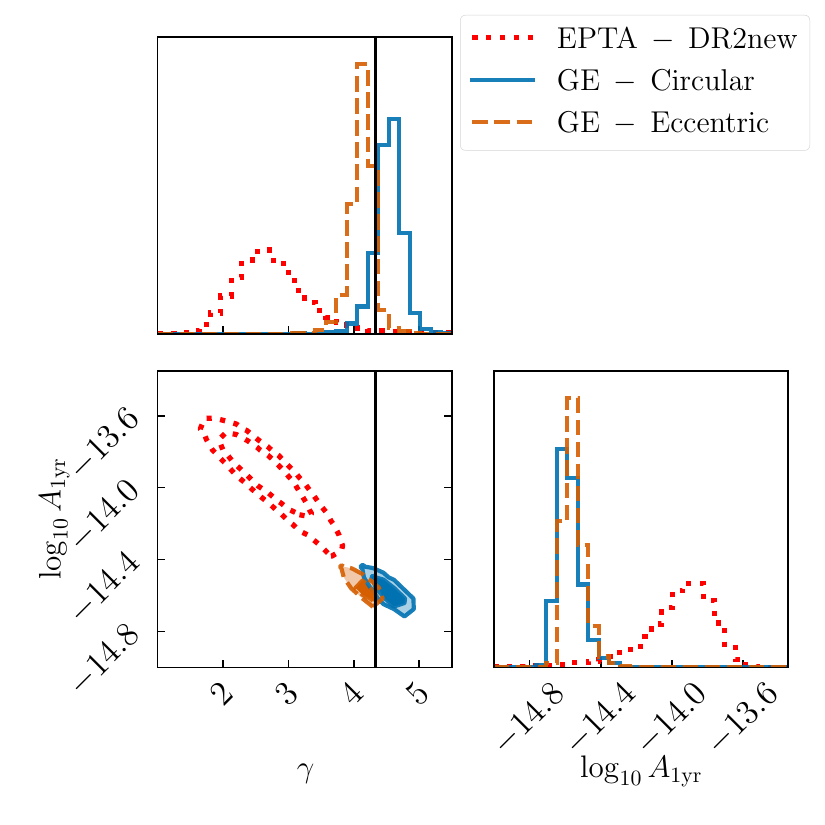}
   \includegraphics[width=0.49\textwidth]{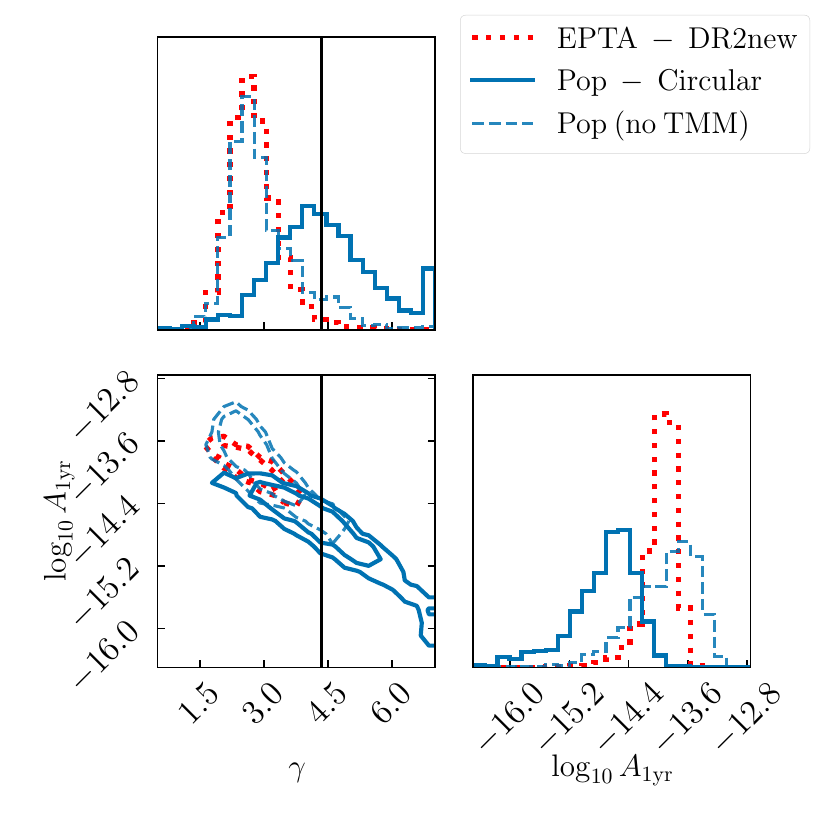}
   \caption{Distributions of the amplitude $\Ayr$ at $\fyr$ and the spectral index $\gamma$, which parameterize the GWB timing residuals PSD, derived from 2\,000 Universe realizations, are compared with the EPTA results (red dotted contours) of \cite{2023A&A...678A..50E}.
   The 68\% and 90\% confidence regions are shown for all distributions. 
   Left panel: 
   Distributions obtained via power-law least-squares fits to spectra from the Gaussian ensemble (GE) method for the circular (blue) and eccentric (orange) populations are shown as filled contours.
  Right panel: 
  Distributions of maximum-a-posteriori power-law parameters from PTA-like inference using the GWB timing residuals from the population (Pop) approach. Solid contours represent the standard data analysis procedure that includes TMM; dashed contours represent the analysis without TMM.
  The effects of spectral leakage on the inference are clearly visible without TMM and remain measurable even when TMM is included.
   In both panels, the expected spectral index, $\gamma = 13/3$, of the average spectrum from a population of circular binaries driven by GW emission, is shown as a black vertical line.
   Note: $\gamma$ relates to the characteristic strain spectral index $\alpha$ via $\gamma = 3 - 2\alpha$.}
  \label{Fig:PLfit_comp}%
\end{figure*}

The obtained GWB noise PSD is used to compute the CW candidate {optimal} match filtering S/N \citep[see e.g.,][]{2011gwpa.book.....C} for the PTA with a noise covariance matrix, $C_{\rm PTA}$, as
\begin{equation}
    \label{eq: SNR formula}
    \mathrm{S/N}^2 = \textbf{r}_{\rm CW}^{\rm T} C_{\rm PTA}^{-1} \textbf{r}_{\rm CW},
\end{equation}
where the CW residuals vector $\textbf{r}_{\rm CW}$ is the concatenation of the CW residuals for the different pulsars of the PTA. 
The noise covariance matrix $C_{\rm PTA}$ of the PTA is constructed using the \texttt{enterprise} software \citep{ellis_2020_4059815}. 
The noise matrix contains four contributions. The first one is the result of the
timing model marginalization (TMM) procedure and can be seen as a transmission function
reducing the sensitivity at low frequencies and removing 1/year frequency from the analysis \citep[see discussion above and][]{2019PhRvD.100j4028H}. 
The second contribution is an intrinsic timing red noise component for each pulsar of the PTA with a PSD modeled as a power law, $S_{\rm r}^{\rm (iRN)} \propto A_{\rm yr} (f/\fyr)^{-\gamma_{\rm iRN}}$.  
We set $\log_{10} A_{\rm yr} = -13.5$ and $\gamma_{\rm iRN} = 3$, for all the pulsars of our toy PTAs; this corresponds to the typical noise process properties found in \cite{2023A&A...678A..49E}.  
The third component is the white noise characterized by $S^{(\rm WN)}_{\rm r}(f) = 2\sigma_{\rm PTA}^2 \Delta t_{\rm obs}$, where (i) $\sigma_{\rm PTA}$ is the typical residuals RMS due to the radio telescope noise and pulse jitter \citep{2012ApJ...761...64S} defined above (ii) $\Delta t_{\rm obs}$ is the observation cadence, which we assume to be constant and equal across pulsars. 
Finally, we include the GWB, which is correlated among pulsars, using Eq. \ref{eq: GWB PSD estimation} and the Hellings-Downs correlation function \citep{1983ApJ...265L..39H}. 
We note that this assumes that the GWB noise PSD is the same for all the pulsars in the toy PTA, which is a reasonable assumption for an isotropic stochastic background.

\begin{table}[h!]
    \centering
    \renewcommand{\arraystretch}{1.25}
    \caption{Configuration of the two toy PTAs used to infer the detectability of individual MBH binary.}
    \begin{tabular}{c||c|c}
     & \texttt{EPTA} & \texttt{SKA} \\
    \hline
    \hline
    $T_{\rm obs}$ [yr] & 10.33 & 10.33 \\
    $N_{\rm psr}$ & 25 & 80 \\
    $\Delta t_{\rm obs}$ [day] & $4$ & $7$ \\
    $\sigma_{\rm PTA}$ [s] & $10^{-6}$ & $10^{-7}$ \\
    \end{tabular}
    
    \label{tab: toy PTAs}
\end{table}

The simulated PTAs and noise parameters are given in \autoref{tab: toy PTAs}. The \texttt{EPTA} configuration is based on the properties of the EPTA \texttt{DR2new} dataset presented in \cite{2023A&A...678A..48E}, while the \texttt{SKA} represents an expected configuration of the future SKA dataset \citep{2015aska.confE..37J}, choosing a similar observation duration as for \texttt{EPTA}. 
For each toy PTA and each Universe realization, we randomly assign sky positions to pulsars, uniformly distributed across the sky, and draw their distances from Earth from a normal distribution with a mean of 1 kpc and a standard deviation of 0.1 kpc.
Assuming an isotropic pulsar distribution reduces the dependence of the detectability of a CW candidate on the pulsars' positions and distances.

\begin{figure}
   \centering
   \includegraphics[width=0.495\textwidth]{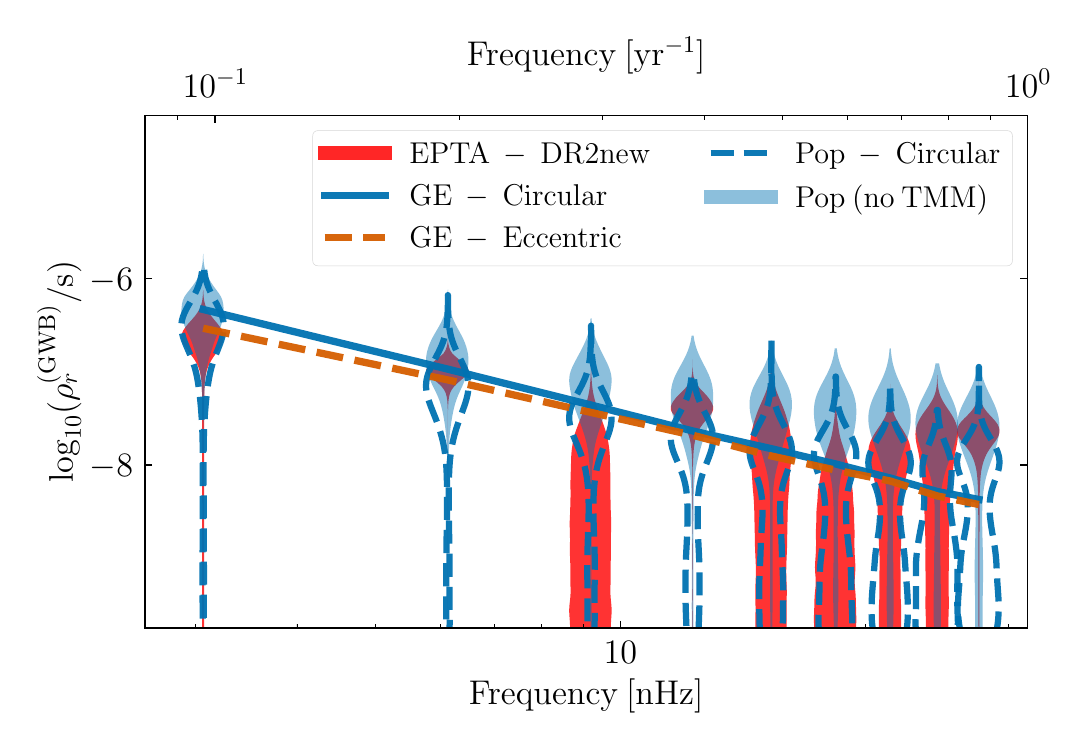}
   \caption{Comparison of the GW residuals RMS induced by the $\hagn$ population, using different modeling and populations, with the spectrum measured by the EPTA collaboration, shown in red. 
   The average GE spectra of both circular and eccentric ensembles are depicted as solid and dashed lines, respectively. 
   The distribution of the maximum-a-posteriori samples for the GWB spectra of the 2\,000 Universe realizations obtained using the  population (Pop)  method and a realistic PTA-like inference with and without TMM, are shown in dashed contours and light blue color, respectively, for the circular population.
   The effect of spectral leakage on the GWB inference results in a shallower spectral shape, significantly biasing the inference when TMM is not included. 
   For some Universe realizations, spectral leakage still has appreciable effects at high frequencies even when TMM is applied.}
  \label{Fig:FS_comp}
\end{figure}

\subsection{Results}

\subsubsection{Comparison with EPTA results}

We now examine the properties of the timing residuals induced by the GW signal from our MBHB population. 
We compare our theoretical expectations, based on the Gaussian ensemble (GE) and population approaches described above, with the current EPTA results obtained with the \texttt{DR2new} dataset, which includes residuals from 25 millisecond pulsars observed over a duration of $T_{\rm EPTA} \simeq 10.3$ years \citep{2023A&A...678A..48E}. 
PTA collaborations are providing the estimated residuals PSD using a power law and a free spectrum model. 
For the population approach, we compare the GWB spectral inference with and without the TMM procedure described above to {quantify its performance on reducing spectral leakage effects.}

Our comparison for the power-law PSD model is shown in Fig. \ref{Fig:PLfit_comp}, where the EPTA results are shown as red dotted lines in both panels.
The results using the GE approach are given in the left panel.
We show the power-law parameters obtained from a least-squares fit using a power law (assuming Eq. \ref{eq:power-law-strain} and using Eq. \ref{eq: Gaussian strain to residual PSD}) on the PSD produced by the population for each of the 2\,000 Universe realizations.  
We apply the same weight to each observer frequency $f_k$ for the least-squares fit.
For the population approach, we combine the maximum-a-posteriori samples obtained from the PTA Bayesian spectral inference (using a power-law PSD model) on the timing residuals derived from our 2\,000 Universe realizations.

Using the GE approach, we find that circular MBHBs would induce a slightly steeper residual PSD than expected from the Phinney average power law, $\gamma = 13/3 \simeq 4.33$ ($\alpha = -2/3$ in terms of $h_c$), with a 90\% confidence interval of $\gamma = 4.55^{+0.29}_{-0.37}$.
For the associated characteristic strain at $\fyr$, we find $\log_{10}\Ayr = -14.58^{+0.13}_{-0.08}$.
The eccentric population produces shallower spectra with a 90\% confidence interval of $\gamma = 4.14^{+0.23}_{-0.33}$, and induces a similar amplitude at $\fyr$, of $\log_{10}\Ayr = -14.56^{+0.12}_{-0.07}$. 
For circular and eccentric ensembles, when using the GE approach, we recover a background that is steeper and fainter (at $\fyr$) compared to the observations from the EPTA.

We compare results using the population approach and the PTA Bayesian spectral inference method in the right panel of Fig. \ref{Fig:PLfit_comp}. 
First, as shown by the blue solid contours, the PTA inference which includes TMM leads to a distribution of power-law parameters that is much broader than the GE approach.
{This is expected because the population approach accounts for both the realistic pulsar response to binary signals -- which reduces the effective number of binaries contributing to the pulsar timing residuals -- and the limited sensitivity of the PTA, unlike the GE approach}.
We find 90\% confidence intervals of $\gamma = 4.28^{+2.49}_{-1.74}$ and $\log_{10}\Ayr =-14.64^{+0.49}_{-1.06}$ for our 2\,000 Universe realizations.
The extension of the spectral index posterior toward lower values, around $3$, along with an increase in the background amplitude at $\fyr$, is likely due to residual spectral leakage effects.

Let us say a few more words about the importance of spectral leakage in the inference of the GWB.
The dashed blue contours give the posterior distribution of the inferred GWB spectra without TMM. As mentioned above, TMM works as a kind of low-pass filter, which reduces the spectral dynamical range and, therefore, reduces the leakage. The effect of GWB spectral leakage is obvious in this case: the posterior is significantly biased toward higher amplitudes and shallower spectra, with a 90\% confidence interval $\gamma = 2.77^{+1.94}_{-0.80}$ and $\log_{10} \Ayr = -13.83^{+0.48}_{-0.90}$.  If spectral leakage is improperly handled, each individual binary, as expressed in Eq. \ref{eq: r(f) ET finite duration}, contributes to the GWB spectral inference with a decaying tail $\propto 1/f^2$ at high frequencies. As a consequence, the more massive (and more numerous) binaries orbiting at low frequencies artificially contribute to the GWB spectrum at much higher frequencies than their actual GW frequency. Although the TMM procedure significantly filters out low-frequency contribution it does not eliminate the leakage completely, which still contributes to the shallower spectra part of the posterior depicted by the solid line.
We note that this effect is more evident for the more sensitive \texttt{SKA} toy PTA, where we find a 90\% confidence interval of $\gamma = 4.30^{+1.06}_{-1.24}$. The asymmetry in the confidence interval toward shallower spectra is highlighting the impact of spectral leakage on the PTA spectral inference, even when the TMM is included.\footnote{{After the submission of this work, \citet{2025arXiv250613866C} appeared on the arXiv, drawing similar conclusions. In addition, the authors propose a ready-to-use method to mitigate the spectral leakage effects by introducing the expected inter-frequency correlation in Eq. \ref{eq:GP-prior}, due to the finite duration of the data.}}
Both the increase in amplitude at $\fyr$ and the decrease in spectral index induced by spectral leakage tend to align the inferred properties of the GWB signal from the $\hagn$ MBHB population with EPTA observations.
We claim that this should be further investigated as a potential bias in the PTA spectral inference pipeline.

To have a more comprehensive view of what is happening across the PTA band, we compare our results with observations, using the model-agnostic free spectrum approach for the residuals PSD. 
The results are presented in Fig. \ref{Fig:FS_comp}.
For clarity and because the spectral uncertainties are much smaller for the GE approach, we show only the average spectrum for the circular (solid blue line) and eccentric (dashed orange line) ensembles, omitting their corresponding confidence regions.
The spectrum produced by the circular population is slightly steeper than the free spectrum suggests, while the eccentric population  better fits the observed EPTA free spectrum (red violins). For the population approach, we evaluated the free spectrum with (dashed blue violins) and without TMM (solid light blue violins). The analysis without TMM again clearly exhibits the spectral leakage at high frequencies as a flattening tail. And once again, we do observe that TMM does not completely eliminate the leakage in some Universe realizations. This remains a concern, as overestimating the GWB amplitude at $\fyr$ can skew the astrophysical interpretation of the signal.

\subsubsection{Likelihood of individually resolvable binaries}
\label{ind_res_sources}

\begin{figure}
    \centering
    \includegraphics[width=0.99\linewidth]{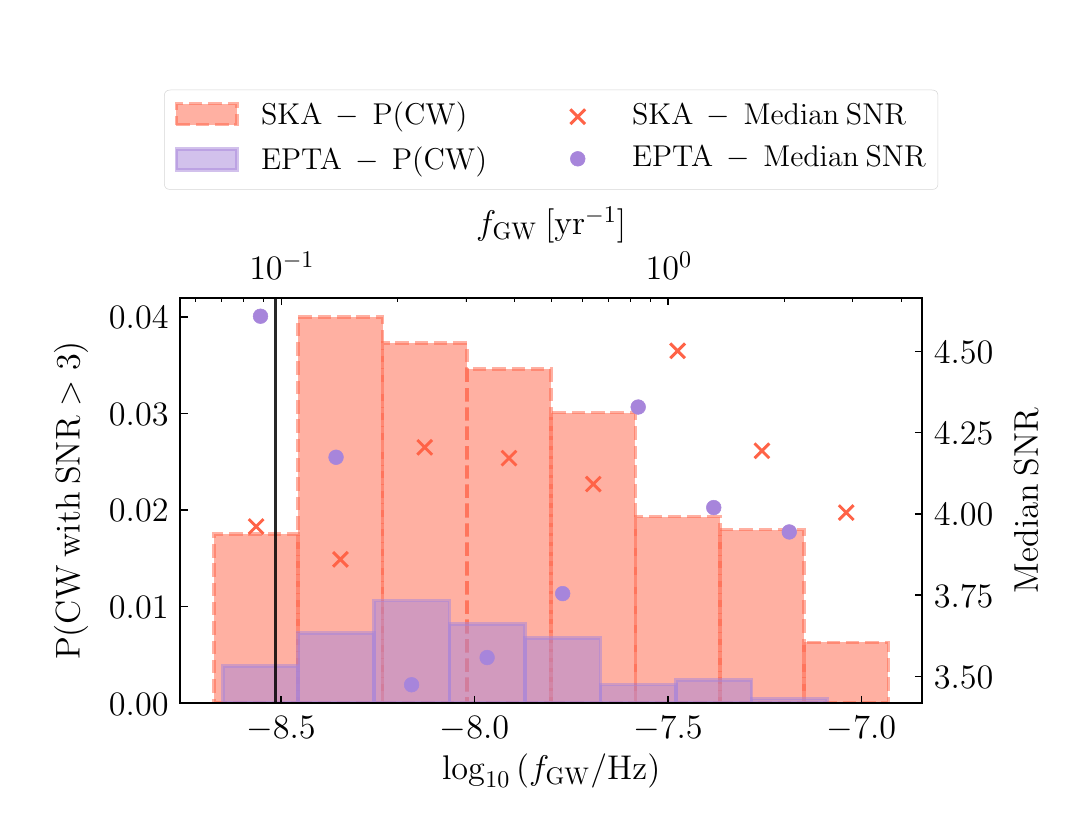}
    \caption{Probability of detecting a Continuous Wave (CW) signal from an individual MBH binary with a GW frequency in a given range, with an S/N greater than 3 for two toy PTAs representing the EPTA and the future SKA. 
    Probabilities, shown as histograms, are derived from 2\,000 Universe realizations. 
    The medians of the S/N distributions in each GW frequency band are overlaid, shown with dots (EPTA) and crosses (SKA).
    The frequency associated with the observation duration of the PTAs (10.33 years) is shown as a black vertical line.}
    \label{Fig:CW_comp}
\end{figure}

Using the method described in Sect.~\ref{sec: CW methodo}, we go on to assess the presence of binaries with an S/N greater than 3.  We refer to these as CW candidates, attributed to each of our 2\,000 Universe realizations and for the two toy PTAs.

The results are summarized in Fig. \ref{Fig:CW_comp}, where we group the CW candidates by their GW frequency in the observer frame.
Overall, we find that the toy \texttt{EPTA} (resp. \texttt{SKA}) experiment has at least one CW candidate in approximately\footnote{Using the actual sky positions of the 25 EPTA pulsars \citep{2023A&A...678A..48E}, we also find a probability of 4\%, indicating that after all, the nonuniform pulsar distribution over the sky does not have a strong impact on our results.} 4\% (resp. 20\%) of our Universe realizations.
For \texttt{EPTA}, there is a 4\% probability of having one and a 0.1\% probability of having two CWs with S/N > 3 across the frequency band. 
For the \texttt{SKA}, we find probabilities of 18\%, 2\%, and 0.3\% for having 1, 2, and 3 CW candidates, respectively.
These candidates are predominantly found in a frequency range between the lowest frequency of PTA, $1/\Tobs$, and a few times this frequency. Indeed, at lower GW frequencies, CW candidates compete with a greater number of binaries and are also affected by the timing model. 
On the other hand, bright CW sources at higher frequencies are also rare, as massive binaries evolve fast and are unlikely to be found at these frequencies, as shown earlier. 
Furthermore, due to white noise and the PTA response to the GW signal, the sensitivity drops significantly at higher frequencies \citep{2019PhRvD.100j4028H}. 
These findings are in agreement with previous studies \citep{2015MNRAS.451.2417R, 2022ApJ...941..119B}. 
However, we note that the probabilities we obtain are only lower bounds for the presence of CW candidates because our criteria for selecting CW candidates before computing their S/N are not optimal, as they do not account for the geometry of the PTA.
As a result, the MBHB with the highest residual amplitude $R_0$ may not be the one that maximizes the S/N for a particular PTA configuration.

In Fig. \ref{Fig:CW_comp}, we analyze the dependence of the median S/N of these CW candidates on their GW frequency. 
The median values in the frequency bins associated with a probability below 1\% are not statistically robust, as they include fewer than 20 CW candidates with S/N > 3 and thus could be subject to significant variance. 
Therefore, based on the \texttt{SKA} results, we find that the median S/N tends to increase with the GW frequency of the CW candidate, in agreement with previous findings \citep{2024A&A...685A..94E}.
The properties of these CW candidates will be investigated in the next sections, along with the electromagnetic properties of their host galaxies.

 \begin{figure}
   \centering
   \includegraphics[width=\columnwidth]{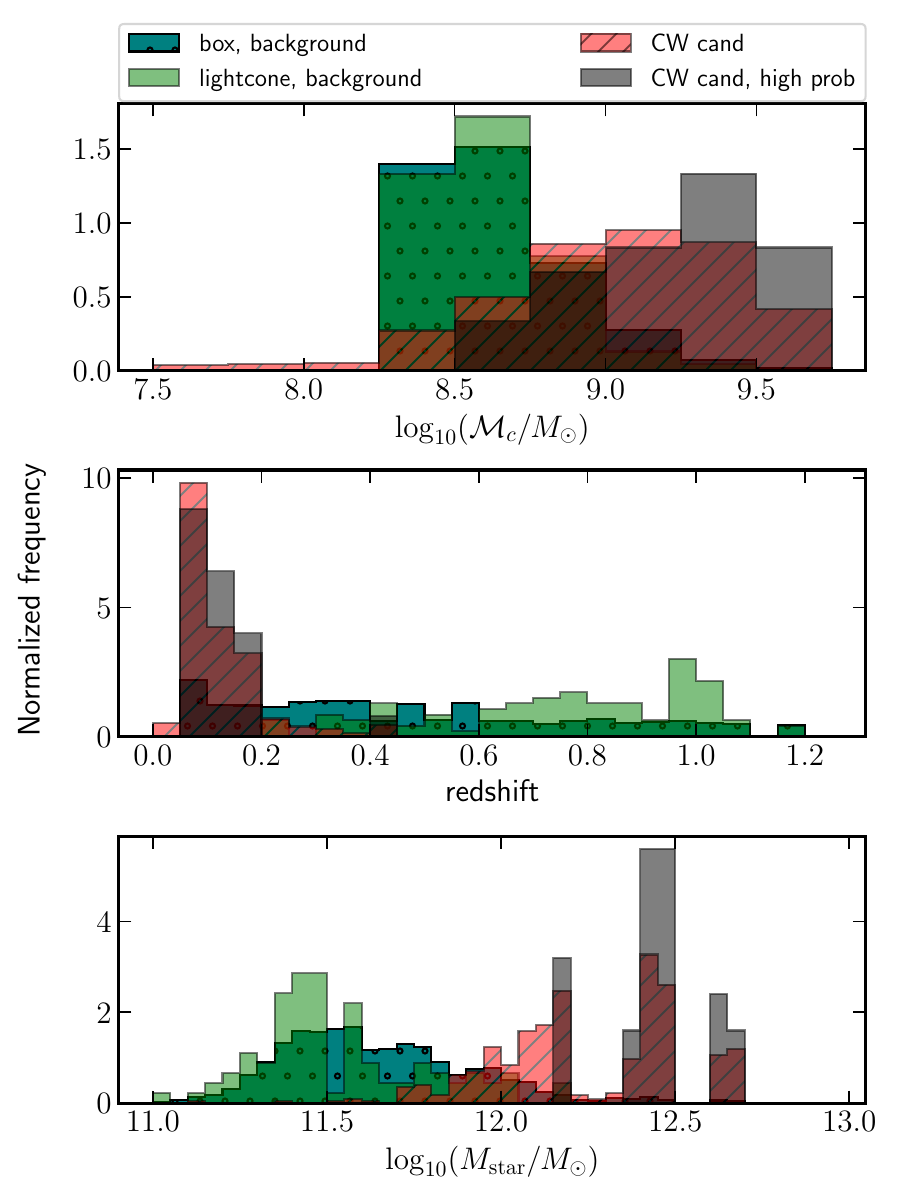}
   \caption{Properties of MBH binaries and their host galaxies, for systems dominating the background and CW candidates. From top to bottom,  we show distributions of chirp mass, redshift and stellar mass of the host galaxy. The teal dotted histogram shows the distribution in the box volume (140 cMpc)$^3$, the green histogram shows the simulation lightcone that models the galaxy distribution as seen by an observer, while the red hatched histogram shows CW candidates (`CW cand'), with the gray histogram highlighting the high probability ones. 
   In these distributions, we include CW candidates accounting for their multiplicity across Universe realizations.}
              \label{Fig:bck_vs_ind}
    \end{figure}

   \begin{figure}
   \centering
   \includegraphics[width=\columnwidth]{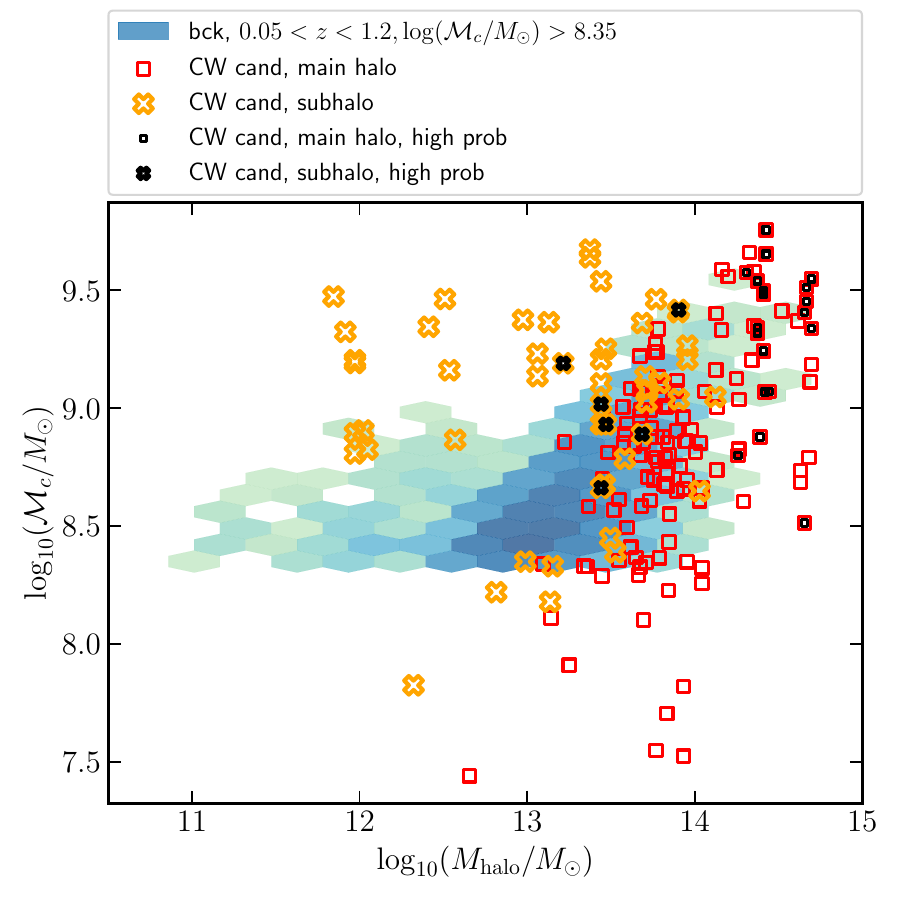}
   \caption{Environmental properties of MBH binaries and their host galaxies, for systems dominating the background (bck) and CW candidates (CW cand). From top to bottom, we show the chirp mass as a function of the stellar mass of the host galaxy and of the mass of the halo hosting the galaxy. Binaries hosted in central halos are shown as red squares; binaries hosted in subhalos are shown as orange crosses. An inner black marking highlights the high probability CW candidates.}
              \label{Fig:gal_halo}%
    \end{figure}

   \begin{figure}
   \centering
   \includegraphics[width=\columnwidth]{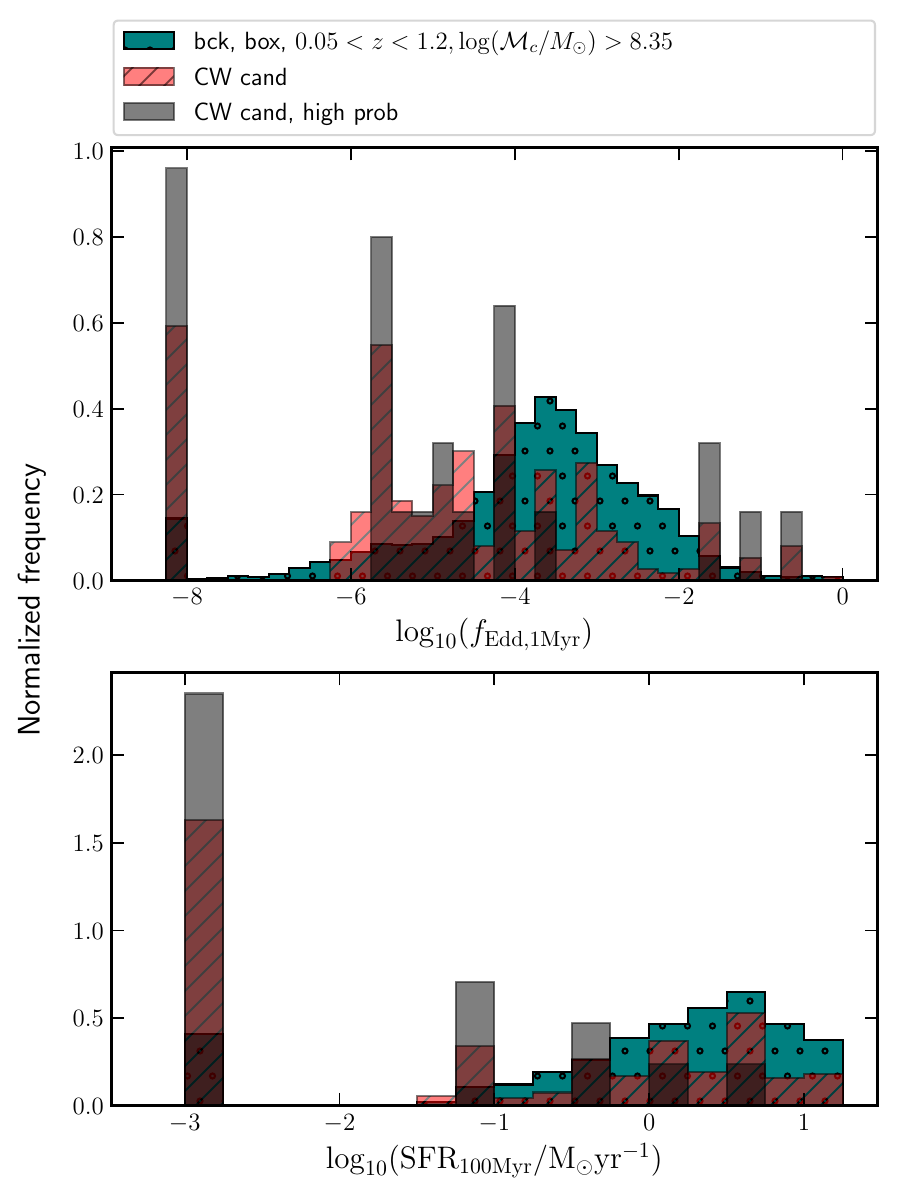}
   \caption{Properties of MBH binaries and their host galaxies, for systems dominating the background and CW candidates. 
   Top: Eddington ratio of the binary averaged over 1~Myr around the time when the MBHs merge. 
   Bottom: SFR averaged over 100~Myr. The teal dotted histogram shows the MBHBs dominating the background, selected in the simulation box.  
   The red-hatched histogram shows CW candidates, with the gray histogram highlighting the high probability ones. MBHs with $f_{\rm Edd}=0$ are shown at $f_{\rm Edd}=10^{-8}$ and galaxies with $\rm{SFR}_{\rm 100Myr}=0$ are shown at $\rm{SFR}_{\rm 100Myr}=10^{-3}$. 
   In these distributions we include CW candidates accounting for their multiplicity across Universe realizations.}
              \label{Fig:props}
    \end{figure}

 \begin{figure*}[h!]
   \centering
   \includegraphics[width=0.48\textwidth]{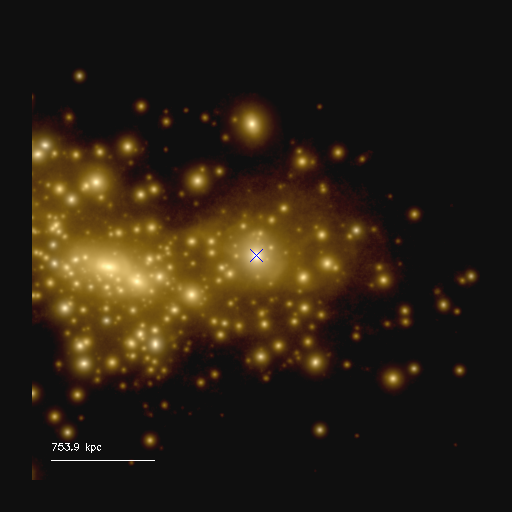}
   \includegraphics[width=0.48\textwidth]{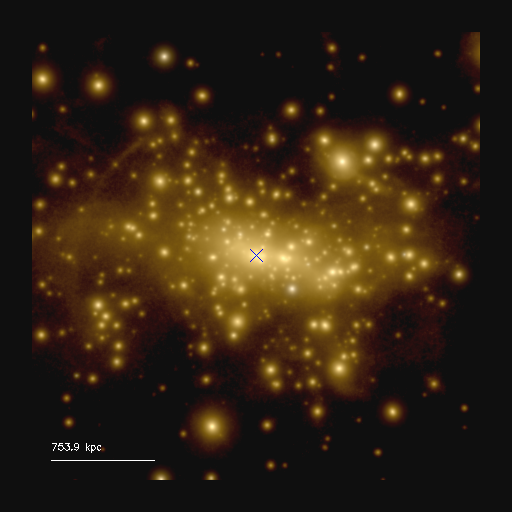}
    \caption{Left: Environment of a CW candidate binary with $\log_{10}(\mathcal{M}_{\rm c}/{\rm M_\odot})=9.02$ and total mass $\log_{10}({M}_{\rm BH}/{\rm M_\odot})=9.97$ at $z=0.025$, hosted in a galaxy with $\log_{10}(M_{\star}/{\rm M_\odot})=12.19$ in a halo of mass $\log_{10}(M_{\rm halo}/{\rm M_\odot})=13.44$. Right: Environment of a CW candidate binary with $\log_{10}(\mathcal{M}_{\rm c}/{\rm M_\odot})=9.55$ and total mass $\log_{10}({M}_{\rm BH}/{\rm M_\odot})=10.04$ at $z=0.026$, hosted in a galaxy with $\log_{10}(M_{\star}/{\rm M_\odot})=12.69$ in a halo of mass $\log_{10}(M_{\rm halo}/{\rm M_\odot})=14.70$.}
              \label{Fig:2_761}
    \end{figure*}

 \begin{figure*}[h!]
   \centering
   \includegraphics[width=0.48\textwidth]{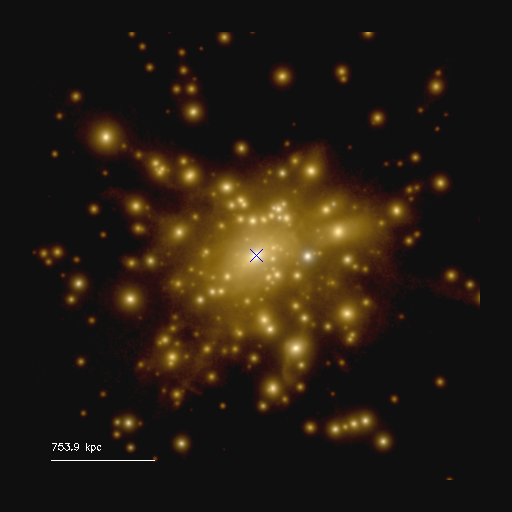}
   \includegraphics[width=0.48\textwidth]{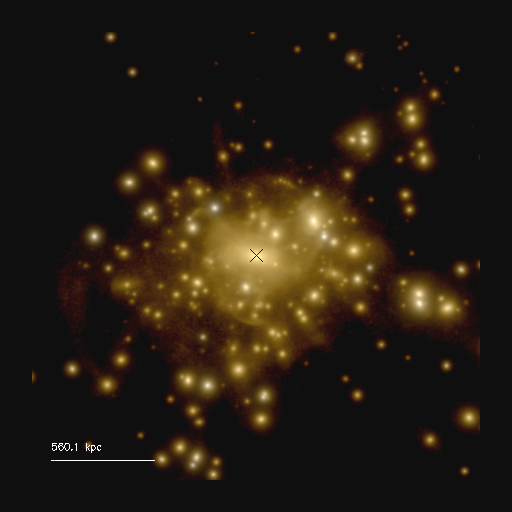}
    \caption{Left: Environment of a CW candidate binary with $\log_{10}(\mathcal{M}_{\rm c}/{\rm M_\odot})=9.07$ and total mass $\log_{10}({M}_{\rm BH}/{\rm M_\odot})=10.16$ at $z=0.032$, hosted in a galaxy with $\log_{10}(M_{\star}/{\rm M_\odot})=12.48$ in a halo of mass $\log_{10}(M_{\rm halo}/{\rm M_\odot})=14.44$. 
    Right: Environment of a CW candidate binary with $\log_{10}(\mathcal{M}_{\rm c}/{\rm M_\odot})=9.57$ and total mass $\log_{10}({M}_{\rm BH}/{\rm M_\odot})=10.06$ at $z=0.42$, hosted in a galaxy with $\log_{10}(M_{\star}/{\rm M_\odot})=12.40$ in a halo of mass $\log_{10}(M_{\rm halo}/{\rm M_\odot})=14.30$. This is one of the highest redshift CW candidates in our analysis, and it is the progenitor of the MBH shown at later redshift in the left panel.}
              \label{Fig:49_761}
    \end{figure*}

\section{Properties of MBHs and host galaxies}
\label{MBH_gal_props}

Based on the analysis presented in the previous sections, here we focus on binaries with chirp mass $>10^{8.35} \msun$ and $0.05<z<1.2$ as the dominant component of the background (Sect.~\ref{Props_contr_MBHBs}) and the selected individual CW candidates in population modeling for circular binaries (Sect.~\ref{ind_res_sources}) as representative of the population. As a general point of information, a given MBH may appear in the catalog multiple times with multiple redshifts as it merges with different MBHs. 
This is the case for the most massive MBHs, hosted in very massive galaxies, since they experience numerous mergers in their lifetime. 
Physically, this does not mean that a given binary in the Universe would be detected multiple times. However, if a simulation is a realistic representation (of a finite volume) of our Universe, the MBH binaries populating a given redshift layer will have similar properties as those merging at the corresponding redshift in the simulation. 
For a sufficiently large simulation box, which probes the large scale structure of the Universe, this is a reasonable assumption. 

First, we consider what it is that makes an MBHB a good CW candidate physically. Besides its position in the sky, near pulsars, and its inclination, which we assumed to be both random in our analysis, the only physical quantities from the simulation are chirp mass and redshift. 
We consider binaries that make the best CW candidates as those that appear most often in the various Universe realizations. 
The best CW candidates are not necessarily those that have the largest S/Ns. 
The largest S/Ns are obtained for the largest chirp masses and the lowest redshifts (luminosity distance). 
The largest chirp masses have however the shortest residence times in the PTA band \citep{2017NatAs...1..886M}, making it harder to catch them in the act. 
The lowest redshift sources are rarer because of the small volume probed. 
As a result, the binaries associated with the very high S/N are also appearing very rarely in the Universe realizations. 
We therefore consider as "high probability" CW candidates those that appear multiple times (i.e., at least five times in this work) across our $2\,000$ realizations.

In the following, we  consider both the simulation box, which is an evolving patch of the Universe that we can analyze at different times, and the simulation lightcone, which mimics what and how an observational survey would ``see'' the simulation. When we move from the simulation box to the lightcone generated for the simulation \citep{laigle19} a given MBH appears only once, and the cosmic time at which the MBH appears in the lightcone may or may not coincide with the time of an MBH merger. This means that only a small fraction of merging binaries are contained in the lightcone. 

In Fig. \ref{Fig:bck_vs_ind}, we compare the properties of  mergers dominating the background in the simulation box which has a fixed co-moving volume (teal), in the simulation lightcone (light green) which simulates the source distribution seen by an observer, and the CW candidates (red). 
In these distributions, we include CW candidates accounting for their multiplicity across Universe realizations, as the multiplicity of a given candidate in the various realizations provides information on how strong a candidate is as it provides information on its likelihood. 
In a cosmological volume, and in a lightcone that has increasing volume as redshift increases, the background is dominated by the more abundant MBHs with mass $\sim 10^{8.5} \msun$, rather than the stronger but rarer GW sources with larger mass. 
CW candidates are instead mainly found at low redshift and have higher MBH masses; some can also have relatively low masses, down to $\sim 3\times 10^7 \msun$, but they are all at $z<0.1$. 
The redshift distribution extends to higher redshifts in the lightcone, because of the larger volume probed with increasing redshift, while it is limited to lower redshifts for CW candidates, because the strength of the GW signal decreases with increasing luminosity distance, hence redshift. 
The degeneracy between luminosity distance (redshift) and chirp mass (see Eqs.~\ref{eq: GW luminosity circ} and~\ref{eq: single binary hc2}) allows for the presence of CW candidates up to $z=0.4$, provided that they have chirp mass $\mathcal{M}_{\rm c}\sim 10^{9.5} \, \msun$. 
The distribution of the masses of the host galaxies reflects the  $\mathcal{M}_{\rm c}$ and $z$ distributions for the background, while for CW candidates it is dominated by massive galaxies, in line with the larger MBH masses with respect to the background.  {These results are in agreement with previous and concurrent investigations \citep[e.g.,][]{2009MNRAS.394.2255S,2025CQGra..42b5021C,2025arXiv250401074T}}. 

We examine the relation between binaries and environment in Fig. \ref{Fig:gal_halo}. Here, $\mathcal{M}_{\rm c}$ is shown as a function of the halo mass of the host. We note some ``undermassive binaries,'' with very low chirp mass with respect to the host galaxy/halo: these are cases where the mass ratio in the binary is very small, and therefore the chirp mass differs significantly from the binary mass, which is what scales with host properties in the simulation \citep[shown in][]{2020MNRAS.498.2219V}.    
The binaries dominating the background in the simulation box are primarily hosted in galaxies and halos that are more massive than the Milky Way, while CW candidates are, for the most part, hosted in group-like main halos (red squares) with $M_{\rm halo}>10^{13} \msun$. 
CW candidates that are hosted in lower mass halos are all located in subhalos (orange crosses), meaning that the subhalo has been affected by stripping and it is part of a larger halo\footnote{The plume of binaries in the background with halo mass $<10^{12} \msun$ is also caused by subhalos.}; the high chirp mass with respect to halo mass is explained by stripping of the halo, which has therefore lost part of its mass \citep[see the discussion in][for the general MBH population]{2016MNRAS.460.2979V}. 
This implies that CW candidates are best looked for in galaxy groups and clusters, even more so than for the general population of MBH binaries \citep{2014MNRAS.439.3986R,2023MNRAS.519.2083I,2024MNRAS.529.4295S}. Using groups/clusters as priors for where to look for CW candidates in the Bayesian search of PTA data could be therefore a possible strategy. 

Furthermore, the MBHs in the central galaxies of overdensities experience a large number of mergers, up to $\sim 20-25$ for the MBHBs that dominate the nHz background. This is a manifestation of the significant fraction of MBH mass gained through MBH mergers for these galaxies \citep{2013ApJ...768...29V,2014MNRAS.440.1590D,2015ApJ...799..178K}. This occurs because of a combination of enhanced galaxy mergers and decreased gas availability for accretion for high-mass MBHs hosted in massive central galaxies of groups and clusters.

In Fig. \ref{Fig:props}, we compare the Eddington ratios and star formation rate (SFR) of background sources in the box vs CW candidates. 
We average SFR over 100~Myr since most star formation diagnostics probe  SFR on such timescales and we average $f_{\rm Edd}$ over 1~Myr to smooth the variability. 
Also in this case we include CW candidates accounting for their multiplicity across Universe realizations. 
The Eddington ratio of the binaries is generally low, $<10^{-2}$, typical for MBHs of this mass and redshift in the simulation as well as in observations \citep{2016MNRAS.460.2979V}. 
Accretion is even lower for CW candidates, since they are at very low redshift, in very massive galaxies. 
The same considerations apply to SFR, in agreement with previous results {which find that the host has low SFR or specific SFR} \citep{2024MNRAS.529.4295S,2025arXiv250401074T,2025CQGra..42b5021C}. Since we are considering  delayed mergers rather than numerical mergers, the time elapsed from galaxy merger to binary observation is sufficiently long that any galaxy merger-driven increase in accretion rate or SFR has since decayed  \citep{2023A&A...673A.120D}.

Examples of hosts of CW candidates in false gri colors are shown in Fig. \ref{Fig:2_761} and Fig. \ref{Fig:49_761}. The MBHB is located at the center of the image. The first three images (Fig. \ref{Fig:2_761} and Fig. \ref{Fig:49_761} left) show two low-redshift cases consistently appearing in the realizations, while Fig. \ref{Fig:49_761}, right, shows a rarer high-redshift ($z=0.42$) case. The images highlight that CW candidates are preferentially found in the center of galaxy groups and clusters, in agreement with the results shown in Fig. \ref{Fig:gal_halo}.

\section{Electromagnetic emission from galaxies and MBHs}
\label{EM}

  \begin{figure*}
   \centering
   \includegraphics[width=0.49\textwidth]{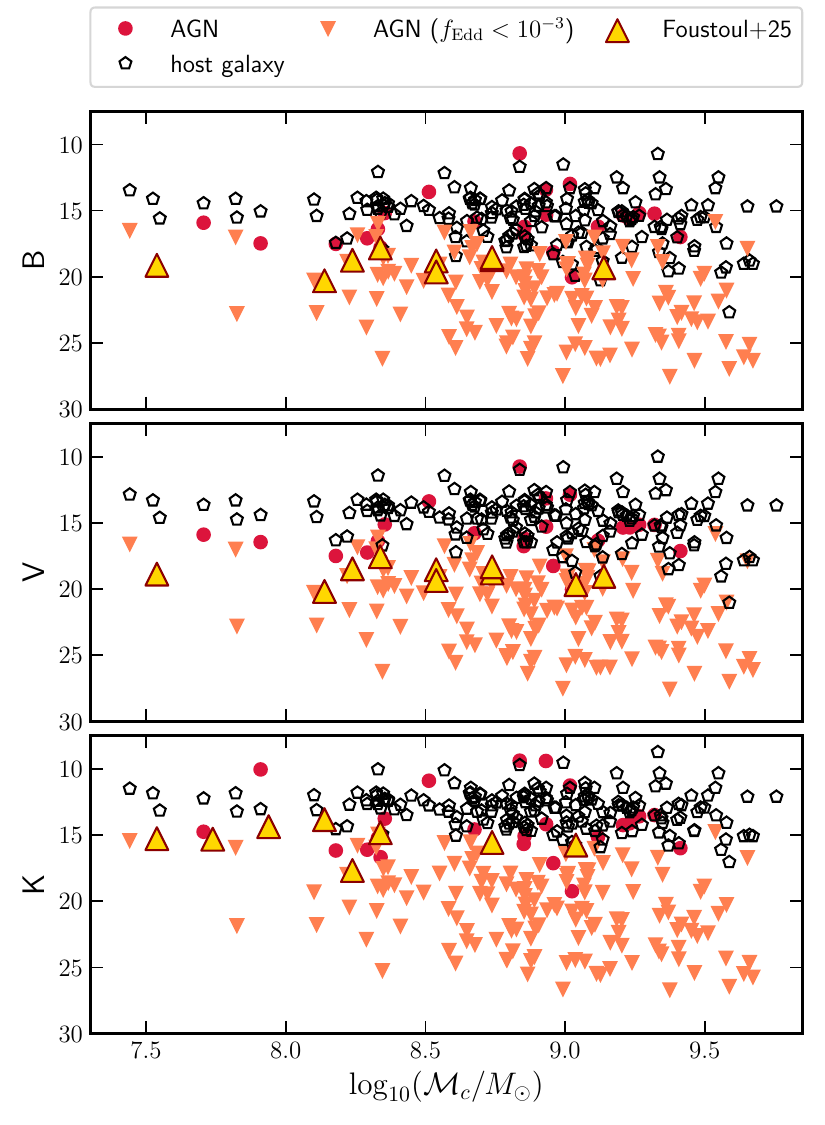}
   \includegraphics[width=0.49\textwidth]{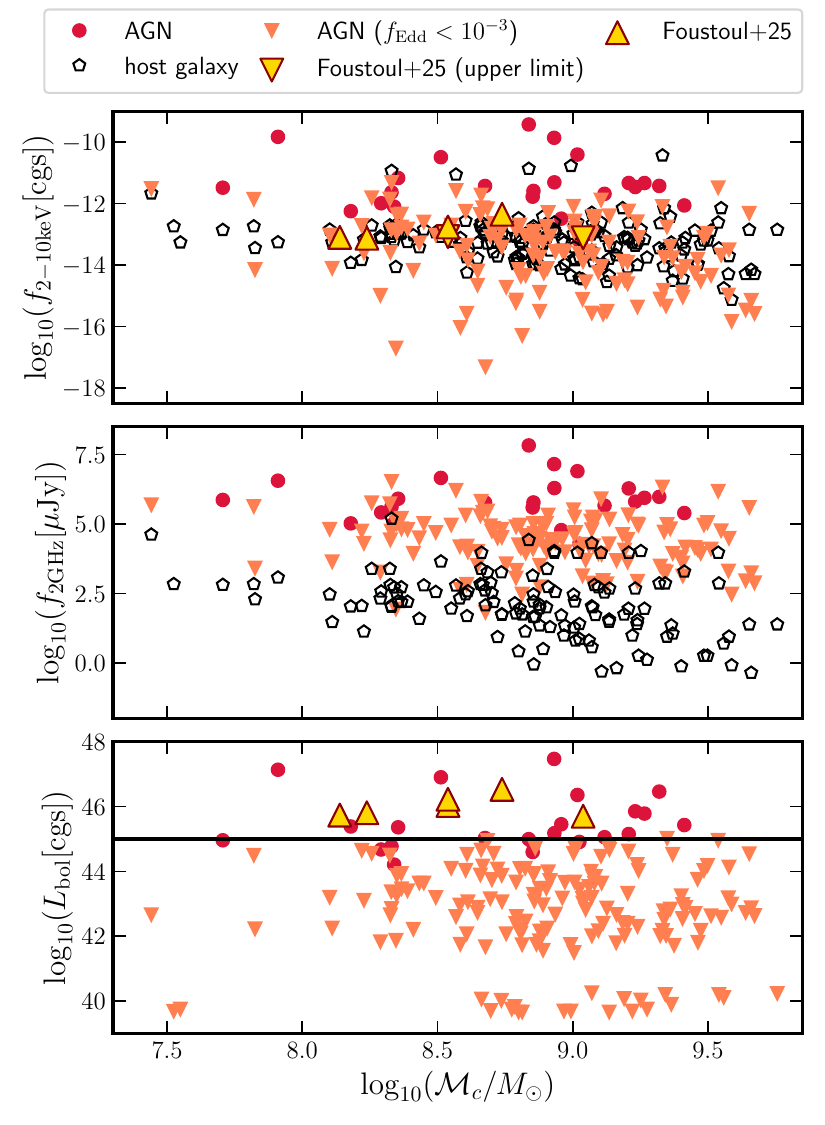}
   \caption{Electromagnetic properties of MBHBs and their host galaxies for CW candidates. The left panels show B, V, and K magnitudes; the right panels show X-ray flux at [2-10] keV, radio flux at 2~GHz and bolometric luminosity (the horizontal line marks $L_{\rm bol}=10^{45}\, \rm{erg}\, {\rm s^{-1}} $), as a typical value for quasar luminosities. For MBHs with $f_{\rm Edd}<10^{-3}$ we show an optimistic upper limit to the brightness ignoring the correction to radiative efficiency.}
              \label{Fig:EMflux}%
    \end{figure*}
 
\subsection{Methodology}
\label{EM_modeling}

In this paper we do not model specific binary signatures and we instead focus on the sheer detectability of AGN and host galaxies, addressing two points. The first is whether the galaxy/AGN are within the reach of observational surveys. The second is whether the AGN emission can outshine the emission from the host galaxy. 

Galaxies magnitudes are computed as described in \cite{laigle19}. In brief, we compute the galaxy rest-frame spectral energy distribution by summing up the contribution of  all stellar particles (knowing their ages and metallicities), assumed to behave as single stellar populations. We use the stellar population synthesis model of \cite{bc03} assuming a \cite{chabrier03} initial mass function. The dust column density and optical depth along the line of sight of  each star particle is computed assuming that 30\% of metals in the galaxy gas distribution are locked into dust grains. We use the $R_{\rm V}=3.1$ Milky Way dust grain model by \cite{WeingartnerDraine2001}. Apparent total magnitudes are computed on the redshifted spectra, using the $\rm B$, $\rm V$ and $\rm K_{\rm s}$ filters.  

In X-rays, the galaxy emission is dominated by X-ray binaries composed of a compact object accreting from a companion star.  Empirical models allow  for the   combined effect of low-mass X-ray binaries to be estimated, where the companion is a low-mass star and the emission depends on the total galaxy mass, and high-mass X-ray binaries, where the companion is a high-mass star and the emission depends on the SFR. These models are  based either on individual observations of nearby galaxies \citep[e.g., $<50 \rm \, Mpc$,][]{2019ApJS..243....3L} or on stacked galaxies observed up to high redshift \citep[$z<5$]{Fornasini2018}. We calculate the combined X-ray luminosity of X-ray binaries in galaxies  using the relations of \cite{Fornasini2018}. 

We further estimate the radio emission from stellar populations in galaxies based on the relation between SFR and radio luminosity derived by \cite{2003ApJ...586..794B}. This relation fits SFR as a function of radio luminosity, rather than vice versa, which is what we need. \citet{2009MNRAS.397.1101G}, however, show that the relation provides a good fit to radio luminosity as a function of SFR. We therefore simply invert the equation to obtain radio luminosity as a function of SFR.

We model the emission from MBHs by considering the total mass of the binary and the accretion rate recorded in the simulation averaged over 1~Myr around the time when the MBHs  merge. 
This is to account for that the MBHs can be observed at any point during the time spent in band, and this time is generally of the order of a Myr (see Fig. \ref{Fig:Dt_evol}). 

We model the AGN emission from X-ray to infrared (IR) using the average spectral energy distribution for AGN developed in \citet{2020MNRAS.495.3252S}. We add a correction for dust by estimating the hydrogen column density within a sphere of radius $4\,\rm kpc$, assuming a dust-to-gas mass ratio of $10^{-2}$ and a Milky Way-like extinction curve. For simplicity we fit the hydrogen column density for the MBH population as
\begin{equation}
\log_{10}\left(\frac{N_{\rm H}}{\rm cm^{-2}}\right)=22.54-0.23(1+z)+[0.65-0.1(1+z)]\log_{10}(f_{\rm Edd}),
\end{equation}
 and then apply this scaling to the binaries. 

The AGN radio luminosity is calculated via the fundamental plane of black hole accretion, an empirical correlation between the MBH mass, the radio and X-ray power-law continuum luminosities \citep[][note that this is the core radio luminosity and not the total luminosity including extended jets]{2019ApJ...871...80G}:
\begin{equation}
    \log_{10} L_{\mathrm{R},5} = 4.8 + 0.78 \log_{10} \left(\frac{M_{\rm BH}}{\msun}\right) + 0.67 \log_{10} L_\mathrm{X},
    \label{eq:FP}
\end{equation}
where $L_{\mathrm{R},5}$ is the radio luminosity at $5\,\rm GHz$ (rest frame), and $L_\mathrm{X}$ is the X-ray luminosity in the rest-frame $2-10\,\rm keV$ energy range, both in $\rm erg\,s^{-1}$. Radio luminosity is calculated at $2\,\rm GHz$ (observer frame), assuming a power-law spectrum with index $-0.7$ \citep{2014ApJ...788L..22G}. 

We should mention important caveats. For MBHs with Eddington ratio $\lesssim 10^{-2}-10^{-3}$, the total luminosity should be rescaled to account for radiatively inefficient accretion discs, and the spectral energy distribution differs overall from the radiatively efficient case. We compared the results with the specific radiatively inefficient solution developed by \citet{2014MNRAS.438.2804N} for a range of masses and accretion rates, and found that rescaling the X-ray to IR spectral energy distribution by a factor $f_{\rm Edd}$ \citep{2008MNRAS.388.1011M} gives an estimate within a factor of about 5 from the \citet{2014MNRAS.438.2804N} solution. Given the large uncertainties, we report the radiatively inefficient sources, which we define as having $f_{\rm Edd}<10^{-3}$, with an optimistic estimate corresponding to the radiatively efficient case, and we discuss the case corresponding to the same multiplied by $f_{\rm Edd}$ to assess the effect of suppressing the radiative efficiency.

\subsection{Electromagnetic properties of MBHs and host galaxies}
\label{EM_props}

We compare the brightness of CW candidates and their host galaxies in Fig. \ref{Fig:EMflux}. The host galaxies are shown as black empty pentagons and the AGN as red filled dots for $f_{\rm Edd}>10^{-3}$ and as orange triangles otherwise, optimistically assuming radiatively efficient accretion even at low Eddington ratios.  Appendix~\ref{EM_bck} compares with radiatively inefficient accretion.  

The first point to note is that the large majority of MBHBs have bolometric luminosities below $10^{45}\,\rm erg\,s^{-1}$, the typical dividing line between quasars (brighter sources) and AGN (fainter sources). 
This is because, although the MBHs are very massive, their typical accretion rates are very low (Fig. \ref{Fig:props}). 
Only 9\% of CW candidates, and 16\% of the high probability CW candidates, have  $L_{\rm bol}>10^{45}\,\rm erg\,s^{-1}$.
More generally, the electromagnetic properties of the CW candidates are very similar to those of high-probability CW candidates.
The fraction is about 3\% for MBHBs dominating the background (see Appendix~\ref{EM_bck}). 
This is in agreement with \citet{2025arXiv250401074T}.  A related, but slightly different question, is the binary fraction in quasars. The fraction of quasars with $L_{\rm bol}>10^{45}\,\rm erg\,s^{-1}$ that host binaries close to coalescence -- without a restriction to nHz frequencies -- is 7\% in our model. \citet{2025ApJ...987..106C} also find a low binary fraction in quasars ($\sim 4$\%).

In the optical and near-IR (NIR) bands (B, V, and K) the host galaxies are generally  bright, with magnitudes brighter than 20, while about half of the AGN have magnitudes fainter than 20. {The results are in qualitative agreement with \citet{2025ApJ...990...46V}, although we do not find a similarly strong dependence of the photometric properties of the binary and its host on the binary mass}.
 Even in this optimistic scenario for the radiative efficiency, none of the MBHBs with $f_{\rm Edd}<10^{-3}$ are brighter than the host galaxies, while for MBHBs with higher accretion rates, 34\%, 17\% and 26\% are brighter than the host in B, V, and K respectively. 
In radio and X-rays, all MBHBs with $f_{\rm Edd}>10^{-3}$ are brighter than the host, while the low accretors can be outshined by the stellar emission from the host galaxy. 
If we optimistically do not include a correction for radiative inefficiency, about 30\% (X-ray) and 80\% (radio) can be brighter than the host. Results are qualitatively similar for binaries dominating the background. 
In Appendix~\ref{EM_bck}, we  explicitly show the ratio of AGN and host galaxy luminosity for this optimistic case as well as the results when applying a correction for radiatively inefficient sources. Fig. \ref{Fig:EMflux_ratio} highlights how MBHBs with $f_{\rm Edd}<10^{-3}$, which are the majority of the sample, are in the more pessimistic case always fainter than the galaxy emission, even in X-ray and radio. 
Overall, however, we consider the radio band the best suited for identifying MBHBs with less contamination from the host galaxy, followed by X-rays.

The preponderance of faint AGN outshined by their host galaxies among CW candidates makes searches for MBHBs harder: variations of the MBHB light curve have to be extracted from the brighter nonvariable (unless there is a rare event like a supernova) galaxy emission. 
Electromagnetic searches for MBHBs are therefore often limited to the brightest MBHBs, which are however a minority with respect to the full MBHB population. 
For instance, \citet{foustoul_2025} searched for MBHB candidates, looking for continuous periodic modulation in the optical light curves (g- and r-bands) of sources coincident with the centers of galaxies. 
Each candidate found in \citet{foustoul_2025} was shown to be a quasar and the luminosity of the central MBHB candidate outshines the host galaxy. 
Plotted in Fig. \ref{Fig:EMflux} are the electromagnetic properties of these MBHB candidates at redshift $0.4<z<1.0$. 
We show all candidates for which there are K-, V- and B-band photometry (left-hand plots) or X-ray detections/upper limits (right-hand plots). 
Considering an orbital origin for the optical variability, the separation of these candidates is between 4 and 13\,mpc, similar to those in the simulations and making them potential PTA CW sources. We calculated their chirp mass, shown in Fig. \ref{Fig:EMflux}, supposing an equal mass binary and with masses estimated from their SDSS optical spectrum in \cite{Masse_estimates_rakshit}. 
The bolometric luminosities were estimated using their X-ray detections/upper limits and the bolometric correction computed for type I AGN in \cite{2020A&A...636A..73D}. 
The MBHB candidates have B, V, and K magnitudes slightly fainter than many of the galaxies and AGN in Fig. \ref{Fig:EMflux} as they are on average at greater distances than the CW candidates in the simulation. 
The X-ray fluxes are similar to the CW candidates, but again slightly fainter than the AGN due to the greater distances. 
Accounting for distance (redshift) by considering the X-ray luminosity, the electromagnetic candidates show very similar luminosities as the AGN and indeed all of them are quasars (see also \citealt{foustoul_2025}). 
Calculating the Eddington ratios for the MBHB candidates, we determined ratio values between 0.05 and 0.3, indicating high accretion rates that would explain the high bolometric luminosities for these candidates. 
Overall, the electromagnetic properties of the MBHB candidates are similar to the expected properties of CW from the simulation with the exception that the candidates discovered through optical observations are generally at higher redshifts. 
This is probably because only $\sim$20\% of MBHBs are expected to outshine their galaxies in the g- and r-bands, so a larger volume must be searched to identify this rarer sample.

\section{Conclusions}\label{Ccl}

This paper is based on the results of the hydrodynamical cosmological simulation \hagn~\citep{Dubois14}, which was used to extract MBHBs and investigate their properties. 
The hardening of MBH pairs identified in \hagn~ was followed through their interaction with the stellar and gaseous environment in post-processing, producing a catalog of inspiralling MBHBs emitting in the PTA band. 
Each MBHB in the simulation was assigned an initial orbital eccentricity motivated by the numerical results of \cite{Roedig2011}.
In addition, we  also considered a purely circular population of MBHBs.
Based on these results, we produced $2\,000$ Universe realizations reflecting the stochasticity {or cosmic variance, inherent to the binary population}. 
The GW signal from an individual MBHB is almost monochromatic over the PTAs observation times for circular binaries. 
The power of the GW signal from eccentric MBHBs is distributed over harmonics of the orbital (mean anomaly) frequencies with a peak shifting to higher modes as eccentricity increases.
The incoherent superposition of GWs from multiple MBHBs creates a stochastic signal (referred to as a GWB) at low frequencies. 
Particularly bright sources that stand above the GWB could be potentially resolved (CW candidates).
We  considered two PTAs: one based on the recent EPTA DR2new dataset and one that reflects our projection of future SKA observations. 
In this paper, we study the properties of MBHBs forming the GWB and CW candidates. Our main findings can be summarized as follows.

\begin{itemize}

\item {Generation of the GW strain from the population of MBHBs.} 
We  considered two approaches to building predictions of the GWB in the PTA band: GE-based and population-based. 
We claim that the population approach is more reliable and closer to the actual observations.
The GE approach is one of the most commonly used in the literature (see e.g., \citealp{2015MNRAS.451.2417R, 2023ApJ...952L..37A, 2024A&A...685A..94E}).
It converts the characteristic GW strain from the population of MBHBs into a timing residuals power spectral density using the pulsar response function averaged over sky position and polarization.
However, this approach has several limitations.
First, it does not account for either the finite duration of observations or the finite sensitivity of the PTA, both of which can significantly impact the inference of the GW background spectrum.
Second, the GE approach underestimates {the cosmic variance of the GW signal observed by PTAs, since it does not account for the reduction in the effective number of binaries contributing to the GW-induced timing residuals caused by the geometrical response of the pulsars.}
We show that the power spectra obtained using this method can deviate significantly from those inferred from PTA observations.
The GW frequencies probed by PTAs are very close to the lowest frequency (defined as the inverse of the observation duration), and the effect of spectral leakage (albeit reduced by the timing model marginalization in the PTA data analysis) could potentially bias the GWB spectral inference toward a shallower power law with a higher amplitude at $f=1/\rm year$. 
This bias propagates in the astrophysical predictions. This should be carefully investigated for both circular and eccentric populations.

\item {Stochasticity of the combined GW signal at low frequencies.} 
Based on $2\,000$ Universe realizations we find that for both populations around an observing frequency of 3 nHz the GW signal is dominated by hundreds to thousands of binaries.
Such high numbers confirm that at low frequencies ($1$--$10$ nHz), the GW signal is consistent with a Gaussian stochastic background.
In addition to the GWB, we find that CW candidates with S/N $> 3$ could be found in $4\%$ ($20\%$) of our Universe realizations using the EPTA (SKA) sensitivity, assuming a ten year-long observing duration.
In most cases, their GW frequency is below 10 nHz.

\item {Characteristics of MBHBs at low GW frequencies.} 
The largest contribution to the GWB at 3 nHz (low frequencies) comes from heavy binaries with the chirp mass in the range $10^{8.5}$–$10^{9.5}\,\msun$ and within the redshift layer $0.05<z<1.2$.
They are hosted primarily in galaxies and halos that are more massive than the Milky Way, with a redshift distribution peaking around $0.3$--$0.4$.
The peak of the mass distribution for the eccentric population is slightly shifted toward higher-mass MBHBs.
These sources are eccentric binaries at very low orbital frequencies that contribute to the $1$--$10$ nHz PTA band by emitting at high orbital harmonics.  
CW candidates are more massive and located at lower redshift.
For the most part, they are hosted in group and cluster-like main halos with masses $10^{13}-10^{15}\; \msun$ at low redshift, up to $z \sim 0.4$. A fraction of them are found to be hosted in lower mass subhalos.

\item {Electromagnetic properties of the MBHBs contributing to GW strain in the PTA band.} 
We find that the binaries contributing most to the background, as well as the CW candidates, will typically appear as AGN rather than quasars, the latter corresponding to only $9$–$16$\% of the CW candidates and just $3$\% of the background binaries.  
This is a consequence of their low Eddington ratio (below $10^{-2}$), which is typical of massive MBHs at low redshift. 
The host galaxies are expected to have low star formation rates, resulting from the time delay between the galaxy merger and the GW emission phase in the PTA band.
However, we find that the CW candidates can outshine their host galaxies, particularly in the X-ray and radio bands, provided that their accretion rate satisfies $f_{\rm Edd} > 10^{-3}$. 
Even for slow-accreting systems, under optimistic assumptions up to $80$\% ($30$\%) of the CW candidates are brighter than their hosts in the radio (X-ray) band.
This suggests that the radio band is the best suited for identifying multi-messenger MBHBs brighter than the host galaxies.
Overall, since CW candidates are typically outshined by their hosts in the optical and NIR bands, searching for binaries using light curve variability in galactic nuclei is expected to be challenging.
The existing MBHB candidates identified using such methods exhibit larger bolometric luminosities than the CW candidates extracted from our simulation.
This indicates a potential selection bias in favor of more luminous (but rarer) MBHB candidates; alternatively,  it suggests that there must be another explanation behind the apparent periodicity.
We also noticed that CW candidates are mostly located in galaxy groups and clusters, suggesting that the CW sky location prior could follow the number density distribution of observed galaxies.

\item {The combined GW strain at high frequencies.} 
At high frequencies, $f_{\rm GW }\ \sim 1/\rm year$, the number of MBHBs contributing to GW strain drops to a few tens (several tens) for the circular (eccentric) population, indicating that the combined GW signal loses stochasticity. 
The GW strain from the eccentric population remains stochastic to a somewhat higher frequency due to the high harmonics contribution of MBHBs with low orbital frequency. 
This small number of binaries implies a larger cosmic variance, especially for the circular population, which is reflected in the properties of the contributing binaries at these frequencies.
High-frequency MBHBs have lower chirp masses (in the range $10^8$--$10^9\,\msun$), compared to those contributing to the low-frequency GWB and are found at higher redshifts, typically in the $0.4$--$1$ range.

\end{itemize}

\section*{Data availability}

The data that support the findings of this study are openly available at \url{https://doi.org/10.5281/zenodo.18391289}.

\begin{acknowledgements}
This work was made possible by funding from the French National Research Agency (grant ANR-21-CE31-0026, project MBH\_waves). RSB acknowledges support from a UKRI Future Leaders Fellowship (grant code: MR/Y015517/1). This work has received funding from the Centre National d’Etudes Spatiales.
Numerical computations were partly performed on the S-CAPAD/DANTE platform, IPGP, France.
\end{acknowledgements}

\bibliographystyle{aa}
\bibliography{bibliography}

@ARTICLE{lescaudronetal23,
       author = {{Lescaudron}, Sandrine and {Dubois}, Yohan and {Beckmann}, Ricarda S. and {Volonteri}, Marta},
        title = "{Dynamical friction of a massive black hole in a turbulent gaseous medium}",
      journal = {\aap},
         year = 2023,
        month = jun,
       volume = {674},
          eid = {A217},
        pages = {A217},
          doi = {10.1051/0004-6361/202243392},
archivePrefix = {arXiv},
       eprint = {2209.13548},
 primaryClass = {astro-ph.GA},
       adsurl = {https://ui.adsabs.harvard.edu/abs/2023A&A...674A.217L}
}

@ARTICLE{2020MNRAS.498.2219V,
       author = {{Volonteri}, Marta and {Pfister}, Hugo and {Beckmann}, Ricarda S. and {Dubois}, Yohan and {Colpi}, Monica and {Conselice}, Christopher J. and {Dotti}, Massimo and {Martin}, Garreth and {Jackson}, Ryan and {Kraljic}, Katarina and {Pichon}, Christophe and {Trebitsch}, Maxime and {Yi}, Sukyoung K. and {Devriendt}, Julien and {Peirani}, S{\'e}bastien},
        title = "{Black hole mergers from dwarf to massive galaxies with the NewHorizon and Horizon-AGN simulations}",
      journal = {\mnras},
         year = 2020,
        month = oct,
       volume = {498},
       number = {2},
        pages = {2219-2238},
          doi = {10.1093/mnras/staa2384},
archivePrefix = {arXiv},
       eprint = {2005.04902},
 primaryClass = {astro-ph.GA},
       adsurl = {https://ui.adsabs.harvard.edu/abs/2020MNRAS.498.2219V}
}

@ARTICLE{2019ApJS..243....3L,
       author = {{Lehmer}, Bret D. and {Eufrasio}, Rafael T. and {Tzanavaris}, Panayiotis and {Basu-Zych}, Antara and {Fragos}, Tassos and {Prestwich}, Andrea and {Yukita}, Mihoko and {Zezas}, Andreas and {Hornschemeier}, Ann E. and {Ptak}, Andrew},
        title = "{X-Ray Binary Luminosity Function Scaling Relations for Local Galaxies Based on Subgalactic Modeling}",
      journal = {\apjs},
         year = 2019,
        month = jul,
       volume = {243},
       number = {1},
          eid = {3},
        pages = {3},
          doi = {10.3847/1538-4365/ab22a8},
archivePrefix = {arXiv},
       eprint = {1905.05197},
 primaryClass = {astro-ph.GA},
       adsurl = {https://ui.adsabs.harvard.edu/abs/2019ApJS..243....3L}
}

@ARTICLE{Fornasini2018,
       author = {{Fornasini}, Francesca M. and {Civano}, Francesca and {Fabbiano}, Giuseppina and {Elvis}, Martin and {Marchesi}, Stefano and {Miyaji}, Takamitsu and {Zezas}, Andreas},
        title = "{Low-luminosity AGN and X-Ray Binary Populations in COSMOS Star-forming Galaxies}",
      journal = {\apj},
         year = 2018,
        month = sep,
       volume = {865},
       number = {1},
          eid = {43},
        pages = {43},
          doi = {10.3847/1538-4357/aada4e},
archivePrefix = {arXiv},
       eprint = {1808.00970},
 primaryClass = {astro-ph.HE},
       adsurl = {https://ui.adsabs.harvard.edu/abs/2018ApJ...865...43F}
}

@ARTICLE{2019ApJ...871...80G,
       author = {{G{\"u}ltekin}, Kayhan and {King}, Ashley L. and {Cackett}, Edward M. and {Nyland}, Kristina and {Miller}, Jon M. and {Di Matteo}, Tiziana and {Markoff}, Sera and {Rupen}, Michael P.},
        title = "{The Fundamental Plane of Black Hole Accretion and Its Use as a Black Hole-Mass Estimator}",
      journal = {\apj},
         year = 2019,
        month = jan,
       volume = {871},
       number = {1},
          eid = {80},
        pages = {80},
          doi = {10.3847/1538-4357/aaf6b9},
archivePrefix = {arXiv},
       eprint = {1901.02530},
 primaryClass = {astro-ph.HE},
       adsurl = {https://ui.adsabs.harvard.edu/abs/2019ApJ...871...80G}
}

@ARTICLE{2014ApJ...788L..22G,
       author = {{G{\"u}ltekin}, Kayhan and {Cackett}, Edward M. and {King}, Ashley L. and {Miller}, Jon M. and {Pinkney}, Jason},
        title = "{Low-mass AGNs and Their Relation to the Fundamental Plane of Black Hole Accretion}",
      journal = {\apjl},
         year = 2014,
        month = jun,
       volume = {788},
       number = {2},
          eid = {L22},
        pages = {L22},
          doi = {10.1088/2041-8205/788/2/L22},
archivePrefix = {arXiv},
       eprint = {1405.6986},
 primaryClass = {astro-ph.HE},
       adsurl = {https://ui.adsabs.harvard.edu/abs/2014ApJ...788L..22G}
}

@ARTICLE{2009MNRAS.397.1101G,
       author = {{Garn}, Timothy and {Green}, David A. and {Riley}, Julia M. and {Alexander}, Paul},
        title = "{The relationship between star formation rate and radio synchrotron luminosity at 0 < z < 2}",
      journal = {\mnras},
         year = 2009,
        month = aug,
       volume = {397},
       number = {2},
        pages = {1101-1112},
          doi = {10.1111/j.1365-2966.2009.15073.x},
archivePrefix = {arXiv},
       eprint = {0905.1218},
 primaryClass = {astro-ph.GA},
       adsurl = {https://ui.adsabs.harvard.edu/abs/2009MNRAS.397.1101G}
}

@ARTICLE{2003ApJ...586..794B,
       author = {{Bell}, Eric F.},
        title = "{Estimating Star Formation Rates from Infrared and Radio Luminosities: The Origin of the Radio-Infrared Correlation}",
      journal = {\apj},
         year = 2003,
        month = apr,
       volume = {586},
       number = {2},
        pages = {794-813},
          doi = {10.1086/367829},
archivePrefix = {arXiv},
       eprint = {astro-ph/0212121},
 primaryClass = {astro-ph},
       adsurl = {https://ui.adsabs.harvard.edu/abs/2003ApJ...586..794B}
}

@ARTICLE{sutherland&dopita93,
   author = {{Sutherland}, R.~S. and {Dopita}, M.~A.},
    title = "{Cooling functions for low-density astrophysical plasmas}",
  journal = {\apjs},
     year = 1993,
    month = sep,
   volume = 88,
    pages = {253-327},
      doi = {10.1086/191823},
   adsurl = {http://adsabs.harvard.edu/abs/1993ApJS...88..253S}
}

@ARTICLE{haardt&madau96,
   author = {{Haardt}, F. and {Madau}, P.},
    title = "{Radiative Transfer in a Clumpy Universe. II. The Ultraviolet Extragalactic Background}",
  journal = {\apj},
     year = 1996,
    month = apr,
   volume = 461,
    pages = {20-+},
      doi = {10.1086/177035},
   adsurl = {http://adsabs.harvard.edu/cgi-bin/nph-bib_query?bibcode=1996ApJ...461...20H&db_key=AST}
}

@ARTICLE{kennicutt98,
   author = {{Kennicutt}, Jr., R.~C.},
    title = "{The Global Schmidt Law in Star-forming Galaxies}",
  journal = {\apj},
   eprint = {arXiv:astro-ph/9712213},
     year = 1998,
    month = may,
   volume = 498,
    pages = {541},
      doi = {10.1086/305588},
   adsurl = {http://cdsads.u-strasbg.fr/abs/1998ApJ...498..541K}
}

@ARTICLE{krumholz&tan07,
   author = {{Krumholz}, M.~R. and {Tan}, J.~C.},
    title = "{Slow Star Formation in Dense Gas: Evidence and Implications}",
  journal = {\apj},
   eprint = {arXiv:astro-ph/0606277},
     year = 2007,
    month = jan,
   volume = 654,
    pages = {304-315},
      doi = {10.1086/509101},
   adsurl = {http://adsabs.harvard.edu/abs/2007ApJ...654..304K}
}

@ARTICLE{rasera&teyssier06,
   author = {{Rasera}, Y. and {Teyssier}, R.},
    title = "{The history of the baryon budget. Cosmic logistics in a hierarchical universe}",
  journal = {\aap},
   eprint = {astro-ph/0505473},
     year = 2006,
    month = jan,
   volume = 445,
    pages = {1-27},
      doi = {10.1051/0004-6361:20053116},
   adsurl = {http://cdsads.u-strasbg.fr/abs/2006A%26A...445....1R}
}

@ARTICLE{dubois&teyssier08winds,
   author = {{Dubois}, Y. and {Teyssier}, R.},
    title = "{On the onset of galactic winds in quiescent star forming galaxies}",
  journal = {\aap},
   eprint = {arXiv:0707.3376},
     year = 2008,
    month = jan,
   volume = 477,
    pages = {79-94},
      doi = {10.1051/0004-6361:20078326},
   adsurl = {http://adsabs.harvard.edu/abs/2008A%26A...477...79D}
}

@ARTICLE{1955ApJ...121..161S,
   author = {{Salpeter}, E.~E.},
    title = "{The Luminosity Function and Stellar Evolution.}",
  journal = {\apj},
     year = 1955,
    month = jan,
   volume = 121,
    pages = {161},
      doi = {10.1086/145971},
   adsurl = {http://adsabs.harvard.edu/abs/1955ApJ...121..161S}
}

@article{Booth2009,
	Adsurl = {http://adsabs.harvard.edu/abs/2009MNRAS.398...53B},
	Archiveprefix = {arXiv},
	Author = {{Booth}, C.~M. and {Schaye}, J.},
	Doi = {10.1111/j.1365-2966.2009.15043.x},
	Eprint = {0904.2572},
	Journal = {\mnras},
	Month = sep,
	Pages = {53-74},
	Primaryclass = {astro-ph.CO},
	Title = {{Cosmological simulations of the growth of supermassive black holes and feedback from active galactic nuclei: method and tests}},
	Volume = 398,
	Year = 2009}

@ARTICLE{Dubois14,
       author = {{Dubois}, Y. and {Pichon}, C. and {Welker}, C. and {Le Borgne}, D. and {Devriendt}, J. and {Laigle}, C. and {Codis}, S. and {Pogosyan}, D. and {Arnouts}, S. and {Benabed}, K. and {Bertin}, E. and {Blaizot}, J. and {Bouchet}, F. and {Cardoso}, J. -F. and {Colombi}, S. and {de Lapparent}, V. and {Desjacques}, V. and {Gavazzi}, R. and {Kassin}, S. and {Kimm}, T. and {McCracken}, H. and {Milliard}, B. and {Peirani}, S. and {Prunet}, S. and {Rouberol}, S. and {Silk}, J. and {Slyz}, A. and {Sousbie}, T. and {Teyssier}, R. and {Tresse}, L. and {Treyer}, M. and {Vibert}, D. and {Volonteri}, M.},
        title = "{Dancing in the dark: galactic properties trace spin swings along the cosmic web}",
      journal = {\mnras},
         year = 2014,
        month = oct,
       volume = {444},
       number = {2},
        pages = {1453-1468},
          doi = {10.1093/mnras/stu1227},
archivePrefix = {arXiv},
       eprint = {1402.1165},
 primaryClass = {astro-ph.CO},
       adsurl = {https://ui.adsabs.harvard.edu/abs/2014MNRAS.444.1453D}
}

@ARTICLE{2012MNRAS.420.2662D,
   author = {{Dubois}, Y. and {Devriendt}, J. and {Slyz}, A. and {Teyssier}, R.
	},
    title = "{Self-regulated growth of supermassive black holes by a dual jet-heating active galactic nucleus feedback mechanism: methods, tests and implications for cosmological simulations}",
  journal = {\mnras},
archivePrefix = "arXiv",
   eprint = {1108.0110},
 primaryClass = "astro-ph.CO",
     year = 2012,
    month = mar,
   volume = 420,
    pages = {2662-2683},
      doi = {10.1111/j.1365-2966.2011.20236.x},
   adsurl = {http://cdsads.u-strasbg.fr/abs/2012MNRAS.420.2662D}
}

@article{Ostriker1999,
	Adsurl = {http://adsabs.harvard.edu/abs/1999ApJ...513..252O},
	Author = {{Ostriker}, E.~C.},
	Doi = {10.1086/306858},
	Eprint = {arXiv:astro-ph/9810324},
	Journal = {ApJ},
	Month = mar,
	Pages = {252-258},
	Title = {{Dynamical Friction in a Gaseous Medium}},
	Volume = 513,
	Year = 1999}

@ARTICLE{chaponetal13,
   author = {{Chapon}, D. and {Mayer}, L. and {Teyssier}, R.},
    title = "{Hydrodynamics of galaxy mergers with supermassive black holes: is there a last parsec problem?}",
  journal = {\mnras},
     year = 2013,
    month = mar,
   volume = 429,
    pages = {3114-3122},
      doi = {10.1093/mnras/sts568},
   adsurl = {http://cdsads.u-strasbg.fr/abs/2013MNRAS.429.3114C}
}

@ARTICLE{2015MNRAS.451.1868T,
   author = {{Tremmel}, M. and {Governato}, F. and {Volonteri}, M. and {Quinn}, T.~R.
	},
    title = "{Off the beaten path: a new approach to realistically model the orbital decay of supermassive black holes in galaxy formation simulations}",
  journal = {\mnras},
archivePrefix = "arXiv",
   eprint = {1501.07609},
     year = 2015,
    month = aug,
   volume = 451,
    pages = {1868-1874},
      doi = {10.1093/mnras/stv1060},
   adsurl = {http://adsabs.harvard.edu/abs/2015MNRAS.451.1868T}
}

@ARTICLE{Dubois2013,
   author = {{Dubois}, Y. and {Pichon}, C. and {Devriendt}, J. and {Silk}, J. and 
	{Haehnelt}, M. and {Kimm}, T. and {Slyz}, A.},
    title = "{Blowing cold flows away: the impact of early AGN activity on the formation of a brightest cluster galaxy progenitor}",
  journal = {\mnras},
archivePrefix = "arXiv",
   eprint = {1206.5838},
 primaryClass = "astro-ph.CO",
     year = 2013,
    month = feb,
   volume = 428,
    pages = {2885-2900},
      doi = {10.1093/mnras/sts224},
   adsurl = {http://adsabs.harvard.edu/abs/2013MNRAS.428.2885D}
}

@ARTICLE{aubertetal04,
   author = {{Aubert}, D. and {Pichon}, C. and {Colombi}, S.},
    title = "{The origin and implications of dark matter anisotropic cosmic infall on \~{}L$_{*}$ haloes}",
  journal = {\mnras},
   eprint = {arXiv:astro-ph/0402405},
     year = 2004,
    month = aug,
   volume = 352,
    pages = {376-398},
      doi = {10.1111/j.1365-2966.2004.07883.x},
   adsurl = {http://cdsads.u-strasbg.fr/abs/2004MNRAS.352..376A}
}

@ARTICLE{poweretal03,
   author = {{Power}, C. and {Navarro}, J.~F. and {Jenkins}, A. and {Frenk}, C.~S. and 
	{White}, S.~D.~M. and {Springel}, V. and {Stadel}, J. and {Quinn}, T.
	},
    title = "{The inner structure of {$\Lambda$}CDM haloes - I. A numerical convergence study}",
  journal = {\mnras},
   eprint = {astro-ph/0201544},
     year = 2003,
    month = jan,
   volume = 338,
    pages = {14-34},
      doi = {10.1046/j.1365-8711.2003.05925.x},
   adsurl = {http://cdsads.u-strasbg.fr/abs/2003MNRAS.338...14P}
}

@ARTICLE{2022MNRAS.510..531C,
       author = {{Chen}, Nianyi and {Ni}, Yueying and {Tremmel}, Michael and {Di Matteo}, Tiziana and {Bird}, Simeon and {DeGraf}, Colin and {Feng}, Yu},
        title = "{Dynamical friction modelling of massive black holes in cosmological simulations and effects on merger rate predictions}",
      journal = {\mnras},
         year = 2022,
        month = feb,
       volume = {510},
       number = {1},
        pages = {531-550},
          doi = {10.1093/mnras/stab3411},
archivePrefix = {arXiv},
       eprint = {2104.00021},
 primaryClass = {astro-ph.GA},
       adsurl = {https://ui.adsabs.harvard.edu/abs/2022MNRAS.510..531C}
}

@ARTICLE{2020ApJ...896..113L,
       author = {{Li}, Kunyang and {Bogdanovi{\'c}}, Tamara and {Ballantyne}, David R.},
        title = "{Pairing of Massive Black Holes in Merger Galaxies Driven by Dynamical Friction}",
      journal = {\apj},
         year = 2020,
        month = jun,
       volume = {896},
       number = {2},
          eid = {113},
        pages = {113},
          doi = {10.3847/1538-4357/ab93c6},
archivePrefix = {arXiv},
       eprint = {2006.08520},
 primaryClass = {astro-ph.GA},
       adsurl = {https://ui.adsabs.harvard.edu/abs/2020ApJ...896..113L}
}

@ARTICLE{2020ApJ...905..123L,
       author = {{Li}, Kunyang and {Bogdanovi{\'c}}, Tamara and {Ballantyne}, David R.},
        title = "{The Pairing Probability of Massive Black Holes in Merger Galaxies in the Presence of Radiative Feedback}",
      journal = {\apj},
         year = 2020,
        month = dec,
       volume = {905},
       number = {2},
          eid = {123},
        pages = {123},
          doi = {10.3847/1538-4357/abc555},
archivePrefix = {arXiv},
       eprint = {2007.02051},
 primaryClass = {astro-ph.GA},
       adsurl = {https://ui.adsabs.harvard.edu/abs/2020ApJ...905..123L}
}

@ARTICLE{2015MNRAS.454L..66S,
   author = {{Sesana}, A. and {Khan}, F.~M.},
    title = "{Scattering experiments meet N-body - I. A practical recipe for the evolution of massive black hole binaries in stellar environments}",
  journal = {\mnras},
archivePrefix = "arXiv",
   eprint = {1505.02062},
     year = 2015,
    month = nov,
   volume = 454,
    pages = {L66-L70},
      doi = {10.1093/mnrasl/slv131},
   adsurl = {http://adsabs.harvard.edu/abs/2015MNRAS.454L..66S}
}

@ARTICLE{2015MNRAS.448.3603D,
   author = {{Dotti}, M. and {Merloni}, A. and {Montuori}, C.},
    title = "{Linking the fate of massive black hole binaries to the active galactic nuclei luminosity function}",
  journal = {\mnras},
archivePrefix = "arXiv",
   eprint = {1502.03101},
 primaryClass = "astro-ph.HE",
     year = 2015,
    month = apr,
   volume = 448,
    pages = {3603-3607},
      doi = {10.1093/mnras/stv291},
   adsurl = {http://adsabs.harvard.edu/abs/2015MNRAS.448.3603D}
}

@ARTICLE{2008MNRAS.388.1011M,
       author = {{Merloni}, Andrea and {Heinz}, Sebastian},
        title = "{A synthesis model for AGN evolution: supermassive black holes growth and feedback modes}",
      journal = {\mnras},
         year = 2008,
        month = aug,
       volume = {388},
       number = {3},
        pages = {1011-1030},
          doi = {10.1111/j.1365-2966.2008.13472.x},
archivePrefix = {arXiv},
       eprint = {0805.2499},
 primaryClass = {astro-ph},
       adsurl = {https://ui.adsabs.harvard.edu/abs/2008MNRAS.388.1011M}
}

@ARTICLE{Q1996,
       author = {{Quinlan}, Gerald D.},
        title = "{The dynamical evolution of massive black hole binaries I. Hardening in a fixed stellar background}",
      journal = {\na},
         year = "1996",
        month = "Jul",
       volume = {1},
        pages = {35-56},
          doi = {10.1016/S1384-1076(96)00003-6},
archivePrefix = {arXiv},
       eprint = {astro-ph/9601092},
 primaryClass = {astro-ph},
       adsurl = {https://ui.adsabs.harvard.edu/abs/1996NewA....1...35Q}
}

@ARTICLE{Sesana2006,
       author = {{Sesana}, Alberto and {Haardt}, Francesco and {Madau}, Piero},
        title = "{Interaction of Massive Black Hole Binaries with Their Stellar Environment. I. Ejection of Hypervelocity Stars}",
      journal = {\apj},
         year = 2006,
        month = nov,
       volume = {651},
       number = {1},
        pages = {392-400},
          doi = {10.1086/507596},
archivePrefix = {arXiv},
       eprint = {astro-ph/0604299},
 primaryClass = {astro-ph},
       adsurl = {https://ui.adsabs.harvard.edu/abs/2006ApJ...651..392S}
}

@ARTICLE{2023JCAP...08..034B,
       author = {{Babak}, Stanislav and {Caprini}, Chiara and {Figueroa}, Daniel G. and {Karnesis}, Nikolaos and {Marcoccia}, Paolo and {Nardini}, Germano and {Pieroni}, Mauro and {Ricciardone}, Angelo and {Sesana}, Alberto and {Torrado}, Jes{\'u}s},
        title = "{Stochastic gravitational wave background from stellar origin binary black holes in LISA}",
      journal = {\jcap},
         year = 2023,
        month = aug,
       volume = {2023},
       number = {8},
          eid = {034},
        pages = {034},
          doi = {10.1088/1475-7516/2023/08/034},
archivePrefix = {arXiv},
       eprint = {2304.06368},
 primaryClass = {astro-ph.CO},
       adsurl = {https://ui.adsabs.harvard.edu/abs/2023JCAP...08..034B}
}

@ARTICLE{SS1973,
       author = {{Shakura}, N.~I. and {Sunyaev}, R.~A.},
        title = "{Reprint of 1973A\&A....24..337S. Black holes in binary systems. Observational appearance.}",
      journal = {\aap},
         year = 1973,
        month = jun,
       volume = {500},
        pages = {33-51},
       adsurl = {https://ui.adsabs.harvard.edu/abs/1973A&A....24..337S}
}

@BOOK{Shapiro1983,
       author = {{Shapiro}, Stuart L. and {Teukolsky}, Saul A.},
        title = "{Black holes, white dwarfs, and neutron stars : the physics of compact objects}",
         year = 1983,
       adsurl = {https://ui.adsabs.harvard.edu/abs/1983bhwd.book.....S}
}

@ARTICLE{Haiman2009,
       author = {{Haiman}, Zolt{\'a}n and {Kocsis}, Bence and {Menou}, Kristen},
        title = "{The Population of Viscosity- and Gravitational Wave-driven Supermassive Black Hole Binaries Among Luminous Active Galactic Nuclei}",
      journal = {\apj},
         year = 2009,
        month = aug,
       volume = {700},
       number = {2},
        pages = {1952-1969},
          doi = {10.1088/0004-637X/700/2/1952},
archivePrefix = {arXiv},
       eprint = {0904.1383},
 primaryClass = {astro-ph.CO},
       adsurl = {https://ui.adsabs.harvard.edu/abs/2009ApJ...700.1952H}
}

@ARTICLE{Roedig2011,
       author = {{Roedig}, C. and {Dotti}, M. and {Sesana}, A. and {Cuadra}, J. and {Colpi}, M.},
        title = "{Limiting eccentricity of subparsec massive black hole binaries surrounded by self-gravitating gas discs}",
      journal = {\mnras},
         year = 2011,
        month = aug,
       volume = {415},
       number = {4},
        pages = {3033-3041},
          doi = {10.1111/j.1365-2966.2011.18927.x},
archivePrefix = {arXiv},
       eprint = {1104.3868},
 primaryClass = {astro-ph.CO},
       adsurl = {https://ui.adsabs.harvard.edu/abs/2011MNRAS.415.3033R}
}

@ARTICLE{laigle19,
       author = {{Laigle}, C. and {Davidzon}, I. and {Ilbert}, O. and {Devriendt}, J. and {Kashino}, D. and {Pichon}, C. and {Capak}, P. and {Arnouts}, S. and {de la Torre}, S. and {Dubois}, Y. and {Gozaliasl}, G. and {Le Borgne}, D. and {Lilly}, S. and {McCracken}, H.~J. and {Salvato}, M. and {Slyz}, A.},
        title = "{Horizon-AGN virtual observatory - 1. SED-fitting performance and forecasts for future imaging surveys}",
      journal = {\mnras},
         year = 2019,
        month = jul,
       volume = {486},
       number = {4},
        pages = {5104-5123},
          doi = {10.1093/mnras/stz1054},
archivePrefix = {arXiv},
       eprint = {1903.10934},
 primaryClass = {astro-ph.GA},
       adsurl = {https://ui.adsabs.harvard.edu/abs/2019MNRAS.486.5104L}
}

@ARTICLE{WeingartnerDraine2001,
       author = {{Weingartner}, Joseph C. and {Draine}, B.~T.},
        title = "{Dust Grain-Size Distributions and Extinction in the Milky Way, Large Magellanic Cloud, and Small Magellanic Cloud}",
      journal = {\apj},
         year = 2001,
        month = feb,
       volume = {548},
       number = {1},
        pages = {296-309},
          doi = {10.1086/318651},
archivePrefix = {arXiv},
       eprint = {astro-ph/0008146},
 primaryClass = {astro-ph},
       adsurl = {https://ui.adsabs.harvard.edu/abs/2001ApJ...548..296W}
}

@ARTICLE{chabrier03,
       author = {{Chabrier}, Gilles},
        title = "{Galactic Stellar and Substellar Initial Mass Function}",
      journal = {\pasp},
         year = 2003,
        month = jul,
       volume = {115},
       number = {809},
        pages = {763-795},
          doi = {10.1086/376392},
archivePrefix = {arXiv},
       eprint = {astro-ph/0304382},
 primaryClass = {astro-ph},
       adsurl = {https://ui.adsabs.harvard.edu/abs/2003PASP..115..763C}
}

@ARTICLE{bc03,
       author = {{Bruzual}, G. and {Charlot}, S.},
        title = "{Stellar population synthesis at the resolution of 2003}",
      journal = {\mnras},
         year = 2003,
        month = oct,
       volume = {344},
       number = {4},
        pages = {1000-1028},
          doi = {10.1046/j.1365-8711.2003.06897.x},
archivePrefix = {arXiv},
       eprint = {astro-ph/0309134},
 primaryClass = {astro-ph},
       adsurl = {https://ui.adsabs.harvard.edu/abs/2003MNRAS.344.1000B}
}

@ARTICLE{Cuadra2009,
       author = {{Cuadra}, J. and {Armitage}, P.~J. and {Alexander}, R.~D. and {Begelman}, M.~C.},
        title = "{Massive black hole binary mergers within subparsec scale gas discs}",
      journal = {\mnras},
         year = 2009,
        month = mar,
       volume = {393},
       number = {4},
        pages = {1423-1432},
          doi = {10.1111/j.1365-2966.2008.14147.x},
archivePrefix = {arXiv},
       eprint = {0809.0311},
 primaryClass = {astro-ph},
       adsurl = {https://ui.adsabs.harvard.edu/abs/2009MNRAS.393.1423C}
}

@ARTICLE{Roedig2012,
       author = {{Roedig}, C. and {Sesana}, A. and {Dotti}, M. and {Cuadra}, J. and {Amaro-Seoane}, P. and {Haardt}, F.},
        title = "{Evolution of binary black holes in self gravitating discs. Dissecting the torques}",
      journal = {\aap},
         year = 2012,
        month = sep,
       volume = {545},
          eid = {A127},
        pages = {A127},
          doi = {10.1051/0004-6361/201219986},
archivePrefix = {arXiv},
       eprint = {1202.6063},
 primaryClass = {astro-ph.CO},
       adsurl = {https://ui.adsabs.harvard.edu/abs/2012A&A...545A.127R}
}

@ARTICLE{Franchini2021,
       author = {{Franchini}, Alessia and {Sesana}, Alberto and {Dotti}, Massimo},
        title = "{Circumbinary disc self-gravity governing supermassive black hole binary mergers}",
      journal = {\mnras},
         year = 2021,
        month = oct,
       volume = {507},
       number = {1},
        pages = {1458-1467},
          doi = {10.1093/mnras/stab2234},
archivePrefix = {arXiv},
       eprint = {2106.13253},
 primaryClass = {astro-ph.HE},
       adsurl = {https://ui.adsabs.harvard.edu/abs/2021MNRAS.507.1458F}
}

@ARTICLE{Amaro2023,
       author = {{Amaro-Seoane}, Pau and {Andrews}, Jeff and {Arca Sedda}, Manuel and {Askar}, Abbas and {Baghi}, Quentin and {Balasov}, Razvan and {Bartos}, Imre and {Bavera}, Simone S. and {Bellovary}, Jillian and {Berry}, Christopher P.~L. and {Berti}, Emanuele and {Bianchi}, Stefano and {Blecha}, Laura and {Blondin}, St{\'e}phane and {Bogdanovi{\'c}}, Tamara and {Boissier}, Samuel and {Bonetti}, Matteo and {Bonoli}, Silvia and {Bortolas}, Elisa and {Breivik}, Katelyn and {Capelo}, Pedro R. and {Caramete}, Laurentiu and {Cattorini}, Federico and {Charisi}, Maria and {Chaty}, Sylvain and {Chen}, Xian and {Chru{\'s}li{\'n}ska}, Martyna and {Chua}, Alvin J.~K. and {Church}, Ross and {Colpi}, Monica and {D'Orazio}, Daniel and {Danielski}, Camilla and {Davies}, Melvyn B. and {Dayal}, Pratika and {De Rosa}, Alessandra and {Derdzinski}, Andrea and {Destounis}, Kyriakos and {Dotti}, Massimo and {Dutan}, Ioana and {Dvorkin}, Irina and {Fabj}, Gaia and {Foglizzo}, Thierry and {Ford}, Saavik and {Fouvry}, Jean-Baptiste and {Franchini}, Alessia and {Fragos}, Tassos and {Fryer}, Chris and {Gaspari}, Massimo and {Gerosa}, Davide and {Graziani}, Luca and {Groot}, Paul and {Habouzit}, Melanie and {Haggard}, Daryl and {Haiman}, Zoltan and {Han}, Wen-Biao and {Istrate}, Alina and {Johansson}, Peter H. and {Khan}, Fazeel Mahmood and {Kimpson}, Tomas and {Kokkotas}, Kostas and {Kong}, Albert and {Korol}, Valeriya and {Kremer}, Kyle and {Kupfer}, Thomas and {Lamberts}, Astrid and {Larson}, Shane and {Lau}, Mike and {Liu}, Dongliang and {Lloyd-Ronning}, Nicole and {Lodato}, Giuseppe and {Lupi}, Alessandro and {Ma}, Chung-Pei and {Maccarone}, Tomas and {Mandel}, Ilya and {Mangiagli}, Alberto and {Mapelli}, Michela and {Mathis}, St{\'e}phane and {Mayer}, Lucio and {McGee}, Sean and {McKernan}, Barry and {Miller}, M. Coleman and {Mota}, David F. and {Mumpower}, Matthew and {Nasim}, Syeda S. and {Nelemans}, Gijs and {Noble}, Scott and {Pacucci}, Fabio and {Panessa}, Francesca and {Paschalidis}, Vasileios and {Pfister}, Hugo and {Porquet}, Delphine and {Quenby}, John and {Ricarte}, Angelo and {R{\"o}pke}, Friedrich K. and {Regan}, John and {Rosswog}, Stephan and {Ruiter}, Ashley and {Ruiz}, Milton and {Runnoe}, Jessie and {Schneider}, Raffaella and {Schnittman}, Jeremy and {Secunda}, Amy and {Sesana}, Alberto and {Seto}, Naoki and {Shao}, Lijing and {Shapiro}, Stuart and {Sopuerta}, Carlos and {Stone}, Nicholas C. and {Suvorov}, Arthur and {Tamanini}, Nicola and {Tamfal}, Tomas and {Tauris}, Thomas and {Temmink}, Karel and {Tomsick}, John and {Toonen}, Silvia and {Torres-Orjuela}, Alejandro and {Toscani}, Martina and {Tsokaros}, Antonios and {Unal}, Caner and {V{\'a}zquez-Aceves}, Ver{\'o}nica and {Valiante}, Rosa and {van Putten}, Maurice and {van Roestel}, Jan and {Vignali}, Christian and {Volonteri}, Marta and {Wu}, Kinwah and {Younsi}, Ziri and {Yu}, Shenghua and {Zane}, Silvia and {Zwick}, Lorenz and {Antonini}, Fabio and {Baibhav}, Vishal and {Barausse}, Enrico and {Bonilla Rivera}, Alexander and {Branchesi}, Marica and {Branduardi-Raymont}, Graziella and {Burdge}, Kevin and {Chakraborty}, Srija and {Cuadra}, Jorge and {Dage}, Kristen and {Davis}, Benjamin and {de Mink}, Selma E. and {Decarli}, Roberto and {Doneva}, Daniela and {Escoffier}, Stephanie and {Gandhi}, Poshak and {Haardt}, Francesco and {Lousto}, Carlos O. and {Nissanke}, Samaya and {Nordhaus}, Jason and {O'Shaughnessy}, Richard and {Portegies Zwart}, Simon and {Pound}, Adam and {Schussler}, Fabian and {Sergijenko}, Olga and {Spallicci}, Alessandro and {Vernieri}, Daniele and {Vigna-G{\'o}mez}, Alejandro},
        title = "{Astrophysics with the Laser Interferometer Space Antenna}",
      journal = {Living Reviews in Relativity},
         year = 2023,
        month = dec,
       volume = {26},
       number = {1},
          eid = {2},
        pages = {2},
          doi = {10.1007/s41114-022-00041-y},
archivePrefix = {arXiv},
       eprint = {2203.06016},
 primaryClass = {gr-qc},
       adsurl = {https://ui.adsabs.harvard.edu/abs/2023LRR....26....2A}
}

@ARTICLE{Bortolas2021,
       author = {{Bortolas}, Elisa and {Franchini}, Alessia and {Bonetti}, Matteo and {Sesana}, Alberto},
        title = "{The Competing Effect of Gas and Stars in the Evolution of Massive Black Hole Binaries}",
      journal = {\apjl},
         year = 2021,
        month = sep,
       volume = {918},
       number = {1},
          eid = {L15},
        pages = {L15},
          doi = {10.3847/2041-8213/ac1c0c},
archivePrefix = {arXiv},
       eprint = {2108.13436},
 primaryClass = {astro-ph.HE},
       adsurl = {https://ui.adsabs.harvard.edu/abs/2021ApJ...918L..15B}
}

@ARTICLE{2016MNRAS.460.2979V,
       author = {{Volonteri}, M. and {Dubois}, Y. and {Pichon}, C. and {Devriendt}, J.},
        title = "{The cosmic evolution of massive black holes in the Horizon-AGN simulation}",
      journal = {\mnras},
         year = 2016,
        month = aug,
       volume = {460},
       number = {3},
        pages = {2979-2996},
          doi = {10.1093/mnras/stw1123},
archivePrefix = {arXiv},
       eprint = {1602.01941},
 primaryClass = {astro-ph.GA},
       adsurl = {https://ui.adsabs.harvard.edu/abs/2016MNRAS.460.2979V}
}

@ARTICLE{2023A&A...673A.120D,
       author = {{Dong-P{\'a}ez}, Chi An and {Volonteri}, Marta and {Beckmann}, Ricarda S. and {Dubois}, Yohan and {Trebitsch}, Maxime and {Mangiagli}, Alberto and {Vergani}, Susanna D. and {Webb}, Natalie A.},
        title = "{Black hole mergers as tracers of spinning massive black hole and galaxy populations in the OBELISK simulation}",
      journal = {\aap},
         year = 2023,
        month = may,
       volume = {673},
          eid = {A120},
        pages = {A120},
          doi = {10.1051/0004-6361/202346295},
archivePrefix = {arXiv},
       eprint = {2303.00766},
 primaryClass = {astro-ph.GA},
       adsurl = {https://ui.adsabs.harvard.edu/abs/2023A&A...673A.120D}
}

@ARTICLE{2007PThPh.117..241E,
       author = {{Enoki}, M. and {Nagashima}, M.},
        title = "{The Effect of Orbital Eccentricity on Gravitational Wave Background Radiation from Supermassive Black Hole Binaries}",
      journal = {Progress of Theoretical Physics},
         year = 2007,
        month = feb,
       volume = {117},
       number = {2},
        pages = {241-256},
          doi = {10.1143/PTP.117.241},
archivePrefix = {arXiv},
       eprint = {astro-ph/0609377},
 primaryClass = {astro-ph},
       adsurl = {https://ui.adsabs.harvard.edu/abs/2007PThPh.117..241E}
}

@ARTICLE{2001astro.ph..8028P,
       author = {{Phinney}, E.~S.},
        title = "{A Practical Theorem on Gravitational Wave Backgrounds}",
      journal = {arXiv e-prints},
         year = 2001,
        month = aug,
          eid = {astro-ph/0108028},
        pages = {astro-ph/0108028},
          doi = {10.48550/arXiv.astro-ph/0108028},
archivePrefix = {arXiv},
       eprint = {astro-ph/0108028},
 primaryClass = {astro-ph},
       adsurl = {https://ui.adsabs.harvard.edu/abs/2001astro.ph..8028P}
}

@ARTICLE{1963PhRv..131..435P,
       author = {{Peters}, P.~C. and {Mathews}, J.},
        title = "{Gravitational Radiation from Point Masses in a Keplerian Orbit}",
      journal = {Physical Review},
         year = 1963,
        month = jul,
       volume = {131},
       number = {1},
        pages = {435-440},
          doi = {10.1103/PhysRev.131.435},
       adsurl = {https://ui.adsabs.harvard.edu/abs/1963PhRv..131..435P}
}

@ARTICLE{2008MNRAS.390..192S,
       author = {{Sesana}, A. and {Vecchio}, A. and {Colacino}, C.~N.},
        title = "{The stochastic gravitational-wave background from massive black hole binary systems: implications for observations with Pulsar Timing Arrays}",
      journal = {\mnras},
         year = 2008,
        month = oct,
       volume = {390},
       number = {1},
        pages = {192-209},
          doi = {10.1111/j.1365-2966.2008.13682.x},
archivePrefix = {arXiv},
       eprint = {0804.4476},
 primaryClass = {astro-ph},
       adsurl = {https://ui.adsabs.harvard.edu/abs/2008MNRAS.390..192S}
}

@ARTICLE{1999astro.ph..5116H,
       author = {{Hogg}, David W.},
        title = "{Distance measures in cosmology}",
      journal = {arXiv e-prints},
         year = 1999,
        month = may,
          eid = {astro-ph/9905116},
        pages = {astro-ph/9905116},
          doi = {10.48550/arXiv.astro-ph/9905116},
archivePrefix = {arXiv},
       eprint = {astro-ph/9905116},
 primaryClass = {astro-ph},
       adsurl = {https://ui.adsabs.harvard.edu/abs/1999astro.ph..5116H}
}

@ARTICLE{2003ApJ...590..691W,
       author = {{Wyithe}, J. Stuart B. and {Loeb}, Abraham},
        title = "{Low-Frequency Gravitational Waves from Massive Black Hole Binaries: Predictions for LISA and Pulsar Timing Arrays}",
      journal = {\apj},
         year = 2003,
        month = jun,
       volume = {590},
       number = {2},
        pages = {691-706},
          doi = {10.1086/375187},
archivePrefix = {arXiv},
       eprint = {astro-ph/0211556},
 primaryClass = {astro-ph},
       adsurl = {https://ui.adsabs.harvard.edu/abs/2003ApJ...590..691W}
}

@ARTICLE{2003ApJ...583..616J,
       author = {{Jaffe}, A.~H. and {Backer}, D.~C.},
        title = "{Gravitational Waves Probe the Coalescence Rate of Massive Black Hole Binaries}",
      journal = {\apj},
         year = 2003,
        month = feb,
       volume = {583},
       number = {2},
        pages = {616-631},
          doi = {10.1086/345443},
archivePrefix = {arXiv},
       eprint = {astro-ph/0210148},
 primaryClass = {astro-ph},
       adsurl = {https://ui.adsabs.harvard.edu/abs/2003ApJ...583..616J}
}

@ARTICLE{2006MNRAS.372.1549E,
       author = {{Edwards}, R.~T. and {Hobbs}, G.~B. and {Manchester}, R.~N.},
        title = "{TEMPO2, a new pulsar timing package - II. The timing model and precision estimates}",
      journal = {\mnras},
         year = 2006,
        month = nov,
       volume = {372},
       number = {4},
        pages = {1549-1574},
          doi = {10.1111/j.1365-2966.2006.10870.x},
archivePrefix = {arXiv},
       eprint = {astro-ph/0607664},
 primaryClass = {astro-ph},
       adsurl = {https://ui.adsabs.harvard.edu/abs/2006MNRAS.372.1549E}
}

@ARTICLE{1979ApJ...234.1100D,
       author = {{Detweiler}, S.},
        title = "{Pulsar timing measurements and the search for gravitational waves}",
      journal = {\apj},
         year = 1979,
        month = dec,
       volume = {234},
        pages = {1100-1104},
          doi = {10.1086/157593},
       adsurl = {https://ui.adsabs.harvard.edu/abs/1979ApJ...234.1100D}
}

@ARTICLE{2023ApJ...951L...8A,
       author = {{Agazie}, Gabriella and {Anumarlapudi}, Akash and {Archibald}, Anne M. and {Arzoumanian}, Zaven and {Baker}, Paul T. and {B{\'e}csy}, Bence and {Blecha}, Laura and {Brazier}, Adam and {Brook}, Paul R. and {Burke-Spolaor}, Sarah and {Burnette}, Rand and {Case}, Robin and {Charisi}, Maria and {Chatterjee}, Shami and {Chatziioannou}, Katerina and {Cheeseboro}, Belinda D. and {Chen}, Siyuan and {Cohen}, Tyler and {Cordes}, James M. and {Cornish}, Neil J. and {Crawford}, Fronefield and {Cromartie}, H. Thankful and {Crowter}, Kathryn and {Cutler}, Curt J. and {Decesar}, Megan E. and {Degan}, Dallas and {Demorest}, Paul B. and {Deng}, Heling and {Dolch}, Timothy and {Drachler}, Brendan and {Ellis}, Justin A. and {Ferrara}, Elizabeth C. and {Fiore}, William and {Fonseca}, Emmanuel and {Freedman}, Gabriel E. and {Garver-Daniels}, Nate and {Gentile}, Peter A. and {Gersbach}, Kyle A. and {Glaser}, Joseph and {Good}, Deborah C. and {G{\"u}ltekin}, Kayhan and {Hazboun}, Jeffrey S. and {Hourihane}, Sophie and {Islo}, Kristina and {Jennings}, Ross J. and {Johnson}, Aaron D. and {Jones}, Megan L. and {Kaiser}, Andrew R. and {Kaplan}, David L. and {Kelley}, Luke Zoltan and {Kerr}, Matthew and {Key}, Joey S. and {Klein}, Tonia C. and {Laal}, Nima and {Lam}, Michael T. and {Lamb}, William G. and {Lazio}, T. Joseph W. and {Lewandowska}, Natalia and {Littenberg}, Tyson B. and {Liu}, Tingting and {Lommen}, Andrea and {Lorimer}, Duncan R. and {Luo}, Jing and {Lynch}, Ryan S. and {Ma}, Chung-Pei and {Madison}, Dustin R. and {Mattson}, Margaret A. and {McEwen}, Alexander and {McKee}, James W. and {McLaughlin}, Maura A. and {McMann}, Natasha and {Meyers}, Bradley W. and {Meyers}, Patrick M. and {Mingarelli}, Chiara M.~F. and {Mitridate}, Andrea and {Natarajan}, Priyamvada and {Ng}, Cherry and {Nice}, David J. and {Ocker}, Stella Koch and {Olum}, Ken D. and {Pennucci}, Timothy T. and {Perera}, Benetge B.~P. and {Petrov}, Polina and {Pol}, Nihan S. and {Radovan}, Henri A. and {Ransom}, Scott M. and {Ray}, Paul S. and {Romano}, Joseph D. and {Sardesai}, Shashwat C. and {Schmiedekamp}, Ann and {Schmiedekamp}, Carl and {Schmitz}, Kai and {Schult}, Levi and {Shapiro-Albert}, Brent J. and {Siemens}, Xavier and {Simon}, Joseph and {Siwek}, Magdalena S. and {Stairs}, Ingrid H. and {Stinebring}, Daniel R. and {Stovall}, Kevin and {Sun}, Jerry P. and {Susobhanan}, Abhimanyu and {Swiggum}, Joseph K. and {Taylor}, Jacob and {Taylor}, Stephen R. and {Turner}, Jacob E. and {Unal}, Caner and {Vallisneri}, Michele and {van Haasteren}, Rutger and {Vigeland}, Sarah J. and {Wahl}, Haley M. and {Wang}, Qiaohong and {Witt}, Caitlin A. and {Young}, Olivia and {Nanograv Collaboration}},
        title = "{The NANOGrav 15 yr Data Set: Evidence for a Gravitational-wave Background}",
      journal = {\apjl},
         year = 2023,
        month = jul,
       volume = {951},
       number = {1},
          eid = {L8},
        pages = {L8},
          doi = {10.3847/2041-8213/acdac6},
archivePrefix = {arXiv},
       eprint = {2306.16213},
 primaryClass = {astro-ph.HE},
       adsurl = {https://ui.adsabs.harvard.edu/abs/2023ApJ...951L...8A}
}

@ARTICLE{2023A&A...678A..50E,
       author = {{EPTA Collaboration} and {InPTA Collaboration} and {Antoniadis}, J. and {Arumugam}, P. and {Arumugam}, S. and {Babak}, S. and {Bagchi}, M. and {Bak Nielsen}, A. -S. and {Bassa}, C.~G. and {Bathula}, A. and {Berthereau}, A. and {Bonetti}, M. and {Bortolas}, E. and {Brook}, P.~R. and {Burgay}, M. and {Caballero}, R.~N. and {Chalumeau}, A. and {Champion}, D.~J. and {Chanlaridis}, S. and {Chen}, S. and {Cognard}, I. and {Dandapat}, S. and {Deb}, D. and {Desai}, S. and {Desvignes}, G. and {Dhanda-Batra}, N. and {Dwivedi}, C. and {Falxa}, M. and {Ferdman}, R.~D. and {Franchini}, A. and {Gair}, J.~R. and {Goncharov}, B. and {Gopakumar}, A. and {Graikou}, E. and {Grie{\ss}meier}, J. -M. and {Guillemot}, L. and {Guo}, Y.~J. and {Gupta}, Y. and {Hisano}, S. and {Hu}, H. and {Iraci}, F. and {Izquierdo-Villalba}, D. and {Jang}, J. and {Jawor}, J. and {Janssen}, G.~H. and {Jessner}, A. and {Joshi}, B.~C. and {Kareem}, F. and {Karuppusamy}, R. and {Keane}, E.~F. and {Keith}, M.~J. and {Kharbanda}, D. and {Kikunaga}, T. and {Kolhe}, N. and {Kramer}, M. and {Krishnakumar}, M.~A. and {Lackeos}, K. and {Lee}, K.~J. and {Liu}, K. and {Liu}, Y. and {Lyne}, A.~G. and {McKee}, J.~W. and {Maan}, Y. and {Main}, R.~A. and {Mickaliger}, M.~B. and {Ni{\c{t}}u}, I.~C. and {Nobleson}, K. and {Paladi}, A.~K. and {Parthasarathy}, A. and {Perera}, B.~B.~P. and {Perrodin}, D. and {Petiteau}, A. and {Porayko}, N.~K. and {Possenti}, A. and {Prabu}, T. and {Quelquejay Leclere}, H. and {Rana}, P. and {Samajdar}, A. and {Sanidas}, S.~A. and {Sesana}, A. and {Shaifullah}, G. and {Singha}, J. and {Speri}, L. and {Spiewak}, R. and {Srivastava}, A. and {Stappers}, B.~W. and {Surnis}, M. and {Susarla}, S.~C. and {Susobhanan}, A. and {Takahashi}, K. and {Tarafdar}, P. and {Theureau}, G. and {Tiburzi}, C. and {van der Wateren}, E. and {Vecchio}, A. and {Venkatraman Krishnan}, V. and {Verbiest}, J.~P.~W. and {Wang}, J. and {Wang}, L. and {Wu}, Z.},
        title = "{The second data release from the European Pulsar Timing Array. III. Search for gravitational wave signals}",
      journal = {\aap},
         year = 2023,
        month = oct,
       volume = {678},
          eid = {A50},
        pages = {A50},
          doi = {10.1051/0004-6361/202346844},
archivePrefix = {arXiv},
       eprint = {2306.16214},
 primaryClass = {astro-ph.HE},
       adsurl = {https://ui.adsabs.harvard.edu/abs/2023A&A...678A..50E}
}

@ARTICLE{2023ApJ...951L...6R,
       author = {{Reardon}, Daniel J. and {Zic}, Andrew and {Shannon}, Ryan M. and {Hobbs}, George B. and {Bailes}, Matthew and {Di Marco}, Valentina and {Kapur}, Agastya and {Rogers}, Axl F. and {Thrane}, Eric and {Askew}, Jacob and {Bhat}, N.~D. Ramesh and {Cameron}, Andrew and {Cury{\l}o}, Ma{\l}gorzata and {Coles}, William A. and {Dai}, Shi and {Goncharov}, Boris and {Kerr}, Matthew and {Kulkarni}, Atharva and {Levin}, Yuri and {Lower}, Marcus E. and {Manchester}, Richard N. and {Mandow}, Rami and {Miles}, Matthew T. and {Nathan}, Rowina S. and {Os{\l}owski}, Stefan and {Russell}, Christopher J. and {Spiewak}, Ren{\'e}e and {Zhang}, Songbo and {Zhu}, Xing-Jiang},
        title = "{Search for an Isotropic Gravitational-wave Background with the Parkes Pulsar Timing Array}",
      journal = {\apjl},
         year = 2023,
        month = jul,
       volume = {951},
       number = {1},
          eid = {L6},
        pages = {L6},
          doi = {10.3847/2041-8213/acdd02},
archivePrefix = {arXiv},
       eprint = {2306.16215},
 primaryClass = {astro-ph.HE},
       adsurl = {https://ui.adsabs.harvard.edu/abs/2023ApJ...951L...6R}
}

@ARTICLE{2023RAA....23g5024X,
       author = {{Xu}, Heng and {Chen}, Siyuan and {Guo}, Yanjun and {Jiang}, Jinchen and {Wang}, Bojun and {Xu}, Jiangwei and {Xue}, Zihan and {Nicolas Caballero}, R. and {Yuan}, Jianping and {Xu}, Yonghua and {Wang}, Jingbo and {Hao}, Longfei and {Luo}, Jingtao and {Lee}, Kejia and {Han}, Jinlin and {Jiang}, Peng and {Shen}, Zhiqiang and {Wang}, Min and {Wang}, Na and {Xu}, Renxin and {Wu}, Xiangping and {Manchester}, Richard and {Qian}, Lei and {Guan}, Xin and {Huang}, Menglin and {Sun}, Chun and {Zhu}, Yan},
        title = "{Searching for the Nano-Hertz Stochastic Gravitational Wave Background with the Chinese Pulsar Timing Array Data Release I}",
      journal = {Research in Astronomy and Astrophysics},
         year = 2023,
        month = jul,
       volume = {23},
       number = {7},
          eid = {075024},
        pages = {075024},
          doi = {10.1088/1674-4527/acdfa5},
archivePrefix = {arXiv},
       eprint = {2306.16216},
 primaryClass = {astro-ph.HE},
       adsurl = {https://ui.adsabs.harvard.edu/abs/2023RAA....23g5024X}
}

@article{10.1093/mnras/stae2571,
    author = {Miles, Matthew T and Shannon, Ryan M and Reardon, Daniel J and Bailes, Matthew and Champion, David J and Geyer, Marisa and Gitika, Pratyasha and Grunthal, Kathrin and Keith, Michael J and Kramer, Michael and Kulkarni, Atharva D and Nathan, Rowina S and Parthasarathy, Aditya and Singha, Jaikhomba and Theureau, Gilles and Thrane, Eric and Abbate, Federico and Buchner, Sarah and Cameron, Andrew D and Camilo, Fernando and Moreschi, Beatrice E and Shaifullah, Golam and Shamohammadi, Mohsen and Possenti, Andrea and Krishnan, Vivek Venkatraman},
    title = {The MeerKAT Pulsar Timing Array: the first search for gravitational waves with the MeerKAT radio telescope},
    journal = {Monthly Notices of the Royal Astronomical Society},
    volume = {536},
    number = {2},
    pages = {1489-1500},
    year = {2024},
    month = {12},
    issn = {0035-8711},
    doi = {10.1093/mnras/stae2571},
    url = {https://doi.org/10.1093/mnras/stae2571},
    eprint = {https://academic.oup.com/mnras/article-pdf/536/2/1489/61215196/stae2571.pdf},
}

@ARTICLE{2023ApJ...951L..50A,
       author = {{Agazie}, Gabriella and {Anumarlapudi}, Akash and {Archibald}, Anne M. and {Arzoumanian}, Zaven and {Baker}, Paul T. and {B{\'e}csy}, Bence and {Blecha}, Laura and {Brazier}, Adam and {Brook}, Paul R. and {Burke-Spolaor}, Sarah and {Case}, Robin and {Casey-Clyde}, J. Andrew and {Charisi}, Maria and {Chatterjee}, Shami and {Cohen}, Tyler and {Cordes}, James M. and {Cornish}, Neil J. and {Crawford}, Fronefield and {Cromartie}, H. Thankful and {Crowter}, Kathryn and {Decesar}, Megan E. and {Demorest}, Paul B. and {Digman}, Matthew C. and {Dolch}, Timothy and {Drachler}, Brendan and {Ferrara}, Elizabeth C. and {Fiore}, William and {Fonseca}, Emmanuel and {Freedman}, Gabriel E. and {Garver-Daniels}, Nate and {Gentile}, Peter A. and {Glaser}, Joseph and {Good}, Deborah C. and {G{\"u}ltekin}, Kayhan and {Hazboun}, Jeffrey S. and {Hourihane}, Sophie and {Jennings}, Ross J. and {Johnson}, Aaron D. and {Jones}, Megan L. and {Kaiser}, Andrew R. and {Kaplan}, David L. and {Kelley}, Luke Zoltan and {Kerr}, Matthew and {Key}, Joey S. and {Laal}, Nima and {Lam}, Michael T. and {Lamb}, William G. and {Lazio}, T. Joseph W. and {Lewandowska}, Natalia and {Liu}, Tingting and {Lorimer}, Duncan R. and {Luo}, Jing and {Lynch}, Ryan S. and {Ma}, Chung-Pei and {Madison}, Dustin R. and {McEwen}, Alexander and {McKee}, James W. and {McLaughlin}, Maura A. and {McMann}, Natasha and {Meyers}, Bradley W. and {Meyers}, Patrick M. and {Mingarelli}, Chiara M.~F. and {Mitridate}, Andrea and {Ng}, Cherry and {Nice}, David J. and {Ocker}, Stella Koch and {Olum}, Ken D. and {Pennucci}, Timothy T. and {Perera}, Benetge B.~P. and {Petrov}, Polina and {Pol}, Nihan S. and {Radovan}, Henri A. and {Ransom}, Scott M. and {Ray}, Paul S. and {Romano}, Joseph D. and {Sardesai}, Shashwat C. and {Schmiedekamp}, Ann and {Schmiedekamp}, Carl and {Schmitz}, Kai and {Shapiro-Albert}, Brent J. and {Siemens}, Xavier and {Simon}, Joseph and {Siwek}, Magdalena S. and {Stairs}, Ingrid H. and {Stinebring}, Daniel R. and {Stovall}, Kevin and {Susobhanan}, Abhimanyu and {Swiggum}, Joseph K. and {Taylor}, Jacob and {Taylor}, Stephen R. and {Turner}, Jacob E. and {Unal}, Caner and {Vallisneri}, Michele and {van Haasteren}, Rutger and {Vigeland}, Sarah J. and {Wahl}, Haley M. and {Witt}, Caitlin A. and {Young}, Olivia and {Nanograv Collaboration}},
        title = "{The NANOGrav 15 yr Data Set: Bayesian Limits on Gravitational Waves from Individual Supermassive Black Hole Binaries}",
      journal = {\apjl},
         year = 2023,
        month = jul,
       volume = {951},
       number = {2},
          eid = {L50},
        pages = {L50},
          doi = {10.3847/2041-8213/ace18a},
archivePrefix = {arXiv},
       eprint = {2306.16222},
 primaryClass = {astro-ph.HE},
       adsurl = {https://ui.adsabs.harvard.edu/abs/2023ApJ...951L..50A}
}

@ARTICLE{2023ApJ...952L..37A,
       author = {{Agazie}, Gabriella and {Anumarlapudi}, Akash and {Archibald}, Anne M. and {Baker}, Paul T. and {B{\'e}csy}, Bence and {Blecha}, Laura and {Bonilla}, Alexander and {Brazier}, Adam and {Brook}, Paul R. and {Burke-Spolaor}, Sarah and {Burnette}, Rand and {Case}, Robin and {Casey-Clyde}, J. Andrew and {Charisi}, Maria and {Chatterjee}, Shami and {Chatziioannou}, Katerina and {Cheeseboro}, Belinda D. and {Chen}, Siyuan and {Cohen}, Tyler and {Cordes}, James M. and {Cornish}, Neil J. and {Crawford}, Fronefield and {Cromartie}, H. Thankful and {Crowter}, Kathryn and {Cutler}, Curt J. and {D'Orazio}, Daniel J. and {Decesar}, Megan E. and {Degan}, Dallas and {Demorest}, Paul B. and {Deng}, Heling and {Dolch}, Timothy and {Drachler}, Brendan and {Ferrara}, Elizabeth C. and {Fiore}, William and {Fonseca}, Emmanuel and {Freedman}, Gabriel E. and {Gardiner}, Emiko and {Garver-Daniels}, Nate and {Gentile}, Peter A. and {Gersbach}, Kyle A. and {Glaser}, Joseph and {Good}, Deborah C. and {G{\"u}ltekin}, Kayhan and {Hazboun}, Jeffrey S. and {Hourihane}, Sophie and {Islo}, Kristina and {Jennings}, Ross J. and {Johnson}, Aaron and {Jones}, Megan L. and {Kaiser}, Andrew R. and {Kaplan}, David L. and {Kelley}, Luke Zoltan and {Kerr}, Matthew and {Key}, Joey S. and {Laal}, Nima and {Lam}, Michael T. and {Lamb}, William G. and {Lazio}, T. Joseph W. and {Lewandowska}, Natalia and {Littenberg}, Tyson B. and {Liu}, Tingting and {Luo}, Jing and {Lynch}, Ryan S. and {Ma}, Chung-Pei and {Madison}, Dustin R. and {McEwen}, Alexander and {McKee}, James W. and {McLaughlin}, Maura A. and {McMann}, Natasha and {Meyers}, Bradley W. and {Meyers}, Patrick M. and {Mingarelli}, Chiara M.~F. and {Mitridate}, Andrea and {Natarajan}, Priyamvada and {Ng}, Cherry and {Nice}, David J. and {Ocker}, Stella Koch and {Olum}, Ken D. and {Pennucci}, Timothy T. and {Perera}, Benetge B.~P. and {Petrov}, Polina and {Pol}, Nihan S. and {Radovan}, Henri A. and {Ransom}, Scott M. and {Ray}, Paul S. and {Romano}, Joseph D. and {Runnoe}, Jessie C. and {Sardesai}, Shashwat C. and {Schmiedekamp}, Ann and {Schmiedekamp}, Carl and {Schmitz}, Kai and {Schult}, Levi and {Shapiro-Albert}, Brent J. and {Siemens}, Xavier and {Simon}, Joseph and {Siwek}, Magdalena S. and {Stairs}, Ingrid H. and {Stinebring}, Daniel R. and {Stovall}, Kevin and {Sun}, Jerry P. and {Susobhanan}, Abhimanyu and {Swiggum}, Joseph K. and {Taylor}, Jacob and {Taylor}, Stephen R. and {Turner}, Jacob E. and {Unal}, Caner and {Vallisneri}, Michele and {Vigeland}, Sarah J. and {Wachter}, Jeremy M. and {Wahl}, Haley M. and {Wang}, Qiaohong and {Witt}, Caitlin A. and {Wright}, David and {Young}, Olivia and {Nanograv Collaboration}},
        title = "{The NANOGrav 15 yr Data Set: Constraints on Supermassive Black Hole Binaries from the Gravitational-wave Background}",
      journal = {\apjl},
         year = 2023,
        month = aug,
       volume = {952},
       number = {2},
          eid = {L37},
        pages = {L37},
          doi = {10.3847/2041-8213/ace18b},
archivePrefix = {arXiv},
       eprint = {2306.16220},
 primaryClass = {astro-ph.HE},
       adsurl = {https://ui.adsabs.harvard.edu/abs/2023ApJ...952L..37A}
}

@ARTICLE{2024A&A...685A..94E,
       author = {{EPTA Collaboration} and {InPTA Collaboration} and {Antoniadis}, J. and {Arumugam}, P. and {Arumugam}, S. and {Babak}, S. and {Bagchi}, M. and {Bak Nielsen}, A. -S. and {Bassa}, C.~G. and {Bathula}, A. and {Berthereau}, A. and {Bonetti}, M. and {Bortolas}, E. and {Brook}, P.~R. and {Burgay}, M. and {Caballero}, R.~N. and {Chalumeau}, A. and {Champion}, D.~J. and {Chanlaridis}, S. and {Chen}, S. and {Cognard}, I. and {Dandapat}, S. and {Deb}, D. and {Desai}, S. and {Desvignes}, G. and {Dhanda-Batra}, N. and {Dwivedi}, C. and {Falxa}, M. and {Ferdman}, R.~D. and {Franchini}, A. and {Gair}, J.~R. and {Goncharov}, B. and {Gopakumar}, A. and {Graikou}, E. and {Grie{\ss}meier}, J. -M. and {Gualandris}, A. and {Guillemot}, L. and {Guo}, Y.~J. and {Gupta}, Y. and {Hisano}, S. and {Hu}, H. and {Iraci}, F. and {Izquierdo-Villalba}, D. and {Jang}, J. and {Jawor}, J. and {Janssen}, G.~H. and {Jessner}, A. and {Joshi}, B.~C. and {Kareem}, F. and {Karuppusamy}, R. and {Keane}, E.~F. and {Keith}, M.~J. and {Kharbanda}, D. and {Kikunaga}, T. and {Kolhe}, N. and {Kramer}, M. and {Krishnakumar}, M.~A. and {Lackeos}, K. and {Lee}, K.~J. and {Liu}, K. and {Liu}, Y. and {Lyne}, A.~G. and {McKee}, J.~W. and {Maan}, Y. and {Main}, R.~A. and {Mickaliger}, M.~B. and {Ni{\c{t}}u}, I.~C. and {Nobleson}, K. and {Paladi}, A.~K. and {Parthasarathy}, A. and {Perera}, B.~B.~P. and {Perrodin}, D. and {Petiteau}, A. and {Porayko}, N.~K. and {Possenti}, A. and {Prabu}, T. and {Quelquejay Leclere}, H. and {Rana}, P. and {Samajdar}, A. and {Sanidas}, S.~A. and {Sesana}, A. and {Shaifullah}, G. and {Singha}, J. and {Speri}, L. and {Spiewak}, R. and {Srivastava}, A. and {Stappers}, B.~W. and {Surnis}, M. and {Susarla}, S.~C. and {Susobhanan}, A. and {Takahashi}, K. and {Tarafdar}, P. and {Theureau}, G. and {Tiburzi}, C. and {van der Wateren}, E. and {Vecchio}, A. and {Venkatraman Krishnan}, V. and {Verbiest}, J.~P.~W. and {Wang}, J. and {Wang}, L. and {Wu}, Z. and {Auclair}, P. and {Barausse}, E. and {Caprini}, C. and {Crisostomi}, M. and {Fastidio}, F. and {Khizriev}, T. and {Middleton}, H. and {Neronov}, A. and {Postnov}, K. and {Roper Pol}, A. and {Semikoz}, D. and {Smarra}, C. and {Steer}, D.~A. and {Truant}, R.~J. and {Valtolina}, S.},
        title = "{The second data release from the European Pulsar Timing Array. IV. Implications for massive black holes, dark matter, and the early Universe}",
      journal = {\aap},
         year = 2024,
        month = may,
       volume = {685},
          eid = {A94},
        pages = {A94},
          doi = {10.1051/0004-6361/202347433},
archivePrefix = {arXiv},
       eprint = {2306.16227},
 primaryClass = {astro-ph.CO},
       adsurl = {https://ui.adsabs.harvard.edu/abs/2024A&A...685A..94E}
}

@ARTICLE{2019PhRvD.100j4028H,
       author = {{Hazboun}, Jeffrey S. and {Romano}, Joseph D. and {Smith}, Tristan L.},
        title = "{Realistic sensitivity curves for pulsar timing arrays}",
      journal = {\prd},
         year = 2019,
        month = nov,
       volume = {100},
       number = {10},
          eid = {104028},
        pages = {104028},
          doi = {10.1103/PhysRevD.100.104028},
archivePrefix = {arXiv},
       eprint = {1907.04341},
 primaryClass = {gr-qc},
       adsurl = {https://ui.adsabs.harvard.edu/abs/2019PhRvD.100j4028H}
}

@ARTICLE{2013PhRvD..87j4021L,
       author = {{Lentati}, Lindley and {Alexander}, P. and {Hobson}, M.~P. and {Taylor}, S. and {Gair}, J. and {Balan}, S.~T. and {van Haasteren}, R.},
        title = "{Hyper-efficient model-independent Bayesian method for the analysis of pulsar timing data}",
      journal = {\prd},
         year = 2013,
        month = may,
       volume = {87},
       number = {10},
          eid = {104021},
        pages = {104021},
          doi = {10.1103/PhysRevD.87.104021},
archivePrefix = {arXiv},
       eprint = {1210.3578},
 primaryClass = {astro-ph.IM},
       adsurl = {https://ui.adsabs.harvard.edu/abs/2013PhRvD..87j4021L}
}

@ARTICLE{2024ApJ...966..105A,
       author = {{Agazie}, G. and {Antoniadis}, J. and {Anumarlapudi}, A. and {Archibald}, A.~M. and {Arumugam}, P. and {Arumugam}, S. and {Arzoumanian}, Z. and {Askew}, J. and {Babak}, S. and {Bagchi}, M. and {Bailes}, M. and {Bak Nielsen}, A. -S. and {Baker}, P.~T. and {Bassa}, C.~G. and {Bathula}, A. and {B{\'e}csy}, B. and {Berthereau}, A. and {Bhat}, N.~D.~R. and {Blecha}, L. and {Bonetti}, M. and {Bortolas}, E. and {Brazier}, A. and {Brook}, P.~R. and {Burgay}, M. and {Burke-Spolaor}, S. and {Burnette}, R. and {Caballero}, R.~N. and {Cameron}, A. and {Case}, R. and {Chalumeau}, A. and {Champion}, D.~J. and {Chanlaridis}, S. and {Charisi}, M. and {Chatterjee}, S. and {Chatziioannou}, K. and {Cheeseboro}, B.~D. and {Chen}, S. and {Chen}, Z. -C. and {Cognard}, I. and {Cohen}, T. and {Coles}, W.~A. and {Cordes}, J.~M. and {Cornish}, N.~J. and {Crawford}, F. and {Cromartie}, H.~T. and {Crowter}, K. and {Cury{\l}o}, M. and {Cutler}, C.~J. and {Dai}, S. and {Dandapat}, S. and {Deb}, D. and {DeCesar}, M.~E. and {DeGan}, D. and {Demorest}, P.~B. and {Deng}, H. and {Desai}, S. and {Desvignes}, G. and {Dey}, L. and {Dhanda-Batra}, N. and {Di Marco}, V. and {Dolch}, T. and {Drachler}, B. and {Dwivedi}, C. and {Ellis}, J.~A. and {Falxa}, M. and {Feng}, Y. and {Ferdman}, R.~D. and {Ferrara}, E.~C. and {Fiore}, W. and {Fonseca}, E. and {Franchini}, A. and {Freedman}, G.~E. and {Gair}, J.~R. and {Garver-Daniels}, N. and {Gentile}, P.~A. and {Gersbach}, K.~A. and {Glaser}, J. and {Good}, D.~C. and {Goncharov}, B. and {Gopakumar}, A. and {Graikou}, E. and {Griessmeier}, J. -M. and {Guillemot}, L. and {G{\"u}ltekin}, K. and {Guo}, Y.~J. and {Gupta}, Y. and {Grunthal}, K. and {Hazboun}, J.~S. and {Hisano}, S. and {Hobbs}, G.~B. and {Hourihane}, S. and {Hu}, H. and {Iraci}, F. and {Islo}, K. and {Izquierdo-Villalba}, D. and {Jang}, J. and {Jawor}, J. and {Janssen}, G.~H. and {Jennings}, R.~J. and {Jessner}, A. and {Johnson}, A.~D. and {Jones}, M.~L. and {Joshi}, B.~C. and {Kaiser}, A.~R. and {Kaplan}, D.~L. and {Kapur}, A. and {Kareem}, F. and {Karuppusamy}, R. and {Keane}, E.~F. and {Keith}, M.~J. and {Kelley}, L.~Z. and {Kerr}, M. and {Key}, J.~S. and {Kharbanda}, D. and {Kikunaga}, T. and {Klein}, T.~C. and {Kolhe}, N. and {Kramer}, M. and {Krishnakumar}, M.~A. and {Kulkarni}, A. and {Laal}, N. and {Lackeos}, K. and {Lam}, M.~T. and {Lamb}, W.~G. and {Larsen}, B.~B. and {Lazio}, T.~J.~W. and {Lee}, K.~J. and {Levin}, Y. and {Lewandowska}, N. and {Littenberg}, T.~B. and {Liu}, K. and {Liu}, T. and {Liu}, Y. and {Lommen}, A. and {Lorimer}, D.~R. and {Lower}, M.~E. and {Luo}, J. and {Luo}, R. and {Lynch}, R.~S. and {Lyne}, A.~G. and {Ma}, C. -P. and {Maan}, Y. and {Madison}, D.~R. and {Main}, R.~A. and {Manchester}, R.~N. and {Mandow}, R. and {Mattson}, M.~A. and {McEwen}, A. and {McKee}, J.~W. and {McLaughlin}, M.~A. and {McMann}, N. and {Meyers}, B.~W. and {Meyers}, P.~M. and {Mickaliger}, M.~B. and {Miles}, M. and {Mingarelli}, C.~M.~F. and {Mitridate}, A. and {Natarajan}, P. and {Nathan}, R.~S. and {Ng}, C. and {Nice}, D.~J. and {Ni{\c{t}}u}, I.~C. and {Nobleson}, K. and {Ocker}, S.~K. and {Olum}, K.~D. and {Os{\l}owski}, S. and {Paladi}, A.~K. and {Parthasarathy}, A. and {Pennucci}, T.~T. and {Perera}, B.~B.~P. and {Perrodin}, D. and {Petiteau}, A. and {Petrov}, P. and {Pol}, N.~S. and {Porayko}, N.~K. and {Possenti}, A. and {Prabu}, T. and {Quelquejay Leclere}, H. and {Radovan}, H.~A. and {Rana}, P. and {Ransom}, S.~M. and {Ray}, P.~S. and {Reardon}, D.~J. and {Rogers}, A.~F. and {Romano}, J.~D. and {Russell}, C.~J. and {Samajdar}, A. and {Sanidas}, S.~A. and {Sardesai}, S.~C. and {Schmiedekamp}, A. and {Schmiedekamp}, C. and {Schmitz}, K. and {Schult}, L. and {Sesana}, A. and {Shaifullah}, G. and {Shannon}, R.~M. and {Shapiro-Albert}, B.~J. and {Siemens}, X. and {Simon}, J. and {Singha}, J.},
        title = "{Comparing Recent Pulsar Timing Array Results on the Nanohertz Stochastic Gravitational-wave Background}",
      journal = {\apj},
         year = 2024,
        month = may,
       volume = {966},
       number = {1},
          eid = {105},
        pages = {105},
          doi = {10.3847/1538-4357/ad36be},
archivePrefix = {arXiv},
       eprint = {2309.00693},
 primaryClass = {astro-ph.HE},
       adsurl = {https://ui.adsabs.harvard.edu/abs/2024ApJ...966..105A}
}

@ARTICLE{2012ApJ...756..175E,
       author = {{Ellis}, J.~A. and {Siemens}, X. and {Creighton}, J.~D.~E.},
        title = "{Optimal Strategies for Continuous Gravitational Wave Detection in Pulsar Timing Arrays}",
      journal = {\apj},
         year = 2012,
        month = sep,
       volume = {756},
       number = {2},
          eid = {175},
        pages = {175},
          doi = {10.1088/0004-637X/756/2/175},
archivePrefix = {arXiv},
       eprint = {1204.4218},
 primaryClass = {astro-ph.IM},
       adsurl = {https://ui.adsabs.harvard.edu/abs/2012ApJ...756..175E}
}

@article{ESS,
author = {Elvira, Víctor and Martino, Luca and Robert, Christian P.},
title = {Rethinking the Effective Sample Size},
journal = {International Statistical Review},
volume = {90},
number = {3},
pages = {525-550},
doi = {https://doi.org/10.1111/insr.12500},
url = {https://onlinelibrary.wiley.com/doi/abs/10.1111/insr.12500},
eprint = {https://onlinelibrary.wiley.com/doi/pdf/10.1111/insr.12500},
year = {2022}
}

@ARTICLE{2021CSF...14510790M,
       author = {{Mitra}, Tushar and {Hossain}, Tomal and {Banerjee}, Santo and {Hassan}, Md. Kamrul},
        title = "{Similarity and self-similarity in random walk with fixed, random and shrinking steps}",
      journal = {Chaos Solitons and Fractals},
         year = 2021,
        month = apr,
       volume = {145},
          eid = {110790},
        pages = {110790},
          doi = {10.1016/j.chaos.2021.110790},
archivePrefix = {arXiv},
       eprint = {2010.02579},
 primaryClass = {cond-mat.stat-mech},
       adsurl = {https://ui.adsabs.harvard.edu/abs/2021CSF...14510790M}
}

@ARTICLE{2015MNRAS.451.2417R,
       author = {{Rosado}, Pablo A. and {Sesana}, Alberto and {Gair}, Jonathan},
        title = "{Expected properties of the first gravitational wave signal detected with pulsar timing arrays}",
      journal = {\mnras},
         year = 2015,
        month = aug,
       volume = {451},
       number = {3},
        pages = {2417-2433},
          doi = {10.1093/mnras/stv1098},
archivePrefix = {arXiv},
       eprint = {1503.04803},
 primaryClass = {astro-ph.HE},
       adsurl = {https://ui.adsabs.harvard.edu/abs/2015MNRAS.451.2417R}
}

@ARTICLE{2017LRR....20....2R,
       author = {{Romano}, Joseph D. and {Cornish}, Neil. J.},
        title = "{Detection methods for stochastic gravitational-wave backgrounds: a unified treatment}",
      journal = {Living Reviews in Relativity},
         year = 2017,
        month = dec,
       volume = {20},
       number = {1},
          eid = {2},
        pages = {2},
          doi = {10.1007/s41114-017-0004-1},
archivePrefix = {arXiv},
       eprint = {1608.06889},
 primaryClass = {gr-qc},
       adsurl = {https://ui.adsabs.harvard.edu/abs/2017LRR....20....2R}
}

@ARTICLE{2023A&A...678A..49E,
       author = {{EPTA Collaboration} and {InPTA Collaboration} and {Antoniadis}, J. and {Arumugam}, P. and {Arumugam}, S. and {Babak}, S. and {Bagchi}, M. and {Nielsen}, A. -S. Bak and {Bassa}, C.~G. and {Bathula}, A. and {Berthereau}, A. and {Bonetti}, M. and {Bortolas}, E. and {Brook}, P.~R. and {Burgay}, M. and {Caballero}, R.~N. and {Chalumeau}, A. and {Champion}, D.~J. and {Chanlaridis}, S. and {Chen}, S. and {Cognard}, I. and {Dandapat}, S. and {Deb}, D. and {Desai}, S. and {Desvignes}, G. and {Dhanda-Batra}, N. and {Dwivedi}, C. and {Falxa}, M. and {Ferdman}, R.~D. and {Franchini}, A. and {Gair}, J.~R. and {Goncharov}, B. and {Gopakumar}, A. and {Graikou}, E. and {Grie{\ss}meier}, J. -M. and {Guillemot}, L. and {Guo}, Y.~J. and {Gupta}, Y. and {Hisano}, S. and {Hu}, H. and {Iraci}, F. and {Izquierdo-Villalba}, D. and {Jang}, J. and {Jawor}, J. and {Janssen}, G.~H. and {Jessner}, A. and {Joshi}, B.~C. and {Kareem}, F. and {Karuppusamy}, R. and {Keane}, E.~F. and {Keith}, M.~J. and {Kharbanda}, D. and {Kikunaga}, T. and {Kolhe}, N. and {Kramer}, M. and {Krishnakumar}, M.~A. and {Lackeos}, K. and {Lee}, K.~J. and {Liu}, K. and {Liu}, Y. and {Lyne}, A.~G. and {McKee}, J.~W. and {Maan}, Y. and {Main}, R.~A. and {Mickaliger}, M.~B. and {Ni{\c{t}}u}, I.~C. and {Nobleson}, K. and {Paladi}, A.~K. and {Parthasarathy}, A. and {Perera}, B.~B.~P. and {Perrodin}, D. and {Petiteau}, A. and {Porayko}, N.~K. and {Possenti}, A. and {Prabu}, T. and {Leclere}, H. Quelquejay and {Rana}, P. and {Samajdar}, A. and {Sanidas}, S.~A. and {Sesana}, A. and {Shaifullah}, G. and {Singha}, J. and {Speri}, L. and {Spiewak}, R. and {Srivastava}, A. and {Stappers}, B.~W. and {Surnis}, M. and {Susarla}, S.~C. and {Susobhanan}, A. and {Takahashi}, K. and {Tarafdar}, P. and {Theureau}, G. and {Tiburzi}, C. and {van der Wateren}, E. and {Vecchio}, A. and {Krishnan}, V. Venkatraman and {Verbiest}, J.~P.~W. and {Wang}, J. and {Wang}, L. and {Wu}, Z.},
        title = "{The second data release from the European Pulsar Timing Array. II. Customised pulsar noise models for spatially correlated gravitational waves}",
      journal = {\aap},
         year = 2023,
        month = oct,
       volume = {678},
          eid = {A49},
        pages = {A49},
          doi = {10.1051/0004-6361/202346842},
archivePrefix = {arXiv},
       eprint = {2306.16225},
 primaryClass = {astro-ph.HE},
       adsurl = {https://ui.adsabs.harvard.edu/abs/2023A&A...678A..49E}
}

@ARTICLE{2023A&A...678A..48E,
       author = {{EPTA Collaboration} and {Antoniadis}, J. and {Babak}, S. and {Bak Nielsen}, A. -S. and {Bassa}, C.~G. and {Berthereau}, A. and {Bonetti}, M. and {Bortolas}, E. and {Brook}, P.~R. and {Burgay}, M. and {Caballero}, R.~N. and {Chalumeau}, A. and {Champion}, D.~J. and {Chanlaridis}, S. and {Chen}, S. and {Cognard}, I. and {Desvignes}, G. and {Falxa}, M. and {Ferdman}, R.~D. and {Franchini}, A. and {Gair}, J.~R. and {Goncharov}, B. and {Graikou}, E. and {Grie{\ss}meier}, J. -M. and {Guillemot}, L. and {Guo}, Y.~J. and {Hu}, H. and {Iraci}, F. and {Izquierdo-Villalba}, D. and {Jang}, J. and {Jawor}, J. and {Janssen}, G.~H. and {Jessner}, A. and {Karuppusamy}, R. and {Keane}, E.~F. and {Keith}, M.~J. and {Kramer}, M. and {Krishnakumar}, M.~A. and {Lackeos}, K. and {Lee}, K.~J. and {Liu}, K. and {Liu}, Y. and {Lyne}, A.~G. and {McKee}, J.~W. and {Main}, R.~A. and {Mickaliger}, M.~B. and {Ni{\c{t}}u}, I.~C. and {Parthasarathy}, A. and {Perera}, B.~B.~P. and {Perrodin}, D. and {Petiteau}, A. and {Porayko}, N.~K. and {Possenti}, A. and {Quelquejay Leclere}, H. and {Samajdar}, A. and {Sanidas}, S.~A. and {Sesana}, A. and {Shaifullah}, G. and {Speri}, L. and {Spiewak}, R. and {Stappers}, B.~W. and {Susarla}, S.~C. and {Theureau}, G. and {Tiburzi}, C. and {van der Wateren}, E. and {Vecchio}, A. and {Venkatraman Krishnan}, V. and {Verbiest}, J.~P.~W. and {Wang}, J. and {Wang}, L. and {Wu}, Z.},
        title = "{The second data release from the European Pulsar Timing Array. I. The dataset and timing analysis}",
      journal = {\aap},
         year = 2023,
        month = oct,
       volume = {678},
          eid = {A48},
        pages = {A48},
          doi = {10.1051/0004-6361/202346841},
archivePrefix = {arXiv},
       eprint = {2306.16224},
 primaryClass = {astro-ph.HE},
       adsurl = {https://ui.adsabs.harvard.edu/abs/2023A&A...678A..48E}
}

@ARTICLE{2012ApJ...761...64S,
       author = {{Shannon}, Ryan M. and {Cordes}, James. M.},
        title = "{Pulse Intensity Modulation and the Timing Stability of Millisecond Pulsars: A Case Study of PSR J1713+0747}",
      journal = {\apj},
         year = 2012,
        month = dec,
       volume = {761},
       number = {1},
          eid = {64},
        pages = {64},
          doi = {10.1088/0004-637X/761/1/64},
archivePrefix = {arXiv},
       eprint = {1210.7021},
 primaryClass = {astro-ph.SR},
       adsurl = {https://ui.adsabs.harvard.edu/abs/2012ApJ...761...64S}
}

@INPROCEEDINGS{2015aska.confE..37J,
       author = {{Janssen}, G. and {Hobbs}, G. and {McLaughlin}, M. and {Bassa}, C. and {Deller}, A. and {Kramer}, M. and {Lee}, K. and {Mingarelli}, C. and {Rosado}, P. and {Sanidas}, S. and {Sesana}, A. and {Shao}, L. and {Stairs}, I. and {Stappers}, B. and {Verbiest}, J.~P.~W.},
        title = "{Gravitational Wave Astronomy with the SKA}",
    booktitle = {Advancing Astrophysics with the Square Kilometre Array (AASKA14)},
         year = 2015,
        month = apr,
          eid = {37},
        pages = {37},
          doi = {10.22323/1.215.0037},
archivePrefix = {arXiv},
       eprint = {1501.00127},
 primaryClass = {astro-ph.IM},
       adsurl = {https://ui.adsabs.harvard.edu/abs/2015aska.confE..37J}
}

@ARTICLE{2012PhRvL.109h1104M,
       author = {{Mingarelli}, C.~M.~F. and {Grover}, K. and {Sidery}, T. and {Smith}, R.~J.~E. and {Vecchio}, A.},
        title = "{Observing the Dynamics of Supermassive Black Hole Binaries with Pulsar Timing Arrays}",
      journal = {\prl},
         year = 2012,
        month = aug,
       volume = {109},
       number = {8},
          eid = {081104},
        pages = {081104},
          doi = {10.1103/PhysRevLett.109.081104},
archivePrefix = {arXiv},
       eprint = {1207.5645},
 primaryClass = {astro-ph.HE},
       adsurl = {https://ui.adsabs.harvard.edu/abs/2012PhRvL.109h1104M}
}

@ARTICLE{2022ApJ...941..119B,
       author = {{B{\'e}csy}, Bence and {Cornish}, Neil J. and {Kelley}, Luke Zoltan},
        title = "{Exploring Realistic Nanohertz Gravitational-wave Backgrounds}",
      journal = {\apj},
         year = 2022,
        month = dec,
       volume = {941},
       number = {2},
          eid = {119},
        pages = {119},
          doi = {10.3847/1538-4357/aca1b2},
archivePrefix = {arXiv},
       eprint = {2207.01607},
 primaryClass = {astro-ph.HE},
       adsurl = {https://ui.adsabs.harvard.edu/abs/2022ApJ...941..119B}
}

@ARTICLE{2024ApJ...971L..10L,
       author = {{Lamb}, William G. and {Taylor}, Stephen R.},
        title = "{Spectral Variance in a Stochastic Gravitational-wave Background from a Binary Population}",
      journal = {\apjl},
         year = 2024,
        month = aug,
       volume = {971},
       number = {1},
          eid = {L10},
        pages = {L10},
          doi = {10.3847/2041-8213/ad654a},
archivePrefix = {arXiv},
       eprint = {2407.06270},
 primaryClass = {gr-qc},
       adsurl = {https://ui.adsabs.harvard.edu/abs/2024ApJ...971L..10L}
}

@article{PhysRevD.111.023047,
  title = {From eccentric binaries to nonstationary gravitational wave backgrounds},
  author = {Falxa, Mikel and Quelquejay Leclere, Hippolyte and Sesana, Alberto},
  journal = {Phys. Rev. D},
  volume = {111},
  issue = {2},
  pages = {023047},
  numpages = {14},
  year = {2025},
  month = {Jan},
  publisher = {American Physical Society},
  doi = {10.1103/PhysRevD.111.023047},
  url = {https://link.aps.org/doi/10.1103/PhysRevD.111.023047}
}

@ARTICLE{2014MNRAS.439.3986R,
       author = {{Rosado}, Pablo A. and {Sesana}, Alberto},
        title = "{Targeting supermassive black hole binaries and gravitational wave sources for the pulsar timing array}",
      journal = {\mnras},
         year = 2014,
        month = apr,
       volume = {439},
       number = {4},
        pages = {3986-4010},
          doi = {10.1093/mnras/stu254},
archivePrefix = {arXiv},
       eprint = {1311.0883},
 primaryClass = {astro-ph.CO},
       adsurl = {https://ui.adsabs.harvard.edu/abs/2014MNRAS.439.3986R}
}

@ARTICLE{2014MNRAS.442...56R,
       author = {{Ravi}, V. and {Wyithe}, J.~S.~B. and {Shannon}, R.~M. and {Hobbs}, G. and {Manchester}, R.~N.},
        title = "{Binary supermassive black hole environments diminish the gravitational wave signal in the pulsar timing band}",
      journal = {\mnras},
         year = 2014,
        month = jul,
       volume = {442},
       number = {1},
        pages = {56-68},
          doi = {10.1093/mnras/stu779},
archivePrefix = {arXiv},
       eprint = {1404.5183},
 primaryClass = {astro-ph.CO},
       adsurl = {https://ui.adsabs.harvard.edu/abs/2014MNRAS.442...56R}
}

@ARTICLE{2025ApJ...978...31A,
       author = {{Agazie}, Gabriella and {Anumarlapudi}, Akash and {Archibald}, Anne M. and {Arzoumanian}, Zaven and {Baier}, Jeremy George and {Baker}, Paul T. and {B{\'e}csy}, Bence and {Blecha}, Laura and {Brazier}, Adam and {Brook}, Paul R. and {Brown}, Lucas and {Burke-Spolaor}, Sarah and {Casey-Clyde}, J. Andrew and {Charisi}, Maria and {Chatterjee}, Shami and {Cohen}, Tyler and {Cordes}, James M. and {Cornish}, Neil J. and {Crawford}, Fronefield and {Cromartie}, H. Thankful and {Crowter}, Kathryn and {DeCesar}, Megan E. and {Demorest}, Paul B. and {Deng}, Heling and {Dolch}, Timothy and {Ferrara}, Elizabeth C. and {Fiore}, William and {Fonseca}, Emmanuel and {Freedman}, Gabriel E. and {Garver-Daniels}, Nate and {Gentile}, Peter A. and {Glaser}, Joseph and {Good}, Deborah C. and {G{\"u}ltekin}, Kayhan and {Hazboun}, Jeffrey S. and {Jennings}, Ross J. and {Johnson}, Aaron D. and {Jones}, Megan L. and {Kaiser}, Andrew R. and {Kaplan}, David L. and {Kelley}, Luke Zoltan and {Kerr}, Matthew and {Key}, Joey S. and {Laal}, Nima and {Lam}, Michael T. and {Lamb}, William G. and {Larsen}, Bjorn and {Lazio}, T. Joseph W. and {Lewandowska}, Natalia and {Liu}, Tingting and {Lorimer}, Duncan R. and {Luo}, Jing and {Lynch}, Ryan S. and {Ma}, Chung-Pei and {Madison}, Dustin R. and {McEwen}, Alexander and {McKee}, James W. and {McLaughlin}, Maura A. and {McMann}, Natasha and {Meyers}, Bradley W. and {Meyers}, Patrick M. and {Mingarelli}, Chiara M.~F. and {Mitridate}, Andrea and {Natarajan}, Priyamvada and {Ng}, Cherry and {Nice}, David J. and {Ocker}, Stella Koch and {Olum}, Ken D. and {Pennucci}, Timothy T. and {Perera}, Benetge B.~P. and {Pol}, Nihan S. and {Radovan}, Henri A. and {Ransom}, Scott M. and {Ray}, Paul S. and {Romano}, Joseph D. and {Runnoe}, Jessie C. and {Sardesai}, Shashwat C. and {Schmiedekamp}, Ann and {Schmiedekamp}, Carl and {Schmitz}, Kai and {Shapiro-Albert}, Brent J. and {Siemens}, Xavier and {Simon}, Joseph and {Siwek}, Magdalena S. and {Sosa Fiscella}, Sophia V. and {Stairs}, Ingrid H. and {Stinebring}, Daniel R. and {Stovall}, Kevin and {Susobhanan}, Abhimanyu and {Swiggum}, Joseph K. and {Taylor}, Stephen R. and {Turner}, Jacob E. and {Unal}, Caner and {Vallisneri}, Michele and {Vigeland}, Sarah J. and {Wahl}, Haley M. and {Willson}, London and {Witt}, Caitlin A. and {Wright}, David and {Young}, Olivia},
        title = "{The NANOGrav 15 yr Data Set: Looking for Signs of Discreteness in the Gravitational-wave Background}",
      journal = {\apj},
         year = 2025,
        month = jan,
       volume = {978},
       number = {1},
          eid = {31},
        pages = {31},
          doi = {10.3847/1538-4357/ad93d5},
archivePrefix = {arXiv},
       eprint = {2404.07020},
 primaryClass = {astro-ph.HE},
       adsurl = {https://ui.adsabs.harvard.edu/abs/2025ApJ...978...31A}
}

@ARTICLE{2024MNRAS.529.4295S,
       author = {{Saeedzadeh}, Vida and {Mukherjee}, Suvodip and {Babul}, Arif and {Tremmel}, Michael and {Quinn}, Thomas R.},
        title = "{Shining light on the hosts of the nano-Hertz gravitational wave sources: a theoretical perspective}",
      journal = {\mnras},
         year = 2024,
        month = apr,
       volume = {529},
       number = {4},
        pages = {4295-4310},
          doi = {10.1093/mnras/stae513},
archivePrefix = {arXiv},
       eprint = {2309.08683},
 primaryClass = {astro-ph.GA},
       adsurl = {https://ui.adsabs.harvard.edu/abs/2024MNRAS.529.4295S}
}

@ARTICLE{2022MNRAS.509.3488I,
       author = {{Izquierdo-Villalba}, David and {Sesana}, Alberto and {Bonoli}, Silvia and {Colpi}, Monica},
        title = "{Massive black hole evolution models confronting the n-Hz amplitude of the stochastic gravitational wave background}",
      journal = {\mnras},
         year = 2022,
        month = jan,
       volume = {509},
       number = {3},
        pages = {3488-3503},
          doi = {10.1093/mnras/stab3239},
archivePrefix = {arXiv},
       eprint = {2108.11671},
 primaryClass = {astro-ph.GA},
       adsurl = {https://ui.adsabs.harvard.edu/abs/2022MNRAS.509.3488I}
}

@ARTICLE{2023MNRAS.519.2083I,
       author = {{Izquierdo-Villalba}, David and {Sesana}, Alberto and {Colpi}, Monica},
        title = "{Unveiling the hosts of parsec-scale massive black hole binaries: morphology and electromagnetic signatures}",
      journal = {\mnras},
         year = 2023,
        month = feb,
       volume = {519},
       number = {2},
        pages = {2083-2100},
          doi = {10.1093/mnras/stac3677},
archivePrefix = {arXiv},
       eprint = {2207.04064},
 primaryClass = {astro-ph.GA},
       adsurl = {https://ui.adsabs.harvard.edu/abs/2023MNRAS.519.2083I}
}

@ARTICLE{2017MNRAS.471.4508K,
       author = {{Kelley}, Luke Zoltan and {Blecha}, Laura and {Hernquist}, Lars and {Sesana}, Alberto and {Taylor}, Stephen R.},
        title = "{The gravitational wave background from massive black hole binaries in Illustris: spectral features and time to detection with pulsar timing arrays}",
      journal = {\mnras},
         year = 2017,
        month = nov,
       volume = {471},
       number = {4},
        pages = {4508-4526},
          doi = {10.1093/mnras/stx1638},
archivePrefix = {arXiv},
       eprint = {1702.02180},
 primaryClass = {astro-ph.HE},
       adsurl = {https://ui.adsabs.harvard.edu/abs/2017MNRAS.471.4508K}
}

@ARTICLE{2020MNRAS.495.3252S,
       author = {{Shen}, Xuejian and {Hopkins}, Philip F. and {Faucher-Gigu{\`e}re}, Claude-Andr{\'e} and {Alexander}, D.~M. and {Richards}, Gordon T. and {Ross}, Nicholas P. and {Hickox}, R.~C.},
        title = "{The bolometric quasar luminosity function at z = 0-7}",
      journal = {\mnras},
         year = 2020,
        month = jan,
       volume = {495},
       number = {3},
        pages = {3252-3275},
          doi = {10.1093/mnras/staa1381},
archivePrefix = {arXiv},
       eprint = {2001.02696},
 primaryClass = {astro-ph.GA},
       adsurl = {https://ui.adsabs.harvard.edu/abs/2020MNRAS.495.3252S}
}

@ARTICLE{2014MNRAS.438.2804N,
       author = {{Nemmen}, Rodrigo S. and {Storchi-Bergmann}, Thaisa and {Eracleous}, Michael},
        title = "{Spectral models for low-luminosity active galactic nuclei in LINERs: the role of advection-dominated accretion and jets}",
      journal = {\mnras},
         year = 2014,
        month = mar,
       volume = {438},
       number = {4},
        pages = {2804-2827},
          doi = {10.1093/mnras/stt2388},
archivePrefix = {arXiv},
       eprint = {1312.1982},
 primaryClass = {astro-ph.HE},
       adsurl = {https://ui.adsabs.harvard.edu/abs/2014MNRAS.438.2804N}
}

@ARTICLE{1983ApJ...265L..39H,
       author = {{Hellings}, R.~W. and {Downs}, G.~S.},
        title = "{Upper limits on the isotropic gravitational radiation background from pulsar timing analysis.}",
      journal = {\apjl},
         year = 1983,
        month = feb,
       volume = {265},
        pages = {L39-L42},
          doi = {10.1086/183954},
       adsurl = {https://ui.adsabs.harvard.edu/abs/1983ApJ...265L..39H}
}

@BOOK{1993sapa.book.....P,
       author = {{Percival}, Donald B. and {Walden}, Andrew T.},
        title = "{Spectral Analysis for Physical Applications}",
         year = 1993,
       adsurl = {https://ui.adsabs.harvard.edu/abs/1993sapa.book.....P}
}

@ARTICLE{2014PhRvD..90j4012V,
       author = {{van Haasteren}, Rutger and {Vallisneri}, Michele},
        title = "{New advances in the Gaussian-process approach to pulsar-timing data analysis}",
      journal = {\prd},
         year = 2014,
        month = nov,
       volume = {90},
       number = {10},
          eid = {104012},
        pages = {104012},
          doi = {10.1103/PhysRevD.90.104012},
archivePrefix = {arXiv},
       eprint = {1407.1838},
 primaryClass = {gr-qc},
       adsurl = {https://ui.adsabs.harvard.edu/abs/2014PhRvD..90j4012V}
}

@software{ellis_2020_4059815,
  author       = {Ellis, Justin A. and
                  Vallisneri, Michele and
                  Taylor, Stephen R. and
                  Baker, Paul T.},
  title        = {ENTERPRISE: Enhanced Numerical Toolbox Enabling a
                   Robust PulsaR Inference SuitE
                  },
  month        = sep,
  year         = 2020,
  publisher    = {Zenodo},
  version      = {v3.0.0},
  doi          = {10.5281/zenodo.4059815},
  url          = {https://doi.org/10.5281/zenodo.4059815},
}

@ARTICLE{1978IEEEP..66...51H,
       author = {{Harris}, Fredric J.},
        title = "{On the Use of Windows for Harmonic Analysis with the Discrete Fourier Transform}",
      journal = {IEEE Proceedings},
         year = 1978,
        month = jan,
       volume = {66},
        pages = {51-83},
       adsurl = {https://ui.adsabs.harvard.edu/abs/1978IEEEP..66...51H}
}

@ARTICLE{2024arXiv241105906C,
       author = {{Chen}, Yifan and {Daniel}, Matthias and {D'Orazio}, Daniel J. and {Mitridate}, Andrea and {Sagunski}, Laura and {Xue}, Xiao and {Agazie}, Gabriella and {Baier}, Jeremy G. and {Baker}, Paul T. and {B{\'e}csy}, Bence and {Blecha}, Laura and {Brazier}, Adam and {Brook}, Paul R. and {Burke-Spolaor}, Sarah and {Burnette}, Rand and {Casey-Clyde}, J. Andrew and {Charisi}, Maria and {Chatterjee}, Shami and {Cohen}, Tyler and {Cordes}, James M. and {Cornish}, Neil J. and {Crawford}, Fronefield and {Cromartie}, H. Thankful and {DeCesar}, Megan E. and {Demorest}, Paul B. and {Deng}, Heling and {Dey}, Lankeswar and {Dolch}, Timothy and {Ferrara}, Elizabeth C. and {Fiore}, William and {Fonseca}, Emmanuel and {Freedman}, Gabriel E. and {Gardiner}, Emiko C. and {Gersbach}, Kyle A. and {Glaser}, Joseph and {Good}, Deborah C. and {G{\"u}ltekin}, Kayhan and {Hazboun}, Jeffrey S. and {Jennings}, Ross J. and {Johnson}, Aaron D. and {Kaplan}, David L. and {Kelley}, Luke Zoltan and {Key}, Joey S. and {Laal}, Nima and {Lam}, Michael T. and {Lamb}, William G. and {Larsen}, Bjorn and {Lazio}, T. Joseph W. and {Lewandowska}, Natalia and {Liu}, Tingting and {Luo}, Jing and {Lynch}, Ryan S. and {Ma}, Chung-Pei and {Madison}, Dustin R. and {McEwen}, Alexander and {McKee}, James W. and {McLaughlin}, Maura A. and {Meyers}, Patrick M. and {Mingarelli}, Chiara M.~F. and {Nice}, David J. and {Ocker}, Stella Koch and {Olum}, Ken D. and {Pennucci}, Timothy T. and {Petrov}, Polina and {Pol}, Nihan S. and {Radovan}, Henri A. and {Ransom}, Scott M. and {Ray}, Paul S. and {Romano}, Joseph D. and {Runnoe}, Jessie C. and {Saffer}, Alexander and {Sardesai}, Shashwat C. and {Schmitz}, Kai and {Siemens}, Xavier and {Simon}, Joseph and {Siwek}, Magdalena S. and {Sosa Fiscella}, Sophia V. and {Stairs}, Ingrid H. and {Stinebring}, Daniel R. and {Susobhanan}, Abhimanyu and {Swiggum}, Joseph K. and {Taylor}, Jacob and {Taylor}, Stephen R. and {Turner}, Jacob E. and {Unal}, Caner and {Vallisneri}, Michele and {van Haasteren}, Rutger and {Verbiest}, Joris and {Vigeland}, Sarah J. and {Witt}, Caitlin A. and {Wright}, David and {Young}, Olivia},
        title = "{Galaxy Tomography with the Gravitational Wave Background from Supermassive Black Hole Binaries}",
      journal = {arXiv e-prints},
         year = 2024,
        month = nov,
          eid = {arXiv:2411.05906},
        pages = {arXiv:2411.05906},
          doi = {10.48550/arXiv.2411.05906},
archivePrefix = {arXiv},
       eprint = {2411.05906},
 primaryClass = {astro-ph.HE},
       adsurl = {https://ui.adsabs.harvard.edu/abs/2024arXiv241105906C}
}

@ARTICLE{2013CQGra..30v4014S,
       author = {{Sesana}, A.},
        title = "{Insights into the astrophysics of supermassive black hole binaries from pulsar timing observations}",
      journal = {Classical and Quantum Gravity},
         year = 2013,
        month = nov,
       volume = {30},
       number = {22},
          eid = {224014},
        pages = {224014},
          doi = {10.1088/0264-9381/30/22/224014},
archivePrefix = {arXiv},
       eprint = {1307.2600},
 primaryClass = {astro-ph.CO},
       adsurl = {https://ui.adsabs.harvard.edu/abs/2013CQGra..30v4014S}
}

@ARTICLE{2013ApJ...768...29V,
       author = {{Volonteri}, Marta and {Ciotti}, Luca},
        title = "{Massive Black Holes in Central Cluster Galaxies}",
      journal = {\apj},
         year = 2013,
        month = may,
       volume = {768},
       number = {1},
          eid = {29},
        pages = {29},
          doi = {10.1088/0004-637X/768/1/29},
archivePrefix = {arXiv},
       eprint = {1211.6840},
 primaryClass = {astro-ph.CO},
       adsurl = {https://ui.adsabs.harvard.edu/abs/2013ApJ...768...29V}
}

@ARTICLE{2015ApJ...799..178K,
       author = {{Kulier}, Andrea and {Ostriker}, Jeremiah P. and {Natarajan}, Priyamvada and {Lackner}, Claire N. and {Cen}, Renyue},
        title = "{Understanding Black Hole Mass Assembly via Accretion and Mergers at Late Times in Cosmological Simulations}",
      journal = {\apj},
         year = 2015,
        month = feb,
       volume = {799},
       number = {2},
          eid = {178},
        pages = {178},
          doi = {10.1088/0004-637X/799/2/178},
archivePrefix = {arXiv},
       eprint = {1307.3684},
 primaryClass = {astro-ph.CO},
       adsurl = {https://ui.adsabs.harvard.edu/abs/2015ApJ...799..178K}
}

@ARTICLE{2014MNRAS.440.1590D,
       author = {{Dubois}, Yohan and {Volonteri}, Marta and {Silk}, Joseph},
        title = "{Black hole evolution - III. Statistical properties of mass growth and spin evolution using large-scale hydrodynamical cosmological simulations}",
      journal = {\mnras},
         year = 2014,
        month = may,
       volume = {440},
       number = {2},
        pages = {1590-1606},
          doi = {10.1093/mnras/stu373},
archivePrefix = {arXiv},
       eprint = {1304.4583},
 primaryClass = {astro-ph.CO},
       adsurl = {https://ui.adsabs.harvard.edu/abs/2014MNRAS.440.1590D}
}

@ARTICLE{2025ApJ...987..106C,
       author = {{Casey-Clyde}, J. Andrew and {Mingarelli}, Chiara M.~F. and {Greene}, Jenny E. and {Goulding}, Andy D. and {Chen}, Siyuan and {Trump}, Jonathan R.},
        title = "{Quasars Can Signpost Supermassive Black Hole Binaries}",
      journal = {\apj},
         year = 2025,
        month = jul,
       volume = {987},
       number = {2},
          eid = {106},
        pages = {106},
          doi = {10.3847/1538-4357/adce05},
archivePrefix = {arXiv},
       eprint = {2405.19406},
 primaryClass = {astro-ph.HE},
       adsurl = {https://ui.adsabs.harvard.edu/abs/2025ApJ...987..106C}
}

@ARTICLE{2025arXiv250401074T,
       author = {{Truant}, Riccardo J. and {Izquierdo-Villalba}, David and {Sesana}, Alberto and {Mohiuddin Shaifullah}, Golam and {Bonetti}, Matteo and {Spinoso}, Daniele and {Bonoli}, Silvia},
        title = "{Lighting up the nano-hertz gravitational wave sky: opportunities and challenges of multimessenger astronomy with PTA experiments}",
      journal = {arXiv e-prints},
         year = 2025,
        month = apr,
          eid = {arXiv:2504.01074},
        pages = {arXiv:2504.01074},
          doi = {10.48550/arXiv.2504.01074},
archivePrefix = {arXiv},
       eprint = {2504.01074},
 primaryClass = {astro-ph.GA},
       adsurl = {https://ui.adsabs.harvard.edu/abs/2025arXiv250401074T}
}

@BOOK{2011gwpa.book.....C,
       author = {{Creighton}, Jolien and {Anderson}, Warren},
        title = "{Gravitational-Wave Physics and Astronomy: An Introduction to Theory, Experiment and Data Analysis.}",
         year = 2011,
       adsurl = {https://ui.adsabs.harvard.edu/abs/2011gwpa.book.....C}
}

@ARTICLE{Masse_estimates_rakshit,
       author = {{Rakshit}, Suvendu and {Stalin}, C.~S. and {Kotilainen}, Jari},
        title = "{Spectral Properties of Quasars from Sloan Digital Sky Survey Data Release 14: The Catalog}",
      journal = {The Astrophysical Journal Supplement Series},
         year = 2020,
        month = jul,
       volume = {249},
       number = {1},
          eid = {17},
        pages = {17},
          doi = {10.3847/1538-4365/ab99c5},
archivePrefix = {arXiv},
       eprint = {1910.10395},
 primaryClass = {astro-ph.GA},
       adsurl = {https://ui.adsabs.harvard.edu/abs/2020ApJS..249...17R}
}

@ARTICLE{2020A&A...636A..73D,
       author = {{Duras}, F. and {Bongiorno}, A. and {Ricci}, F. and {Piconcelli}, E. and {Shankar}, F. and {Lusso}, E. and {Bianchi}, S. and {Fiore}, F. and {Maiolino}, R. and {Marconi}, A. and {Onori}, F. and {Sani}, E. and {Schneider}, R. and {Vignali}, C. and {La Franca}, F.},
        title = "{Universal bolometric corrections for active galactic nuclei over seven luminosity decades}",
      journal = {\aap},
         year = 2020,
        month = apr,
       volume = {636},
          eid = {A73},
        pages = {A73},
          doi = {10.1051/0004-6361/201936817},
archivePrefix = {arXiv},
       eprint = {2001.09984},
 primaryClass = {astro-ph.GA},
       adsurl = {https://ui.adsabs.harvard.edu/abs/2020A&A...636A..73D}
}

@ARTICLE{2010arXiv1010.3785C,
       author = {{Cordes}, J.~M. and {Shannon}, R.~M.},
        title = "{A Measurement Model for Precision Pulsar Timing}",
      journal = {arXiv e-prints},
         year = 2010,
        month = oct,
          eid = {arXiv:1010.3785},
        pages = {arXiv:1010.3785},
          doi = {10.48550/arXiv.1010.3785},
archivePrefix = {arXiv},
       eprint = {1010.3785},
 primaryClass = {astro-ph.IM},
       adsurl = {https://ui.adsabs.harvard.edu/abs/2010arXiv1010.3785C}
}

@ARTICLE{2013MNRAS.428.1147V,
       author = {{van Haasteren}, Rutger and {Levin}, Yuri},
        title = "{Understanding and analysing time-correlated stochastic signals in pulsar timing}",
      journal = {\mnras},
         year = 2013,
        month = jan,
       volume = {428},
       number = {2},
        pages = {1147-1159},
          doi = {10.1093/mnras/sts097},
archivePrefix = {arXiv},
       eprint = {1202.5932},
 primaryClass = {astro-ph.IM},
       adsurl = {https://ui.adsabs.harvard.edu/abs/2013MNRAS.428.1147V}
}

@ARTICLE{2024Natur.627...59M,
       author = {{Maiolino}, Roberto and {Scholtz}, Jan and {Witstok}, Joris and {Carniani}, Stefano and {D'Eugenio}, Francesco and {de Graaff}, Anna and {{\"U}bler}, Hannah and {Tacchella}, Sandro and {Curtis-Lake}, Emma and {Arribas}, Santiago and {Bunker}, Andrew and {Charlot}, St{\'e}phane and {Chevallard}, Jacopo and {Curti}, Mirko and {Looser}, Tobias J. and {Maseda}, Michael V. and {Rawle}, Timothy D. and {Rodr{\'\i}guez del Pino}, Bruno and {Willott}, Chris J. and {Egami}, Eiichi and {Eisenstein}, Daniel J. and {Hainline}, Kevin N. and {Robertson}, Brant and {Williams}, Christina C. and {Willmer}, Christopher N.~A. and {Baker}, William M. and {Boyett}, Kristan and {DeCoursey}, Christa and {Fabian}, Andrew C. and {Helton}, Jakob M. and {Ji}, Zhiyuan and {Jones}, Gareth C. and {Kumari}, Nimisha and {Laporte}, Nicolas and {Nelson}, Erica J. and {Perna}, Michele and {Sandles}, Lester and {Shivaei}, Irene and {Sun}, Fengwu},
        title = "{A small and vigorous black hole in the early Universe}",
      journal = {\nat},
         year = 2024,
        month = mar,
       volume = {627},
       number = {8002},
        pages = {59-63},
          doi = {10.1038/s41586-024-07052-5},
archivePrefix = {arXiv},
       eprint = {2305.12492},
 primaryClass = {astro-ph.GA},
       adsurl = {https://ui.adsabs.harvard.edu/abs/2024Natur.627...59M}
}

@ARTICLE{2021NatRP...3..732V,
       author = {{Volonteri}, Marta and {Habouzit}, M{\'e}lanie and {Colpi}, Monica},
        title = "{The origins of massive black holes}",
      journal = {Nature Reviews Physics},
         year = 2021,
        month = sep,
       volume = {3},
       number = {11},
        pages = {732-743},
          doi = {10.1038/s42254-021-00364-9},
archivePrefix = {arXiv},
       eprint = {2110.10175},
 primaryClass = {astro-ph.GA},
       adsurl = {https://ui.adsabs.harvard.edu/abs/2021NatRP...3..732V}
}

@ARTICLE{2017arXiv170200786A,
       author = {{Amaro-Seoane}, Pau and {Audley}, Heather and {Babak}, Stanislav and {Baker}, John and {Barausse}, Enrico and {Bender}, Peter and {Berti}, Emanuele and {Binetruy}, Pierre and {Born}, Michael and {Bortoluzzi}, Daniele and {Camp}, Jordan and {Caprini}, Chiara and {Cardoso}, Vitor and {Colpi}, Monica and {Conklin}, John and {Cornish}, Neil and {Cutler}, Curt and {Danzmann}, Karsten and {Dolesi}, Rita and {Ferraioli}, Luigi and {Ferroni}, Valerio and {Fitzsimons}, Ewan and {Gair}, Jonathan and {Gesa Bote}, Lluis and {Giardini}, Domenico and {Gibert}, Ferran and {Grimani}, Catia and {Halloin}, Hubert and {Heinzel}, Gerhard and {Hertog}, Thomas and {Hewitson}, Martin and {Holley-Bockelmann}, Kelly and {Hollington}, Daniel and {Hueller}, Mauro and {Inchauspe}, Henri and {Jetzer}, Philippe and {Karnesis}, Nikos and {Killow}, Christian and {Klein}, Antoine and {Klipstein}, Bill and {Korsakova}, Natalia and {Larson}, Shane L and {Livas}, Jeffrey and {Lloro}, Ivan and {Man}, Nary and {Mance}, Davor and {Martino}, Joseph and {Mateos}, Ignacio and {McKenzie}, Kirk and {McWilliams}, Sean T and {Miller}, Cole and {Mueller}, Guido and {Nardini}, Germano and {Nelemans}, Gijs and {Nofrarias}, Miquel and {Petiteau}, Antoine and {Pivato}, Paolo and {Plagnol}, Eric and {Porter}, Ed and {Reiche}, Jens and {Robertson}, David and {Robertson}, Norna and {Rossi}, Elena and {Russano}, Giuliana and {Schutz}, Bernard and {Sesana}, Alberto and {Shoemaker}, David and {Slutsky}, Jacob and {Sopuerta}, Carlos F. and {Sumner}, Tim and {Tamanini}, Nicola and {Thorpe}, Ira and {Troebs}, Michael and {Vallisneri}, Michele and {Vecchio}, Alberto and {Vetrugno}, Daniele and {Vitale}, Stefano and {Volonteri}, Marta and {Wanner}, Gudrun and {Ward}, Harry and {Wass}, Peter and {Weber}, William and {Ziemer}, John and {Zweifel}, Peter},
        title = "{Laser Interferometer Space Antenna}",
      journal = {arXiv e-prints},
         year = 2017,
        month = feb,
          eid = {arXiv:1702.00786},
        pages = {arXiv:1702.00786},
          doi = {10.48550/arXiv.1702.00786},
archivePrefix = {arXiv},
       eprint = {1702.00786},
 primaryClass = {astro-ph.IM},
       adsurl = {https://ui.adsabs.harvard.edu/abs/2017arXiv170200786A}
}

@ARTICLE{1990ApJ...361..300F,
       author = {{Foster}, R.~S. and {Backer}, D.~C.},
        title = "{Constructing a Pulsar Timing Array}",
      journal = {\apj},
         year = 1990,
        month = sep,
       volume = {361},
        pages = {300},
          doi = {10.1086/169195},
       adsurl = {https://ui.adsabs.harvard.edu/abs/1990ApJ...361..300F}
}

@article{Goncharov:2024htb,
    author = "Goncharov, Boris and others",
    title = "{Reading signatures of supermassive binary black holes in pulsar timing array observations}",
    eprint = "2409.03627",
    archivePrefix = "arXiv",
    primaryClass = "astro-ph.HE",
    doi = "10.1038/s41467-025-65450-3",
    journal = "Nature Commun.",
    volume = "16",
    number = "1",
    pages = "9692",
    year = "2025"
}

@ARTICLE{2014ApJ...789..156M,
       author = {{McWilliams}, Sean T. and {Ostriker}, Jeremiah P. and {Pretorius}, Frans},
        title = "{Gravitational Waves and Stalled Satellites from Massive Galaxy Mergers at z <= 1}",
      journal = {\apj},
         year = 2014,
        month = jul,
       volume = {789},
       number = {2},
          eid = {156},
        pages = {156},
          doi = {10.1088/0004-637X/789/2/156},
archivePrefix = {arXiv},
       eprint = {1211.5377},
 primaryClass = {astro-ph.CO},
       adsurl = {https://ui.adsabs.harvard.edu/abs/2014ApJ...789..156M}
}

@ARTICLE{2018MNRAS.477.2599B,
       author = {{Bonetti}, Matteo and {Sesana}, Alberto and {Barausse}, Enrico and {Haardt}, Francesco},
        title = "{Post-Newtonian evolution of massive black hole triplets in galactic nuclei - III. A robust lower limit to the nHz stochastic background of gravitational waves}",
      journal = {\mnras},
         year = 2018,
        month = jun,
       volume = {477},
       number = {2},
        pages = {2599-2612},
          doi = {10.1093/mnras/sty874},
archivePrefix = {arXiv},
       eprint = {1709.06095},
 primaryClass = {astro-ph.GA},
       adsurl = {https://ui.adsabs.harvard.edu/abs/2018MNRAS.477.2599B}
}

@ARTICLE{2025ApJ...991L..19C,
       author = {{Chen}, Nianyi and {Di Matteo}, Tiziana and {Zhou}, Yihao and {Kelley}, Luke Zoltan and {Blecha}, Laura and {Ni}, Yueying and {Bird}, Simeon and {Yang}, Yanhui and {Croft}, Rupert},
        title = "{The Gravitational-wave Background from Massive Black Holes in the ASTRID Simulation}",
      journal = {\apjl},
         year = 2025,
        month = sep,
       volume = {991},
       number = {1},
          eid = {L19},
        pages = {L19},
          doi = {10.3847/2041-8213/adefe2},
archivePrefix = {arXiv},
       eprint = {2502.01024},
 primaryClass = {astro-ph.GA},
       adsurl = {https://ui.adsabs.harvard.edu/abs/2025ApJ...991L..19C}
}

@ARTICLE{2017MNRAS.464.3131K,
       author = {{Kelley}, Luke Zoltan and {Blecha}, Laura and {Hernquist}, Lars},
        title = "{Massive black hole binary mergers in dynamical galactic environments}",
      journal = {\mnras},
         year = 2017,
        month = jan,
       volume = {464},
       number = {3},
        pages = {3131-3157},
          doi = {10.1093/mnras/stw2452},
archivePrefix = {arXiv},
       eprint = {1606.01900},
 primaryClass = {astro-ph.HE},
       adsurl = {https://ui.adsabs.harvard.edu/abs/2017MNRAS.464.3131K}
}

@ARTICLE{2022MNRAS.511.5241S,
       author = {{Sykes}, Bailey and {Middleton}, Hannah and {Melatos}, Andrew and {Di Matteo}, Tiziana and {DeGraf}, Colin and {Bhowmick}, Aklant},
        title = "{An estimate of the stochastic gravitational wave background from the MassiveBlackII simulation}",
      journal = {\mnras},
         year = 2022,
        month = apr,
       volume = {511},
       number = {4},
        pages = {5241-5250},
          doi = {10.1093/mnras/stac388},
archivePrefix = {arXiv},
       eprint = {2202.05410},
 primaryClass = {astro-ph.GA},
       adsurl = {https://ui.adsabs.harvard.edu/abs/2022MNRAS.511.5241S}
}

@ARTICLE{2025CQGra..42b5021C,
       author = {{Cella}, Katharine and {Taylor}, Stephen R. and {Zoltan Kelley}, Luke},
        title = "{Host galaxy demographics of individually detectable supermassive black-hole binaries with pulsar timing arrays}",
      journal = {Classical and Quantum Gravity},
         year = 2025,
        month = jan,
       volume = {42},
       number = {2},
          eid = {025021},
        pages = {025021},
          doi = {10.1088/1361-6382/ad9131},
archivePrefix = {arXiv},
       eprint = {2407.01659},
 primaryClass = {astro-ph.GA},
       adsurl = {https://ui.adsabs.harvard.edu/abs/2025CQGra..42b5021C}
}

@ARTICLE{2017NatAs...1..886M,
       author = {{Mingarelli}, Chiara M.~F. and {Lazio}, T. Joseph W. and {Sesana}, Alberto and {Greene}, Jenny E. and {Ellis}, Justin A. and {Ma}, Chung-Pei and {Croft}, Steve and {Burke-Spolaor}, Sarah and {Taylor}, Stephen R.},
        title = "{The local nanohertz gravitational-wave landscape from supermassive black hole binaries}",
      journal = {Nature Astronomy},
         year = 2017,
        month = nov,
       volume = {1},
        pages = {886-892},
          doi = {10.1038/s41550-017-0299-6},
archivePrefix = {arXiv},
       eprint = {1708.03491},
 primaryClass = {astro-ph.GA},
       adsurl = {https://ui.adsabs.harvard.edu/abs/2017NatAs...1..886M}
}

@ARTICLE{2024ApJ...976..129P,
       author = {{Petrov}, Polina and {Taylor}, Stephen R. and {Charisi}, Maria and {Ma}, Chung-Pei},
        title = "{Identifying the Host Galaxies of Supermassive Black Hole Binaries Found by Pulsar Timing Arrays}",
      journal = {\apj},
         year = 2024,
        month = nov,
       volume = {976},
       number = {1},
          eid = {129},
        pages = {129},
          doi = {10.3847/1538-4357/ad7b14},
archivePrefix = {arXiv},
       eprint = {2406.04409},
 primaryClass = {astro-ph.GA},
       adsurl = {https://ui.adsabs.harvard.edu/abs/2024ApJ...976..129P}
}

@ARTICLE{2021ApJ...921..178L,
       author = {{Liu}, Tingting and {Vigeland}, Sarah J.},
        title = "{Multi-messenger Approaches to Supermassive Black Hole Binary Detection and Parameter Estimation: Implications for Nanohertz Gravitational Wave Searches with Pulsar Timing Arrays}",
      journal = {\apj},
         year = 2021,
        month = nov,
       volume = {921},
       number = {2},
          eid = {178},
        pages = {178},
          doi = {10.3847/1538-4357/ac1da9},
archivePrefix = {arXiv},
       eprint = {2105.08087},
 primaryClass = {astro-ph.HE},
       adsurl = {https://ui.adsabs.harvard.edu/abs/2021ApJ...921..178L}
}

@ARTICLE{2022MNRAS.510.5929C,
       author = {{Charisi}, Maria and {Taylor}, Stephen R. and {Runnoe}, Jessie and {Bogdanovic}, Tamara and {Trump}, Jonathan R.},
        title = "{Multimessenger time-domain signatures of supermassive black hole binaries}",
      journal = {\mnras},
         year = 2022,
        month = mar,
       volume = {510},
       number = {4},
        pages = {5929-5944},
          doi = {10.1093/mnras/stab3713},
archivePrefix = {arXiv},
       eprint = {2110.14661},
 primaryClass = {astro-ph.HE},
       adsurl = {https://ui.adsabs.harvard.edu/abs/2022MNRAS.510.5929C}
}

@ARTICLE{2024NatAs...8..126B,
       author = {{Bogd{\'a}n}, {\'A}kos and {Goulding}, Andy D. and {Natarajan}, Priyamvada and {Kov{\'a}cs}, Orsolya E. and {Tremblay}, Grant R. and {Chadayammuri}, Urmila and {Volonteri}, Marta and {Kraft}, Ralph P. and {Forman}, William R. and {Jones}, Christine and {Churazov}, Eugene and {Zhuravleva}, Irina},
        title = "{Evidence for heavy-seed origin of early supermassive black holes from a z {\ensuremath{\approx}} 10 X-ray quasar}",
      journal = {Nature Astronomy},
         year = 2024,
        month = jan,
       volume = {8},
       number = {1},
        pages = {126-133},
          doi = {10.1038/s41550-023-02111-9},
archivePrefix = {arXiv},
       eprint = {2305.15458},
 primaryClass = {astro-ph.GA},
       adsurl = {https://ui.adsabs.harvard.edu/abs/2024NatAs...8..126B}
}

@ARTICLE{2004MNRAS.354L..37M,
       author = {{Merloni}, Andrea and {Rudnick}, Gregory and {Di Matteo}, Tiziana},
        title = "{Tracing the cosmological assembly of stars and supermassive black holes in galaxies}",
      journal = {\mnras},
         year = 2004,
        month = nov,
       volume = {354},
       number = {3},
        pages = {L37-L42},
          doi = {10.1111/j.1365-2966.2004.08382.x},
archivePrefix = {arXiv},
       eprint = {astro-ph/0409187},
 primaryClass = {astro-ph},
       adsurl = {https://ui.adsabs.harvard.edu/abs/2004MNRAS.354L..37M}
}

@ARTICLE{1998Natur.395A..14R,
       author = {{Richstone}, D. and {Ajhar}, E.~A. and {Bender}, R. and {Bower}, G. and {Dressler}, A. and {Faber}, S.~M. and {Filippenko}, A.~V. and {Gebhardt}, K. and {Green}, R. and {Ho}, L.~C. and {Kormendy}, J. and {Lauer}, T.~R. and {Magorrian}, J. and {Tremaine}, S.},
        title = "{Supermassive black holes and the evolution of galaxies.}",
      journal = {\nat},
         year = 1998,
        month = oct,
       volume = {385},
       number = {6701},
        pages = {A14},
          doi = {10.48550/arXiv.astro-ph/9810378},
archivePrefix = {arXiv},
       eprint = {astro-ph/9810378},
 primaryClass = {astro-ph},
       adsurl = {https://ui.adsabs.harvard.edu/abs/1998Natur.395A..14R}
}

@article{foustoul_2025,
    author = {{Foustoul}, V. and {Webb}, N.~A and {Mignon-Risse}, R. and {Kammoun}, E. and {Volonteri}, M. and {Dong-P{\'a}ez}, Chi An},
    title = {A catalogue of candidate milli-parsec separation massive black hole binaries from long term optical photometric monitoring},
    journal = {\aap},
    year = {2025},
    note= {forthcoming}}

@ARTICLE{2025ApJ...990...46V,
       author = {{Veronesi}, Niccol{\`o} and {Charisi}, Maria and {Taylor}, Stephen R. and {Runnoe}, Jessie and {D'Orazio}, Daniel J.},
        title = "{The Host Galaxies of Pulsar Timing Array Sources: Converting Supermassive Black Hole Binary Parameters into Electromagnetic Observables}",
      journal = {\apj},
         year = 2025,
        month = sep,
       volume = {990},
       number = {1},
          eid = {46},
        pages = {46},
          doi = {10.3847/1538-4357/adf065},
archivePrefix = {arXiv},
       eprint = {2505.11598},
 primaryClass = {astro-ph.HE},
       adsurl = {https://ui.adsabs.harvard.edu/abs/2025ApJ...990...46V}
}

@ARTICLE{2009MNRAS.394.2255S,
       author = {{Sesana}, A. and {Vecchio}, A. and {Volonteri}, M.},
        title = "{Gravitational waves from resolvable massive black hole binary systems and observations with Pulsar Timing Arrays}",
      journal = {\mnras},
         year = 2009,
        month = apr,
       volume = {394},
       number = {4},
        pages = {2255-2265},
          doi = {10.1111/j.1365-2966.2009.14499.x},
archivePrefix = {arXiv},
       eprint = {0809.3412},
 primaryClass = {astro-ph},
       adsurl = {https://ui.adsabs.harvard.edu/abs/2009MNRAS.394.2255S}
}

@ARTICLE{2022MNRAS.516..410Z,
       author = {{Zic}, Andrew and {Hobbs}, George and {Shannon}, R.~M. and {Reardon}, Daniel and {Goncharov}, Boris and {Bhat}, N.~D. Ramesh and {Cameron}, Andrew and {Dai}, Shi and {Dawson}, J.~R. and {Kerr}, Matthew and {Manchester}, R.~N. and {Mandow}, Rami and {Marshman}, Tommy and {Russell}, Christopher J. and {Thyagarajan}, Nithyanandan and {Zhu}, X. -J.},
        title = "{Evaluating the prevalence of spurious correlations in pulsar timing array data sets}",
      journal = {\mnras},
         year = 2022,
        month = oct,
       volume = {516},
       number = {1},
        pages = {410-420},
          doi = {10.1093/mnras/stac2100},
archivePrefix = {arXiv},
       eprint = {2207.12237},
 primaryClass = {astro-ph.HE},
       adsurl = {https://ui.adsabs.harvard.edu/abs/2022MNRAS.516..410Z}
}

@ARTICLE{2024A&A...690A.118E,
       author = {{EPTA Collaboration} and {InPTA Collaboration} and {Antoniadis}, J. and {Arumugam}, P. and {Arumugam}, S. and {Babak}, S. and {Bagchi}, M. and {Bak Nielsen}, A. -S. and {Bassa}, C.~G. and {Bathula}, A. and {Berthereau}, A. and {Bonetti}, M. and {Bortolas}, E. and {Brook}, P.~R. and {Burgay}, M. and {Caballero}, R.~N. and {Chalumeau}, A. and {Champion}, D.~J. and {Chanlaridis}, S. and {Chen}, S. and {Cognard}, I. and {Dandapat}, S. and {Deb}, D. and {Desai}, S. and {Desvignes}, G. and {Dhanda-Batra}, N. and {Dwivedi}, C. and {Falxa}, M. and {Ferranti}, I. and {Ferdman}, R.~D. and {Franchini}, A. and {Gair}, J.~R. and {Goncharov}, B. and {Gopakumar}, A. and {Graikou}, E. and {Grie{\ss}meier}, J. -M. and {Guillemot}, L. and {Guo}, Y.~J. and {Gupta}, Y. and {Hisano}, S. and {Hu}, H. and {Iraci}, F. and {Izquierdo-Villalba}, D. and {Jang}, J. and {Jawor}, J. and {Janssen}, G.~H. and {Jessner}, A. and {Joshi}, B.~C. and {Kareem}, F. and {Karuppusamy}, R. and {Keane}, E.~F. and {Keith}, M.~J. and {Kharbanda}, D. and {Kikunaga}, T. and {Kolhe}, N. and {Kramer}, M. and {Krishnakumar}, M.~A. and {Lackeos}, K. and {Lee}, K.~J. and {Liu}, K. and {Liu}, Y. and {Lyne}, A.~G. and {McKee}, J.~W. and {Maan}, Y. and {Main}, R.~A. and {Manzini}, S. and {Mickaliger}, M.~B. and {Ni{\c{t}}u}, I.~C. and {Nobleson}, K. and {Paladi}, A.~K. and {Parthasarathy}, A. and {Perera}, B.~B.~P. and {Perrodin}, D. and {Petiteau}, A. and {Porayko}, N.~K. and {Possenti}, A. and {Prabu}, T. and {Quelquejay Leclere}, H. and {Rana}, P. and {Samajdar}, A. and {Sanidas}, S.~A. and {Sesana}, A. and {Shaifullah}, G. and {Singha}, J. and {Speri}, L. and {Spiewak}, R. and {Srivastava}, A. and {Stappers}, B.~W. and {Surnis}, M. and {Susarla}, S.~C. and {Susobhanan}, A. and {Takahashi}, K. and {Tarafdar}, P. and {Theureau}, G. and {Tiburzi}, C. and {van der Wateren}, E. and {Vecchio}, A. and {Venkatraman Krishnan}, V. and {Verbiest}, J.~P.~W. and {Wang}, J. and {Wang}, L. and {Wu}, Z.},
        title = "{The second data release from the European Pulsar Timing Array. V. Search for continuous gravitational wave signals}",
      journal = {\aap},
         year = 2024,
        month = oct,
       volume = {690},
          eid = {A118},
        pages = {A118},
          doi = {10.1051/0004-6361/202348568},
archivePrefix = {arXiv},
       eprint = {2306.16226},
 primaryClass = {astro-ph.HE},
       adsurl = {https://ui.adsabs.harvard.edu/abs/2024A&A...690A.118E}
}

@ARTICLE{2025MNRAS.537L...1V,
       author = {{van Haasteren}, Rutger},
        title = "{Use model averaging instead of model selection in pulsar timing}",
      journal = {\mnras},
         year = 2025,
        month = feb,
       volume = {537},
       number = {1},
        pages = {L1-L6},
          doi = {10.1093/mnrasl/slae108},
archivePrefix = {arXiv},
       eprint = {2409.06050},
 primaryClass = {astro-ph.IM},
       adsurl = {https://ui.adsabs.harvard.edu/abs/2025MNRAS.537L...1V}
}

@ARTICLE{2025arXiv250613866C,
       author = {{Crisostomi}, Marco and {van Haasteren}, Rutger and {Meyers}, Patrick M. and {Vallisneri}, Michele},
        title = "{Beyond diagonal approximations: improved covariance modeling for pulsar timing array data analysis}",
      journal = {arXiv e-prints},
         year = 2025,
        month = jun,
          eid = {arXiv:2506.13866},
        pages = {arXiv:2506.13866},
          doi = {10.48550/arXiv.2506.13866},
archivePrefix = {arXiv},
       eprint = {2506.13866},
 primaryClass = {astro-ph.IM},
       adsurl = {https://ui.adsabs.harvard.edu/abs/2025arXiv250613866C}
}

@ARTICLE{2006ApJ...653.1571J,
       author = {{Jenet}, F.~A. and {Hobbs}, G.~B. and {van Straten}, W. and {Manchester}, R.~N. and {Bailes}, M. and {Verbiest}, J.~P.~W. and {Edwards}, R.~T. and {Hotan}, A.~W. and {Sarkissian}, J.~M. and {Ord}, S.~M.},
        title = "{Upper Bounds on the Low-Frequency Stochastic Gravitational Wave Background from Pulsar Timing Observations: Current Limits and Future Prospects}",
      journal = {\apj},
         year = 2006,
        month = dec,
       volume = {653},
       number = {2},
        pages = {1571-1576},
          doi = {10.1086/508702},
archivePrefix = {arXiv},
       eprint = {astro-ph/0609013},
 primaryClass = {astro-ph},
       adsurl = {https://ui.adsabs.harvard.edu/abs/2006ApJ...653.1571J}
}

@ARTICLE{1983MNRAS.203..945B,
       author = {{Bertotti}, B. and {Carr}, B.~J. and {Rees}, M.~J.},
        title = "{Limits from the timing of pulsars on the cosmic gravitational wave background.}",
      journal = {\mnras},
         year = 1983,
        month = jun,
       volume = {203},
        pages = {945-954},
          doi = {10.1093/mnras/203.4.945},
       adsurl = {https://ui.adsabs.harvard.edu/abs/1983MNRAS.203..945B}
}

@ARTICLE{2013PhRvD..87d4035T,
       author = {{Taylor}, Stephen R. and {Gair}, Jonathan R. and {Lentati}, L.},
        title = "{Weighing the evidence for a gravitational-wave background in the first International Pulsar Timing Array data challenge}",
      journal = {\prd},
         year = 2013,
        month = feb,
       volume = {87},
       number = {4},
          eid = {044035},
        pages = {044035},
          doi = {10.1103/PhysRevD.87.044035},
archivePrefix = {arXiv},
       eprint = {1210.6014},
 primaryClass = {astro-ph.IM},
       adsurl = {https://ui.adsabs.harvard.edu/abs/2013PhRvD..87d4035T}
}

\begin{appendix}
\onecolumn

\section{\label{sec: appendix link Ni n} Relating merger density and number of inspiralling MBHBs}

An intuitive way to obtain the characteristic strain, or equivalently the energy density of the GWB, is to add up the energy contribution from each inspiralling binary emitting in the PTA band by introducing ${\rm d}^3 N_{\rm i} / {\rm d}z {\rm d}\vec\xi {\rm d} \ln\forb$, the number of binaries in a layer of comoving volume encompassed between $z$ and $z+{\rm d}z$, with binary parameters in [$\xi$, $\xi +{\rm d}\xi$] and orbital frequency in $[\ln \forb, \ln\forb + {\rm d}\ln\forb]$ (see Eq. \ref{eq: GWB as a sum over sources}).

However, from the cosmological simulation we can only obtain the number of merging binaries $N_{\rm m}$, per comoving volume, per unit cosmological time, given in terms of redshift $z$, $\frac{{\rm d}n}{{\rm d}z}(z)=\frac{{\rm d}^2 N_{\rm m}}{{\rm d} z {\rm d}V_{c}}(z)$. It turns out that assuming that the merger rate is in a steady state, the two can be related. Following \citet{2023JCAP...08..034B}, we first switch from $\ln\forb$ to time to coalescence $\tau_{\rm c}$ to track the dynamics of each inspiralling MBHBs,
\begin{equation}
    \label{eq: ap switch to tauc}
    \frac{{\rm d}^3 N_{\rm i}}{{\rm d}z {\rm d}\vec\xi {\rm d} \ln\forb} = \frac{{\rm d}^3 N_{\rm i}}{{\rm d}z {\rm d}\vec\xi {\rm d} \tau_{\rm c}}\frac{{\rm d}\tau_{\rm c}}{{\rm d}\ln\forb}.
\end{equation}

The quantity $\frac{{\rm d}^3 N_{\rm i}}{{\rm d}z {\rm d}\vec\xi {\rm d} \tau_{\rm c}}(z, \vec\xi, \tau_{\rm c})$ is the rate of inspiralling binaries in redshift layer $z$ with parameters $\vec \xi$ that reach a time to coalescence of $\tau_{\rm c}$. As a result, if one assumes that the population of inspiralling binaries is in a steady state on time scales comparable with the merger timescale of MBHBs emitting in the PTA band, at redshift $z$, this quantity does not depend on $\tau_{\rm c}$, such that $\frac{{\rm d}^3 N_{\rm i}}{{\rm d}z {\rm d}\vec\xi {\rm d} \tau_{\rm c}}(z, \vec\xi, \tau_{\rm c}) = \frac{{\rm d}^3 N_{\rm i}}{{\rm d}z {\rm d}\vec\xi {\rm d} \tau_{\rm c}}(z, \vec\xi, 0)$, which is exactly the rate of merging SMBHBs at redshift $z$, $\frac{{\rm d}^2 N_{\rm m}}{{\rm d}\vec\xi {\rm d}\tau_{\rm c}}$. Hence, one can write
\begin{align}
    \frac{{\rm d}^3 N_{\rm i}}{{\rm d}z {\rm d}\vec\xi {\rm d} \ln\forb} &= \frac{{\rm d}^3 N_{\rm m}}{{\rm d}z {\rm d}\vec\xi {\rm d} \tau_{\rm c}}\frac{{\rm d}\tau_{\rm c}}{{\rm d}\ln\forb}\\
    &= \frac{{\rm d}^2 n}{{\rm d}\vec\xi {\rm d}\tau_{\rm c}} \frac{{\rm d}V_{\rm c}}{{\rm d}z} \frac{{\rm d}\tau_{\rm c}}{{\rm d}\ln\forb}\\
    &= \frac{{\rm d}^2 n}{{\rm d}\vec\xi {\rm d}z} \frac{{\rm d}z}{{\rm d}\tau_{\rm c}}\frac{{\rm d}\tau_{\rm c}}{{\rm d}\ln\forb} \frac{{\rm d}V_{\rm c}}{{\rm d}z},
\end{align}
where in the last equality we converted the measure of time from cosmological time to redshift $z$. 
 The distribution $\frac{{\rm d}^2 n}{{\rm d}z{\rm d}\vec\xi}$ obtained from cosmological simulations is discretized and can be written as a sum over the simulation mergers $j$, $\frac{{\rm d}^2 n}{{\rm d}z{\rm d}\vec\xi} = \sum_j \frac{{\rm d}^2 n^{(j)}}{{\rm d}z{\rm d}\vec\xi} = \frac{1}{\Vsim}\sum_j \delta(z-z_j) \delta(\vec\xi - \vec\xi_j)$, where $\Vsim$ is the comoving volume of the simulation. 

As a result, in a realistic Universe realization, for a given merger $j$, a given observer would, on average, observe a number $\langle N_{\rm i}^{(j)} \rangle (\forb)$ of such a binary with an orbital frequency in the range $[\ln\forb - \Delta \ln\forb /2, \ln\forb + \Delta \ln\forb /2]$, namely,
\begin{align}
    \langle N_{\rm i}^{(j)} \rangle (\forb) &= \int {\rm d}z {\rm d}\vec\xi \int_{\ln\forb - \Delta \ln\forb /2}^{\ln\forb + \Delta \ln\forb /2}{\rm d}\ln\forb \frac{{\rm d}^2 n^{(j)}}{{\rm d}\vec\xi {\rm d}z} \frac{{\rm d}z}{{\rm d}\tau_{\rm c}}\frac{{\rm d}\tau_{\rm c}}{{\rm d}\ln\forb} \frac{{\rm d}V_{\rm c}}{{\rm d}z} ,\\
    &= \int {\rm d}z {\rm d}\vec\xi  \frac{{\rm d}^2 n^{(j)}}{{\rm d}\vec\xi {\rm d}z} \frac{{\rm d}z}{{\rm d}\tau_{\rm c}} \frac{{\rm d}V_{\rm c}}{{\rm d}z} \int_{\ln\forb - \Delta \ln\forb /2}^{\ln\forb + \Delta \ln\forb /2}{\rm d}\ln\forb \frac{{\rm d}\tau_{\rm c}}{{\rm d}\ln\forb} ,\\
    &= \frac{1}{\Vsim} \left[\frac{{\rm d}z}{{\rm d}\tau_{\rm c}} \frac{{\rm d}V_{\rm c}}{{\rm d}z}\right]_{z_j} \Delta \tau_{\rm c}^{(j)}(\forb).
\end{align}
Finally, at the end of this calculation, we recover Eq. \ref{eq: mean occupation number}.

\section{\label{sec: appendix fasten rf sum} Optimization of the GWB computation}

To have a realistic estimate of timing residuals induced by the GWB from a population of inspiralling MBHBs, we must sum up the contribution from each individual binary as presented in Sect.~\ref{sec: population approach}. 

Within our framework, for each binary $j$ of the $\hagn$ catalog ($\sim 35.000$), we have to model an average number of binaries, $\langle N_{\rm i}^{(j)} \rangle (\forb^{(k)})$, per orbital frequency bin $\forb^{(k)}$. In our study, we used $125$ such orbital frequency bins for the circular population. In total, this requires the computation of approximately $100$ million binaries per Universe realization. This is computationally prohibitive if one aims to compute it for $2\,000$ Universe realizations in a reasonable amount of time.

Thus, we used some approximations in order to speed up the GWB timing residuals computation.
First, we consider that for each of the catalog merger $j$, each of the $\langle N_{\rm i}^{(j)} \rangle (\forb^{(k)})$ binaries has the exact same orbital frequency $\forb^{(k)}$.
Then, we consider two cases: (i) if the number of binaries drawn from the Poisson distribution with the mean $\langle N_{\rm i}^{(j)} \rangle (\forb^{(k)})$ is smaller than a threshold number $N_{\rm i}^{\rm (max)} = 80$, we indeed compute each individual waveform, assigning random orbital parameters and sky location to each source.  The choice of $N_{\rm i}^{\rm (max)}$ results from a compromise between selecting a low value to reduce computational cost and a high value to ensure a valid approximation of the random walk described next.

(ii) If the number of binaries to model is greater than $N_{\rm i}^{\rm (max)}$, we can use the following approximation.
We start by introducing $N_{\rm i}^{(j)} (\forb^{(k)})$ ($> N_{\rm i}^{\rm (max)}$ here), which is the number of binaries similar to the catalog binary, $j$, and orbiting at $\forb^{(k)}$ that we have to model for a given Universe realization. Using Eq. \ref{eq: r(f) ET finite duration}, for each individual binary contribution $\tilde{r}_{\rm GW}^{(l)}(f)$, it is straightforward to see that their total contribution to the GWB timing residuals can be written as
\begin{equation}
    \label{eq: appendix sum of large N}
    \sum_{l=1}^{N_{\rm i}^{(j)} \left(\forb^{(k)}\right)} \tilde{r}_{\rm GW}^{(l)}(f) 
        = R_0^{(j)}\left(\forb^{(k)}\right) \left[ 
        \mathcal{S}_{j,k}\delta_T\left(f - \fgw^{(j,k)}\right) + \mathcal{S}_{j,k}^* \delta_T\left(f + \fgw^{(j,k)}\right) \right],
\end{equation}
where we used the fact that all the binaries $l$ have the same binary parameters and thus have the same residuals amplitude $R_0^{(j)}\left(\forb^{(k)}\right)$ and observer GW frequency $\fgw^{(j,k)}= \frac{2\forb^{(k)}}{1 + z_j}$ (recall that we apply here the population approach to the circular ensemble only). 
As a result, the sum over binaries is included in the term
\begin{equation}
    \label{eq: appendix def of big S}
    \mathcal{S}_{j,k} = \sum_{l=1}^{N_{\rm i}^{(j)}\left(\forb^{(k)}\right)} G_{\hat k_l} e^{i\phi_{0,l}}.
\end{equation}
To avoid computing this potentially huge sum for all catalog mergers $j$ and orbital frequency bin $k$, it is useful to rewrite the sum as
\begin{equation}
    \label{eq: appendix approximating large sum}
    \mathcal{S}_{j,k} = \sum_{l=1}^{N_{\rm i}^{\rm (max)}} G_{\hat k_l} e^{i\phi_{0,l}} \times \left(
        1 + \frac{\sum_{l=N_{\rm i}^{\rm (max)}+1}^{N_{\rm i}^{(j)}\left(\forb^{(k)}\right)} G_{\hat k_l} e^{i\phi_{0,l}}}{\sum_{l=1}^{N_{\rm i}^{\rm (max)}} G_{\hat k_l} e^{i\phi_{0,l}}}
    \right).
\end{equation}
The key idea is to compute a fast estimate of the ratio of complex numbers appearing in the parentheses on the right-hand side.
To achieve this, we approximate both sums as random walks in the complex plane, where each step has a random phase and an amplitude following the distribution of $\left|G_{\hat{k}}(\psi, \cos \iota)\right|$.
Here, we assume a uniform distribution for the inclination angle, polarization angle, and binary sky location $\hat k$. 
The modulus of the ratio can thus be estimated by the ratio of the sums of the modulus standard deviations (their mean being zero). 
Following \citet{2021CSF...14510790M}, it gives
\begin{align}
    \label{eq: appendix approx of ratio module}
    \left|\frac{\sum_{l=N_{\rm i}^{\rm (max)}+1}^{N_{\rm i}^{(j)}(\forb^{(k)})} G_{\hat k_l} e^{i\phi_{0,l}}}{\sum_{l=1}^{N_{\rm i}^{\rm (max)}} G_{\hat k_l} e^{i\phi_{0,l}}}\right|
    &\sim \frac{\sqrt{\left(N_{\rm i}^{(j)}(\forb^{(k)}) - N_{\rm i}^{\rm (max)}\right)\langle\left|G_{\hat{k}}\right|^2\rangle}}{\sqrt{N_{\rm i}^{\rm (max)}\langle\left|G_{\hat{k}}\right|^2\rangle}} \\
    &\sim \sqrt{\frac{N_{\rm i}^{(j)}(\forb^{(k)})}{N_{\rm i}^{\rm (max)}} - 1} \: .
\end{align}
We can then assign a random complex phase to the modulus of Eq. \ref{eq: appendix approx of ratio module}, to estimate the ratio in Eq. \ref{eq: appendix approximating large sum}. 
This way, the cost of the GWB timing residuals computation is greatly reduced without suffering much in accuracy, allowing us to compute thousands of Universe realizations.

We note that instead of using this approximation, one can directly estimate Eq. \ref{eq: appendix def of big S} in the case \( N_{\rm i}^{(j)}\left(\forb^{(k)}\right) \geq N_{\rm i}^{\rm (max)} \) by invoking the central limit theorem.  
The total sum can be estimated by drawing two real numbers, \( \hat{X}_1 \) and \( \hat{X}_2 \), from two independent and identically distributed Gaussian variables  
\( X_1, X_2 \sim \mathcal{N}\left(0, \sqrt{N_{\rm i}^{(j)}\left(\forb^{(k)}\right) \times \left\langle\left|G_{\hat{k}}\right|^2\right\rangle /2}\right) \),  
and computing \( \hat{X}_1 + i \hat{X}_2 \).
For the isotropic/uniform prior that we use here for $\hat k, \cos \iota$ and $\psi$, we find $\langle\left|G_{\hat{k}}\right|^2\rangle = 1/30$.
We verified that the two methods yield highly consistent results, with deviations at the percentage level.

\section{\label{EM_bck} EM properties of MBHBs}

In complement to Fig. \ref{Fig:EMflux}, where we show the electromagnetic properties of CW candidates, we report in Fig. \ref{Fig:EMflux_bck} an analogous figure for the MBHBs dominating the background. We model and analyze only the MBH binaries that are in the $\hagn$ lightcone \citep{laigle19}. The sources are overall fainter than CW candidates, because they are located at higher redshift and characterized by lighter MBHs and galaxies (cf. Fig. \ref{Fig:bck_vs_ind}), but the results are in broad agreement, in the sense that generally the host galaxies outshine the MBHB emission at optical/NIR wavelengths (cf. Fig. \ref{Fig:EMflux}, left) while for radiatively efficient accretors the MBHB emission is comparable to or larger than the host galaxy's in radio and X-rays. Bolometric luminosities tend to remain also in this case below quasars' typical luminosities. 

In Fig. \ref{Fig:EMflux_ratio} we show the ratio of AGN and galaxy brightness for CW candidates.  Results are qualitatively similar for binaries dominating the background.  Red filled dots assume radiatively efficient accretion at all Eddington ratios, while empty red dots show a correction for radiatively inefficient emission for $f_{\rm Edd}<10^{-3}$. 
In the optical and NIR bands (B, V, and K), the host galaxies are always brighter than  MBHB-powered AGNs for radiatively inefficient MBHs, and only a few of the efficient accretors have magnitudes comparable to or, in a very few cases, brighter than, their hosts {with a preference for longer wavelengths (e.g., K), similarly to \citet{2025ApJ...990...46V}.} In radio and X-rays all efficient accretors ($f_{\rm Edd}>10^{-3}$) are brighter than the host, while the inefficient accretors are always outshined by the stellar emission from the host galaxy in X-rays when including a correction for radiative efficiency. In radio, the fraction of inefficient accretors brighter than the hosts is 61\%.

   \begin{figure}[h!]
   \centering
   \includegraphics[width=0.49\textwidth]{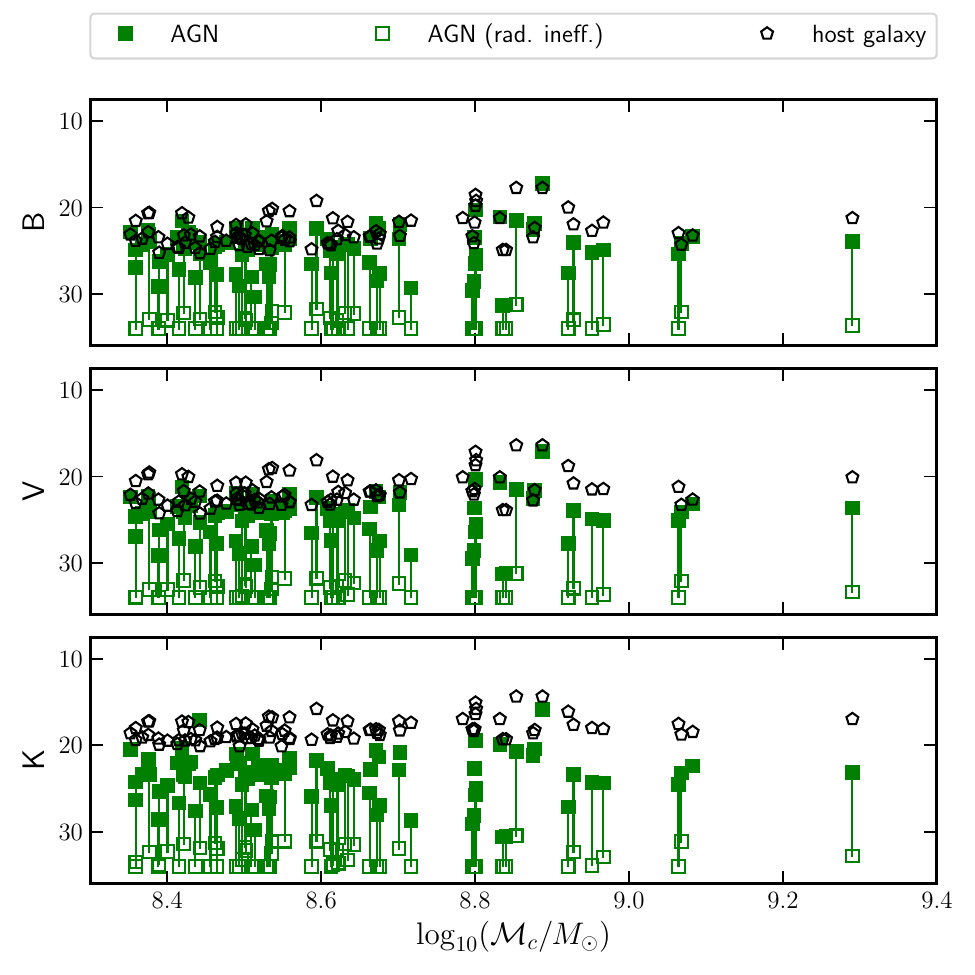}
   \includegraphics[width=0.49\textwidth]{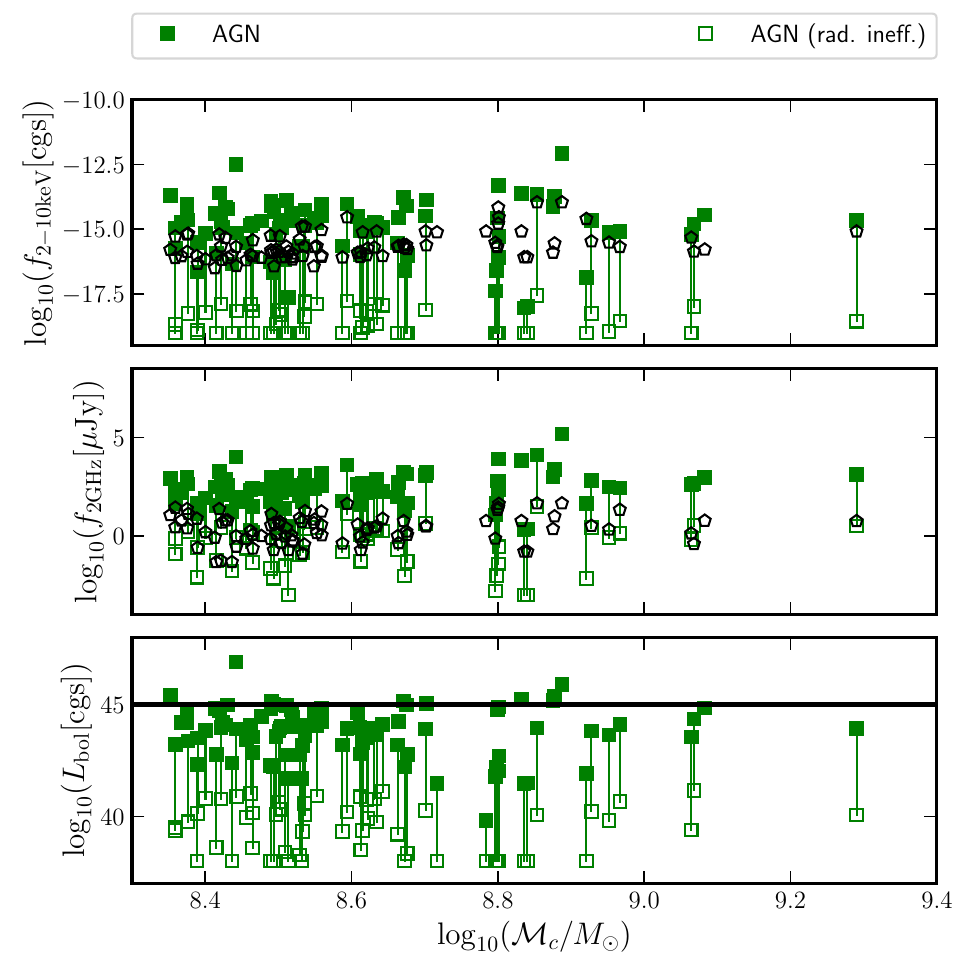}
   \caption{Electromagnetic properties of MBHBs and their host galaxies for MBHBs contributing 90\% of the background. The left panels show B, V, and K magnitudes; the right panels radio flux at 2\,GHz, X-ray flux at [2-10]\,keV and bolometric luminosity, where the horizontal line marks $L_{\rm bol}=10^{45}\, \rm{erg}\, {\rm s^{-1}} $, as a typical value for quasar luminosities. Sources with magnitude fainter than 34 are set at 34, and similarly sources with X-ray flux $<10^{-19}$ and radio flux $<10^{-3}\,\rm {\mu Jy}$ are set at these values for clarity. For MBHs with $f_{\rm Edd}<10^{-3}$ we show an upper limit to the brightness ignoring the correction to radiative efficiency (filled squares) and a lower limit including the correction (empty squares connected to the filled squares).}
              \label{Fig:EMflux_bck}%
    \end{figure}

 \begin{figure*}[h!]
   \centering
   \includegraphics[width=0.49\textwidth]{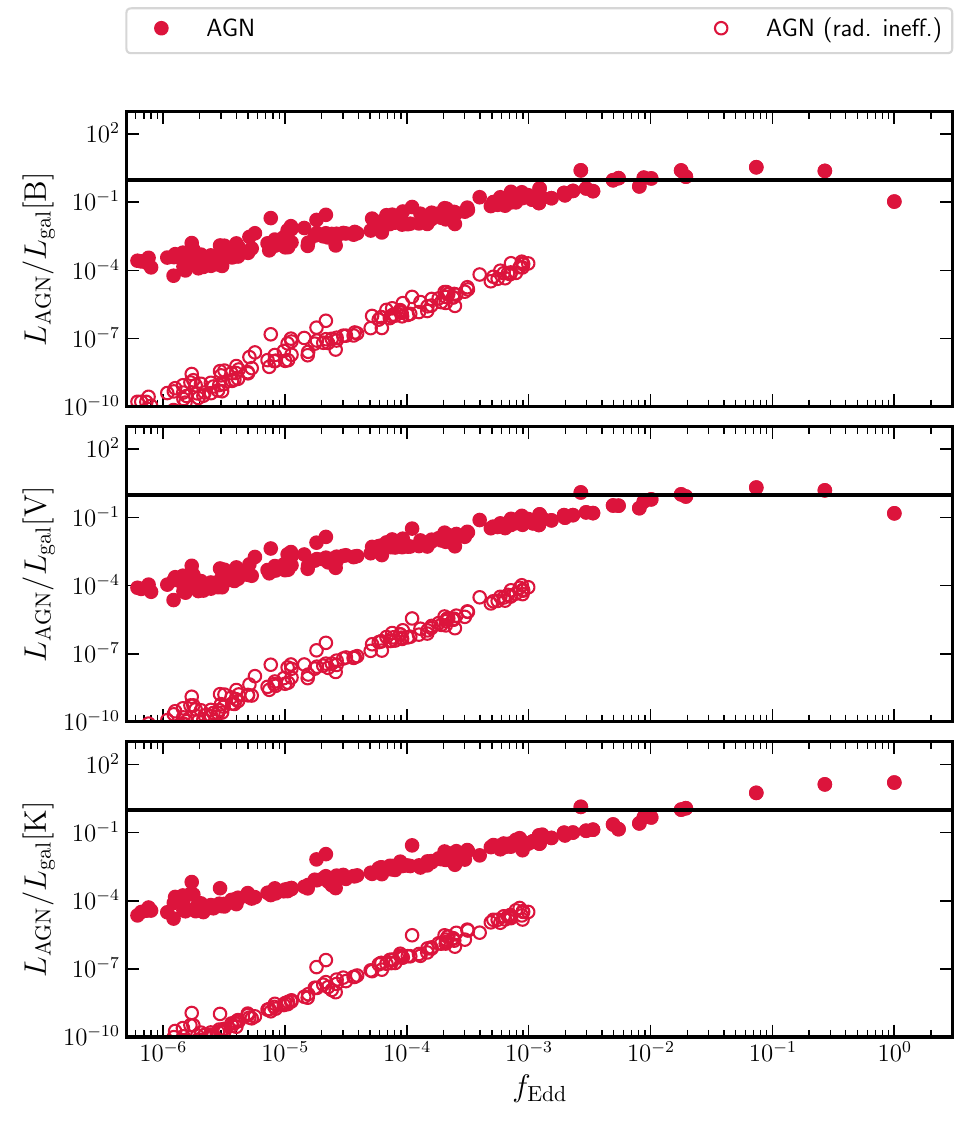}
   \includegraphics[width=0.49\textwidth]{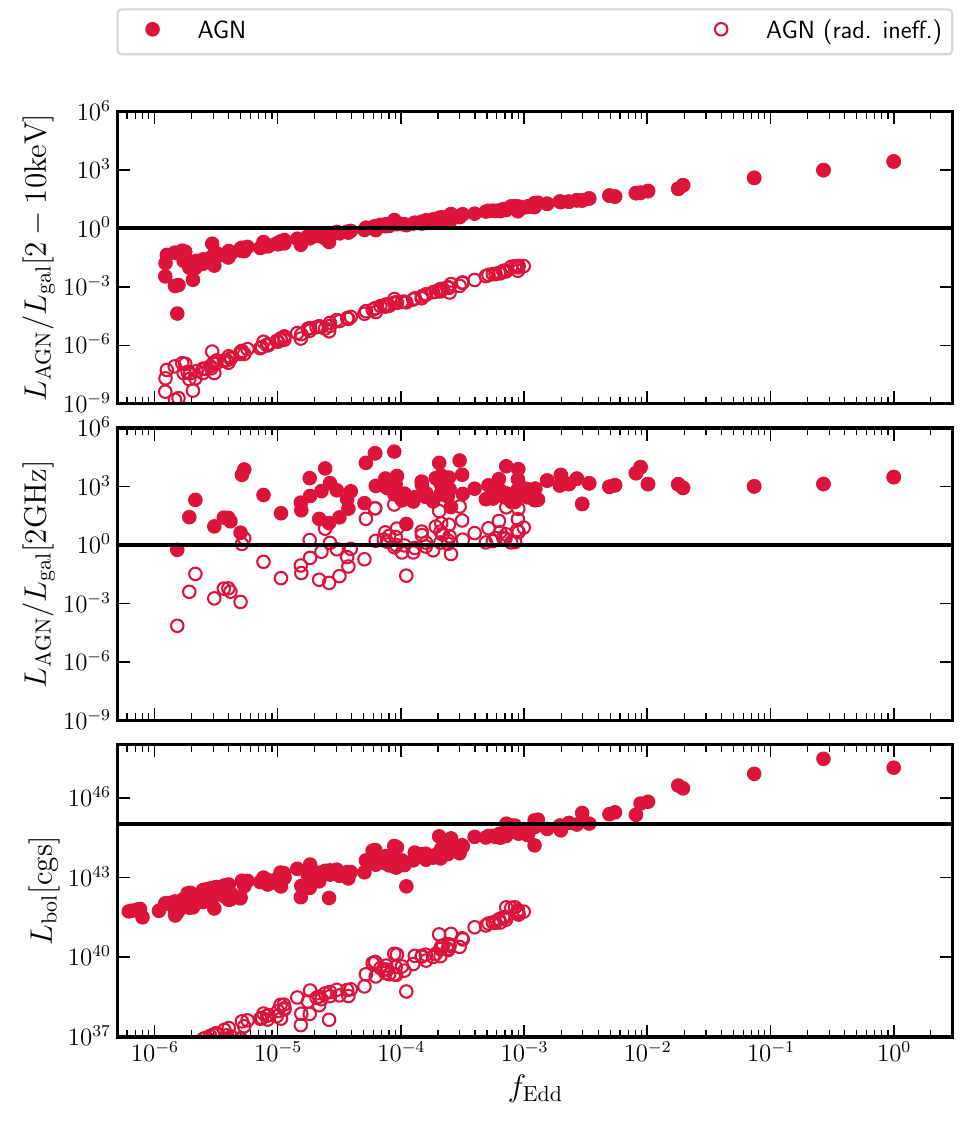}
   \caption{Ratio of AGN and galaxy luminosity for CW candidates. The left panels show the ratios in B, V, and K; the top and middle right panels the ratio in X-rays at [2-10] keV, and in radio at 2~GHz; the horizontal lines mark $L_{\rm AGN}/L_{\rm gal}=1$. The bottom-right panel shows the bolometric luminosity; the horizontal line marks $L_{\rm bol}=10^{45}\, \rm{erg}\, {\rm s^{-1}}$. For MBHs with $f_{\rm Edd}<10^{-3}$ we show an upper limit to the brightness ignoring the correction to radiative efficiency (filled dots) and a lower limit including the correction (empty dots).}
              \label{Fig:EMflux_ratio}
    \end{figure*}

\end{appendix}

\end{document}